\documentclass[prd,aps,
nofootinbib,
floatfix,
superscriptaddress]{revtex4}
\usepackage{graphicx}
\usepackage{epsfig}
\usepackage{rotating}
\usepackage{amssymb}
\usepackage{subfigure}
\usepackage{dsfont}
\usepackage{psfrag}
\usepackage{amsmath,euscript,array,mathrsfs}
\usepackage{bbold}
\usepackage{bm}
\usepackage{multirow}
\usepackage{dcolumn}
\usepackage{xcolor}

\newcommand{\beq}{\begin{equation}}
\newcommand{\eeq}{\end{equation}}
\newcommand{\beqs}{\begin{eqnarray}}
\newcommand{\eeqs}{\end{eqnarray}}

\newcommand{\gsim}{\mathrel{\raisebox{-
.6ex}{$\stackrel{\textstyle>}{\sim}$}}}
\newcommand{\Tr}{{\rm Tr}}

\def\hbar{\hspace{0pt}\raisebox{1pt}{$-$} \hspace{-7pt} h}

\newcommand{\be}{\begin{equation}}
\newcommand{\ee}{\end{equation}}

\newcommand{\bea}{\begin{eqnarray}}
\newcommand{\eea}{\end{eqnarray}}
\newcommand{\nn}{\nonumber}

\def\lbldef#1#2{\expandafter\gdef\csname #1\endcsname {#2}}

\def\href#1#2{#2}

\newcommand{\ber}{\begin{eqnarray}}
\newcommand{\eer}{\end{eqnarray}}

\newcommand{\beqar}{\begin{eqnarray}}

\newcommand{\eeqar}{\end{eqnarray}}

\newcommand{\dsl}
  {\kern.06em\hbox{\raise.15ex\hbox{$/$}\kern-.56em\hbox{$\partial$}}}

\newcommand{\eeqarr}{\end{eqnarray}}
\newcommand{\ZZ}{{\rm \kern 0.275em Z \kern -0.92em Z}\;}

\def\CC{{\mathchoice
{\rm C\mkern-8mu\vrule height1.45ex depth-.05ex
width.05em\mkern9mu\kern-.05em}
{\rm C\mkern-8mu\vrule height1.45ex depth-.05ex
width.05em\mkern9mu\kern-.05em}
{\rm C\mkern-8mu\vrule height1ex depth-.07ex
width.035em\mkern9mu\kern-.035em}
{\rm C\mkern-8mu\vrule height.65ex depth-.1ex
width.025em\mkern8mu\kern-.025em}}}

\def\RR{{\rm I\kern-1.6pt {\rm R}}}

\def\ZZ{{\rm Z}\kern-3.8pt {\rm Z} \kern2pt}
\def\IB{\relax{\rm I\kern-.18em B}}
\def\ID{\relax{\rm I\kern-.18em D}}
\def\II{\relax{\rm I\kern-.18em I}}
\def\IP{\relax{\rm I\kern-.18em P}}

\newcommand{\bear}{\begin{eqnarray}}
\newcommand{\eear}{\end{eqnarray}}

\def\6{\partial}

\def\bea{\begin{eqnarray}}
\def\eea{\end{eqnarray}}

\def\beqx{\begin{displaymath}}
\def\eeqx{\end{displaymath}}

\newcommand{\bmat}{\left(\begin{array}}
\newcommand{\emat}{\end{array}\right)}

\newcommand\Sec[1]{Section~\ref{Sec:#1}}
\newcommand\App[1]{Appendix~\ref{Sec:#1}}
\newcommand\Tab[1]{Table~\ref{tab:#1}}
\newcommand\Fig[1]{Fig.~\ref{fig:#1}}
\newcommand\Eq[1]{Eq.~(\ref{eq:#1})}

\def\bo{{\raise-.3ex\hbox{\large$\Box$}}}               
\def\face{{\raise.2ex\hbox{$\displaystyle \bigodot$}\mskip-2.2mu \llap {$\ddot
        \smile$}}}                                   
\def\>{\rangle}                                      
\def\<{\langle}                                      

\def\leftrightarrowfill{$\mathsurround=0pt \mathord\leftarrow \mkern-6mu
        \cleaders\hbox{$\mkern-2mu \mathord- \mkern-2mu$}\hfill
        \mkern-6mu \mathord\rightarrow$}        
\def\dvec#1{\vbox{\ialign{##\crcr
        \leftrightarrowfill\crcr\noalign{\kern-1pt\nointerlineskip}
        $\hfil\displaystyle{#1}\hfil$\crcr}}}           
\def\Tr{{\rm Tr \,}}                                    

\def\-{\hphantom{-}}

\begin{document}

\preprint{PNUTP-21/A01}

\title{Lattice studies of the  $Sp(4)$ gauge theory with 
two fundamental and three antisymmetric Dirac fermions}

\vspace{6mm}

\begin{abstract}
We consider the $Sp(4)$ gauge theory coupled to $N_f=2$
 fundamental and $n_f=3$ antisymmetric flavours of Dirac fermions 
in four dimensions. 
This theory serves as the microscopic origin for composite Higgs models
 with $SU(4)/Sp(4)$ coset, supplemented by
partial top compositeness. 
We study numerically its lattice realisation, and couple
 the fundamental  plaquette action
to Wilson-Dirac fermions in mixed representations, 
by adopting a (rational) hybrid Monte Carlo method, to 
perform non-trivial tests of the 
properties of the resulting lattice theory.

We find evidence of a surface (with boundaries) of first-order bulk phase transitions
in the three-dimensional space of bare parameters (one coupling and two masses).
Explicit evaluation of  the Dirac eigenvalues confirms the expected patterns of global symmetry breaking.
After investigating finite volume effects in the weak-coupling phase of the theory,
for the largest available lattice
we study the mass spectra of the lightest spin-0 and spin-1 flavoured mesons 
composed of fermions in each representation, and of
the lightest half-integer spin composite particle made of fermions in different representations---the chimera baryon. 
This work sets the stage for future systematical studies of the non-perturbative dynamics in phenomenologically
relevant regions of parameter space.
\end{abstract}

\author{Ed Bennett}
\email{e.j.bennett@swansea.ac.uk}
\affiliation{Swansea Academy of Advanced Computing, Swansea University,
Fabian Way, SA1 8EN Swansea, Wales, UK}

\author{Deog Ki Hong}
\email{dkhong@pusan.ac.kr}
\affiliation{Department of Physics, Pusan National University, Busan 46241, Korea}

\author{Ho Hsiao}
\email{thepaulxiao@gmail.com}
\affiliation{Institute of Physics, National Yang Ming Chiao Tung University, 1001 Ta-Hsueh 
Road, Hsinchu 30010, Taiwan}

\author{Jong-Wan Lee}
\email{jwlee823@pusan.ac.kr}
\affiliation{Department of Physics, Pusan National University, Busan 46241, Korea}
\affiliation{Extreme Physics Institute, Pusan National University, Busan 46241, Korea}

\author{C.-J. David Lin}
\email{dlin@nycu.edu.tw}
\affiliation{Institute of Physics, National Yang Ming Chiao Tung University, 1001 Ta-Hsueh 
Road, Hsinchu 30010, Taiwan}
\affiliation{Center for High Energy Physics, Chung-Yuan Christian University,
Chung-Li 32023, Taiwan}
\affiliation{Centre for Theoretical and Computational Physics, National Yang Ming Chiao Tung University, 1001 Ta-Hsueh Road, Hsinchu 30010, Taiwan}

\author{Biagio Lucini}
\email{b.lucini@swansea.ac.uk}
\affiliation{Department of Mathematics, Faculty  of Science and Engineering,
Swansea University, Fabian Way, SA1 8EN Swansea, Wales, UK}
\affiliation{Swansea Academy of Advanced Computing, Swansea University,
Fabian Way, SA1 8EN Swansea, Wales, UK}

\author{Michele Mesiti}
\email{michele.mesiti@swansea.ac.uk}
\affiliation{Swansea Academy of Advanced Computing, Swansea University,
Fabian Way, SA1 8EN Swansea, Wales, UK}

\author{Maurizio Piai}
\email{m.piai@swansea.ac.uk}
\affiliation{Department of Physics, Faculty  of Science and Engineering,
Swansea University, Singleton Park, SA2 8PP, Swansea, Wales, UK}

\author{Davide Vadacchino}
\email{davide.vadacchino@plymouth.ac.uk}
\affiliation{School of Mathematics and Hamilton Mathematics Institute, Trinity
College, Dublin 2, Ireland}
\affiliation{Centre for Mathematical Sciences, University of Plymouth, Plymouth, PL4 8AA, United Kingdom}

\date{\today}

\maketitle
\flushbottom
\tableofcontents



\section{Introduction}
\label{Sec:intro}

The Standard Model (SM) of particle physics is an astonishing achievement,
as it provides an outstanding 
wealth of correct predictions and (in selected cases) with uncommonly high   accuracy. 
Yet, it is unlikely to be 
the complete and final description of fundamental physics,
given, for example,  that it does not include gravity, that many of its interactions are not asymptotically safe
at short distances (the $U(1)_Y$ coupling, all the Yukawa couplings, and the scalar self-coupling
have positive beta function), that it does not provide a compelling explanation for dark matter,
for inflationary cosmology, and for the observed baryon asymmetry of our universe.
Hence, the theoretical and experimental search for new physics extending beyond the standard model (BSM)
is as active a field today as ever. 

It is a remarkable fact,  suggestive of promising new search directions, that both the two latest additions
to the SM spectrum of particles have properties somewhat unusual for---though
not inconsistent with---the  low energy effective field theory (EFT) paradigm,
according to which the SM would be accurate  only up to a new 
physics scale $\Lambda$, higher than the electroweak scale $v_W\simeq 246$ GeV.
The mass of the top quark ($m_t\sim 173$ GeV) is orders of magnitude larger than that of other fermions, 
which in the SM context implies that its Yukawa coupling is comparatively large---so much so that its
effects in radiative (quantum) corrections might be invoked as 
a possible cause for the vacuum instability that triggers electroweak symmetry breaking (EWSB).
Conversely, naive dimensional analysis (NDA) arguments suggest the mass $m_h$ of the Higgs boson
should be sensitive to $\Lambda$, which
indirect and direct searches at the Large Hadron Collider (LHC)
put in the multi-TeV range. But, experimentally,
 $m_h\simeq 125$ GeV~\cite{Aad:2012tfa,Chatrchyan:2012xdj}, 
leading to the little hierarchy $m_h \ll \Lambda$.
These two observations suggest that Higgs and top physics might 
be sensitive to new physics, and motivate  many proposals for
extensions of the standard model, including the one
we will focus on in the following.

This paper is inspired by the theoretical proposal in Ref.~\cite{Barnard:2013zea}, 
which postulates the existence a new, strongly-coupled fundamental theory with $Sp(4)$ gauge group,
interprets the SM Higgs-doublet fields 
in terms  of the composite pseudo-Nambu-Goldstone bosons (PNGBs)
describing the spontaneous breaking of an approximate $SU(4)$ symmetry 
(acting on $N_f=2$ Dirac fermions transforming on the fundamental representation of $Sp(4)$)
to its $Sp(4)$ subgroup,
and furthermore reinterprets the SM top quark as a partially composite object,
resulting from the mixing with composite fermions, dubbed
chimera baryons (the constituents of which are an admixture of fermions 
transforming in the fundamental and anisymmetric representation of $Sp(4)$).
In the rest of this introduction, we explain why this model is particularly interesting,
standing out in the
BSM literature. The body of the paper is devoted to reporting 
a set of lattice results demonstrating that our collaboration
has put in place and tested successfully all the lattice field theory tools that are necessary to 
perform a systematic, quantitative analysis of the non-perturbative features of 
this strongly-coupled theory.

The common feature to composite Higgs models (CHMs)
is that  scalar fields 
originate as PNGBs in the underlying dynamics~\cite{Kaplan:1983fs,Georgi:1984af,Dugan:1984hq}.
Symmetry arguments constrain their potential, suppressing  masses and couplings.
Reviews  can be found  in Refs.~\cite{Panico:2015jxa,Witzel:2019jbe,Cacciapaglia:2020kgq},
and it may be helpful to the reader to use the summary tables in Refs.~\cite{Ferretti:2013kya,Ferretti:2016upr,Cacciapaglia:2019bqz}.
A selection of interesting studies focusing on model-building, perturbative studies
 and phenomenological applications
includes Refs.~\cite{Katz:2005au,Barbieri:2007bh,Lodone:2008yy,Gripaios:2009pe,Mrazek:2011iu,
Marzocca:2012zn,Grojean:2013qca,Cacciapaglia:2014uja,Ferretti:2014qta,Arbey:2015exa,
Cacciapaglia:2015eqa,Feruglio:2016zvt,DeGrand:2016pgq,Fichet:2016xvs,Galloway:2016fuo,
Agugliaro:2016clv,Belyaev:2016ftv,Csaki:2017cep,
Chala:2017sjk,Golterman:2017vdj,Csaki:2017jby,Alanne:2017rrs,Alanne:2017ymh,Sannino:2017utc,
Alanne:2018wtp,Bizot:2018tds,Cai:2018tet,Agugliaro:2018vsu,
Cacciapaglia:2018avr,Gertov:2019yqo,Ayyar:2019exp,
Cacciapaglia:2019ixa,BuarqueFranzosi:2019eee,Cacciapaglia:2019dsq,
Dong:2020eqy,Cacciapaglia:2020vyf,
Cacciapaglia:2021uqh,Banerjee:2022izw}.
In these studies, EFT (and perturbative) arguments and guidance from the experiment 
are combined to constraint the strongly coupled dynamics,
but its detailed description is accessible only with non-perturbative 
instruments.
There is a rich literature on the topic coming from gauge-gravity dualities,
in the context of bottom-up holographic models~\cite{Agashe:2004rs,Contino:2006qr,Falkowski:2008fz},
with a recent resurgence of interest~\cite{Erdmenger:2020lvq,Erdmenger:2020flu,
Elander:2020nyd,Elander:2021bmt}, including a first attempt
at identifying a complete top-down model~\cite{Elander:2021kxk}.
Alternative ways to approach the dynamics have also been proposed in
Ref.~\cite{Bizot:2016zyu}.

Lattice field theory is the most direct, first principle 
way to approach  non-perturbative dynamics.
Detailed lattice studies of theories leading to symmetry breaking described by the 
 $SU(4)/Sp(4)\sim SO(6)/SO(5)$ coset have focused on the simplest $SU(2)$ gauge theories
 coupled to fundamental fermions~\cite{Hietanen:2014xca,
Detmold:2014kba,Arthur:2016dir,Arthur:2016ozw,Pica:2016zst,Lee:2017uvl,Drach:2017btk,Drach:2020wux,
Drach:2021uhl}, but these models cannot realise top compositeness.
Explorations of  $SU(4)$ gauge theories with multiple representations~\cite{Ayyar:2017qdf,
Ayyar:2018zuk, Ayyar:2018ppa,Ayyar:2018glg,Cossu:2019hse,Shamir:2021frg} aim at 
gathering non-perturbative information about Ferretti's $SU(5)/SO(5)$ 
model~\cite{Ferretti:2014qta}, though
the fermionic field contents do not match.
An alternative route to studying models yielding both composite Higgs and partial top compositeness
 has been proposed by Vecchi in Ref.~\cite{Vecchi:2015fma} (see also
 Refs.~\cite{Ma:2015gra,BuarqueFranzosi:2018eaj}), by exploiting the fact that in $SU(3)$ theories
 the antisymmetric representation is the conjugate of the fundamental,
 so that one can use the lattice information made available over the years by the 
 LatKMI~\cite{Aoki:2014oha,Aoki:2016wnc} and LSD~\cite{Appelquist:2016viq,
 Gasbarro:2017fmi,LSD:2018inr,Appelquist:2018yqe,LatticeStrongDynamicsLSD:2021gmp}  collaborations
to test  the viability of CHMs based on the  $SU(N_f)\times SU(N_f)/SU(N_f)$ cosets
(as done explicitly  in Ref.~\cite{Appelquist:2020bqj}).

Our collaboration announced in 2017 the intention to carry out a systematic study of 
confining, lattice  gauge theories in the $Sp(2N)$ sequence, coupled to various
types of fermion matter fields~\cite{Bennett:2017kga}. We have published results 
for the $Sp(4)$ gauge theory coupled to $N_f=2$
dynamical fermions transforming in the fundamental representation 
of the group~\cite{Lee:2018ztv,Bennett:2019jzz}, and for
quenched fermions in mixed (fundamental and antisymmetric)
 representations~\cite{Bennett:2019cxd}.
We have calculated  the spectra of  glueballs and strings in 
the $Sp(2N)$ Yang-Mills theories~\cite{Bennett:2020hqd,Bennett:2020qtj}---reaching far beyond
the pioneering lattice work
for $N=2,3$ in Ref.~\cite{Holland:2003kg}.
Besides the ambitious  applications in the CHM context,
an equally important physics motivation relates to models of dark matter with 
strong-coupling origin~\cite{Hochberg:2014dra,Hochberg:2014kqa,Hochberg:2015vrg} 
(see also the more recent Refs.~\cite{Bernal:2017mqb,Berlin:2018tvf,Bernal:2019uqr,Cai:2020njb,
Tsai:2020vpi,Maas:2021gbf,Zierler:2021cfa,Kulkarni:2022bvh}).
On more general grounds, we aim at putting our numerical understanding of these theories on
a level comparable to that achieved for  the $SU(N_c)$ theories,  in
reference to the approach to the large-$N_c$ limit~\cite{Lucini:2001ej,
Lucini:2004my,Lucini:2010nv,Lucini:2012gg,Athenodorou:2015nba,Lau:2017aom,
Hong:2017suj,Yamanaka:2021xqh,Athenodorou:2021qvs,Hernandez:2020tbc},
but also for the purposes of  determining the boundaries of the
conformal window~\cite{Sannino:2009aw,Ryttov:2017dhd,Kim:2020yvr,
Lee:2020ihn}, and of testing their EFT description~\cite{Appelquist:1999dq}.
We will deliver further publications on the topology of 
$Sp(2N)$ gauge theories, and their  quenched meson spectra, as well as on the
(partially quenched) dynamical theory with $n_f=3$ dynamical antisymmetric fermions---preliminary results 
have been presented in Ref.~\cite{Lucini:2021xke,Bennett:2021mbw}.

Our  diversified lattice strategy combines
exploratory as well as precision studies, moving in  different directions in 
the space of $Sp(2N)$ theories. Aside from the aforementioned desire
to explore other applications of these theories,
even when we restrict attention to the CHM context, there are still two good reasons
to adopt this gradual approach.
First, the CHM candidate proposed in Ref.~\cite{Barnard:2013zea} is rather unusual, 
and there are no reference results in the literature
for comparable theories.
 It is hence important to build a whole portfolio of related theories,
against which we can benchmark our results.
The pragmatic reason why this benchmarking is needed, 
is that lattice studies with fermions in mixed representations
 are technically challenging and resource intensive.
Most of the existing, publicly available lattice 
 codes developed for other purposes
 do not implement multiple dynamical representations---we mentioned above some very recent examples
 for the $SU(4)$ theories.
Even after the code becomes available, and after testing the correctness of the
 behaviour of the algorithms used in the calculations---as we shall demonstrate shortly---one still must
explore the phase space of the lattice theory.
In our case, this is controlled by three bare parameters (the gauge coupling and the two fermion masses),
 besides the lattice size, making the mapping of phase transitions quite non-trivial.
 Finally,  the number of elementary degrees of freedom of the $Sp(4)$ theory with
  $N_f=2$ and $n_f=3$ is large, and hence,
  while the theory is still asymptotically free, 
   one expects slow running of the couplings, and possibly the emergence of
  large anomalous dimensions, making it more challenging to  characterise the theory.
We will
  provide evidence of the fact that we can address  all of these challenges, 
  and we can start production
 of  ensembles giving access to physically relevant regions of parameter space.

The paper is organised as follows.
We start by presenting essential  information about the continuum theory in Sect.~\ref{Sec:model}.
This exercise makes the paper self-contained, and allows us to connect to potential applications, prominently to CHMs.
We then describe the lattice theory in Sect.~\ref{Sec:latticesetup}, by providing enough details about the
algorithms we use to allow reproducibility of our results. Sect.~\ref{Sec:observables} defines the main
 observable quantities we use to probe
  our lattice theory. Out numerical results  for these observables
   are presented in Sect.~\ref{Sec:results}. We conclude with the summary and outlook in Sect.~\ref{Sec:outlook}.
We supplement the paper by \App{A}, detailing some of the conventions we adopted throughout the paper,
\App{B}, which displays an additional 
technical test we performed on 2-point functions
involving chimera baryon operators,
 and \App{C}, containing
 summary tables characterising the 
 numerical data used for the analysis.

\section{The model}
\label{Sec:model}

The model we study has been  proposed in Ref.~\cite{Barnard:2013zea}.
We  adapt and improve the conventions in Ref.~\cite{Bennett:2019cxd},  
 to make both the presentation in the paper self-contained
and the notation
precise enough to make contact with the lattice.
 We hence review the field content and symmetries of the 
continuum theory defining its short-distance dynamics,
and review  its low-energy EFT description.
We supplement  the list of interpolating operators
used for the study of mesons (already published elsewhere)
by presenting original material  
detailing the operators used for  chimera baryons.

\subsection{Short distance dynamics}
\label{Sec:short}

The $Sp(4)$ gauge theory has field content consisting of $N_f=2$ Dirac fermions $Q^{i\,a}$ transforming
in the fundamental, (f)  representation of the gauge group, and $n_f=3$ Dirac fermions $\Psi^{k\,ab}$
transforming in the 2-index antisymmetric, (as) representation. Here and in the following,
 $a,\,b=1,\,\cdots,\,4$ 
denote color indices, while $i=1,\,2$ and $k=1,\,2,\,3$ denote flavour indices.
\begin{table}
\caption{Field content of the microscopic theory.
$Sp(4)$ is the gauge group, and $SU(4)\times SU(6)$ (ignoring Abelian factors) the global one. 
The elementary fields $V_{\mu}$ are gauge bosons, and $q$ and $\psi$ are 2-component spinors,
described in the main text.
}
\begin{center}
\begin{tabular}{|c|c|c|c|c|}
\hline
{\rm ~~~Fields~~~} & ~~~$Sp(4)$~~~  &  ~~~$SU(4)$~~~ & ~~~$SU(6)$~~~ \cr
\hline
$V_{\mu}$ & $10$ & $1$ & $1$  \cr
$q$ & $4$ & $4$ & $1$ \cr
$\psi$ & $5$ & $1$ & $6$\cr
\hline
\end{tabular}
\end{center}
\label{Fig:fields}
\end{table}
The Lagrangian density is 
\beqs
{\cal L}&=& -\frac{1}{2} \Tr V_{\mu\nu} V^{\mu\nu}
\,+\,\frac{1}{2}\sum_{i=1}^{2}\left(i\overline{Q^{i}}_a 
\gamma^{\mu}\left(D_{\mu} Q^i\right)^a
\,-\,i\overline{D_{\mu}Q^{i}}_a \gamma^{\mu}Q^{i\,a}\right)\,
-\,m^{f}\sum_{i=1}^{2}\overline{Q^i}_a Q^{i\,a}+\nonumber\\
&&
\,+\,\frac{1}{2}\sum_{k=1}^{3}\left(i\overline{\Psi^{k}}_{ab} \gamma^{\mu}\left(D_{\mu} \Psi^k\right)^{ab}
\,-\,i\overline{D_{\mu}\Psi^{k}}_{ab} \gamma^{\mu}\Psi^{k\,ab}\right)\,
-\,m^{as}\sum_{k=1}^{3}\overline{\Psi^k}_{ab} \Psi^{k\,ab}\,,
\label{eq:lagrangian}
\eeqs
where summations over color and Lorentz indices are understood,
while spinor indices are implicit. 
$m^{f}$ and $m^{as}$ are the (degenerate) masses of $Q$ and $\Psi$, respectively. 
The covariant derivatives are defined by making use of the transformation 
properties under the action of an element $U$ of the 
$Sp(4)$ gauge group---$Q\rightarrow U Q$ and $\Psi \rightarrow U \Psi U^{\mathrm{T}}$---so that
\beqs
V_{\mu\nu}&\equiv& \partial_{\mu}V_{\nu}-\partial_{\nu}V_{\mu} + i g \left[V_{\mu}\,,\,V_{\nu}\right]\,,\\
D_{\mu} Q^i&=& \partial_{\mu} Q^i \,+\,i g V_{\mu} Q^{i}\,,\\ 
D_{\mu} \Psi^k&=& \partial_{\mu} \Psi^k \,+\,i g V_{\mu} \Psi^{k}\,+\,i g \Psi^{k} V_{\mu}^{\mathrm{T}}\,,
\eeqs
where $g$ is the gauge coupling.

Because of the pseudo-real nature of the representations of $Sp(4)$, it is convenient to 
split each Dirac fermion into 2-component spinors $q^{m\,a}$ and $\psi^{n\,ab}$, 
for the (f) and (as) representation, respectively. The flavour indices $m=1,\,\cdots,\,4$ and
$n=1,\,\cdots,\,6$ denote the components of a fundamental representation of the global
symmetry groups $SU(4)$ acting on $q^{m\,a}$ and $SU(6)$ acting on $\psi^{n\,ab}$.
Here and in the following we ignore the $U(1)$ factors in the
symmetry group.
The field content is summarised in Table~\ref{Fig:fields}. 
To make the symmetries manifest, we borrow Eqs.~(5) and~(6) from Ref.~\cite{Bennett:2019cxd}, and introduce
the symplectic matrix $\Omega$ and the symmetric matrix $\omega$, that are defined by
\beqs
\Omega&=&\Omega_{mn}\,=\,\Omega^{mn}\,\equiv\,
\left(\begin{array}{cccc}
0 & 0 & 1 & 0\cr
0 & 0 & 0 & 1\cr
-1 & 0 & 0 & 0\cr
0 & -1 & 0 & 0\cr
\end{array}\right)\,,
\label{Eq:symplecticmatrix}
~~~~
\omega\,=\,\omega_{mn}\,=\,\omega^{mn}\,\equiv\,
\left(\begin{array}{cccccc}
0 & 0 & 0 & 1 & 0 & 0 \cr
0 & 0 & 0 & 0 & 1 & 0 \cr
0 & 0 & 0 & 0 & 0 & 1  \cr
1 & 0 & 0 & 0 & 0 & 0 \cr
0 & 1 & 0 & 0 & 0 & 0 \cr
0 & 0 & 1 & 0 & 0 & 0 \cr
\end{array}\right)\,.
\eeqs
The two-component notation is related as follows to the four component notation:
\beqs
Q^{i\,a}&=&\left(
\begin{array}{c}
q^{i\,a} \cr \Omega^{ab}(-\tilde{C}q^{i+2\,\ast})_b
\end{array}
\right)\,,
~~~~
\Psi^{k\,ab}\,=\,\left(
\begin{array}{c}
\psi^{k\,ab} \cr
\Omega^{ac}\Omega^{bd} (-\tilde{C}\psi^{k+3\,\ast})_{cd}
\end{array}
\right)\,,
\eeqs
where $\tilde{C}=-i \tau^2$ is the charge-conjugation matrix, and $\tau^2$ the second Pauli matrix.
The Lagrangian density can then be rewritten as follows:
\beqs
{\cal L}&=& -\frac{1}{2} \Tr V_{\mu\nu} V^{\mu\nu}
\,+\,\frac{1}{2}\sum_{m=1}^{4}\left(i(q^{m})^{\dagger}_{\,\,\,a}
\bar{\sigma}^{\mu}\left(D_{\mu} q^{m}\right)^a
\,-\,i(D_{\mu}q^{m})^{\dagger}_{\,\,\,a} \bar{\sigma}^{\mu}q^{m\,a}\right)\,+\,\nonumber\\
&&
\,-\,\frac{1}{2} m^{f} \sum_{m,n=1}^{4}\Omega_{mn}\left( q^{m\,a\, T} \Omega_{ab} \tilde{C} q^{n\,b} 
- (q^{m})^{\dagger}_{\,\,\,a}\Omega^{ab} \tilde{C} (q^{n\,\ast})_b\right)+\nonumber\\
&&
\,+\,\frac{1}{2} \sum_{m=1}^{6} \left(i(\psi^{m})^{\dagger}_{\,\,\,ab} \bar{\sigma}^{\mu}\left(D_{\mu} \psi^{m}\right)^{ab}
\,-\,i(D_{\mu}\psi^{m})^{\dagger}_{\,\,\,ab} \bar{\sigma}^{\mu}\psi^{m\,ab}\right)\,+\,\nonumber\\
&&
\,-\,\frac{1}{2}m^{as} 
\sum_{m,n=1}^{6} \omega_{mn}\left(\psi^{m\,ab\,\mathrm{T}} \Omega_{ac}\Omega_{bd} \tilde{C}\psi^{n\,cd}\,
-\,(\psi^{m\,})^{\dagger}_{\,\,\,ab}\Omega^{ac}\Omega^{bd} \tilde{C}(\psi^{n\,\ast})_{cd}\right)\,,
\eeqs
where the kinetic terms for the 2-component spinors are written by making use of the
$2 \times 2$ matrices $\bar{\sigma}^{\mu}\equiv \left(\mathbb{1}_2,\,\tau^i\right)$.

The structure of the Dirac mass terms, rewritten in this 2-component formalism, 
shows that as long as $m^{f} \neq 0 \neq m^{as}$,
the non-Abelian
 global symmetry groups $SU(4)$ and $SU(6)$ 
 are explicitly broken to their $Sp(4)$ and  $SO(6)$ maximal subgroups, respectively.
Vacuum alignment arguments then imply that, as long as these are the only symmetry-breaking terms
 in the Langrangian density, if fermion bilinear 
 condensates emerge they spontaneously break  the  global symmetries
 according to the same breaking pattern~\cite{Peskin:1980gc}.

\subsection{Long distance dynamics}
\label{Sec:long}

The dynamics of the underlying  theory gives rise to 
 $15-10=5$ PNGBs describing the $SU(4)/Sp(4)$ coset,
 and $35-15=20$ PNGBs spanning the $SU(6)/SO(6)$ coset.
Following Ref.~\cite{Bennett:2019cxd}, we divide the
 15 generators $T^A$ of the global $SU(4)$, and 35 generators $t^B$ of $SU(6)$, in two sets
by denoting with $A=1\,,\,\cdots\,,\,5$ and  with $B=1\,,\,\cdots\,,\,20$ the broken ones,
which obey the following relations:
\beqs
\Omega T^A-T^{A\,\mathrm{T}} \Omega &=&0\,,~~~~\omega t^B-t^{B\,\mathrm{T}} \omega \,=\,0\,.
\eeqs
The unbroken generators have adjoint indices 
$A=6\,,\,\cdots\,,\,15$ and  $B=21\,,\,\cdots\,,\,35$. They satisfy the relations:
\beqs
\Omega T^A+T^{A\,\mathrm{T}} \Omega &=&0\,,~~~~\omega t^B+t^{B\,\mathrm{T}} \omega \,=\,0\,.
\eeqs

As long as the masses $m^{f}$ and $m^{as}$ 
are smaller than the dynamically generated,
chiral symmetry breaking scale of the theory,
one expects long-distance dynamics to be well captured by an EFT providing the
description of the PNGBs as  weakly-coupled scalar fields.
To this purpose, we introduce two non-linear sigma-model fields. 
The matrix-valued $\Sigma_6$ transforms as 
$\Omega_{ab}q^{m\,a\,T}\tilde{C} q^{n\,b}$, in the antisymmetric
representation of the global  $SU(4)$.
 $\Sigma_{21}$ has the quantum numbers 
of $-\Omega_{ab}\Omega_{cd}\psi^{m\,ac\,T}\tilde{C} \psi^{n\,bd}$,
and transforms in the symmetric representation of the $SU(6)$ global symmetry.

In the vacuum, the antisymmetric representation
decomposes as $6=1\oplus 5$ of the unbroken $Sp(4)$,
and the symmetric as $21=1\oplus 20$ of $SO(6)$;
the non-linear sigma-model fields can be parameterised by
the PNGB fields  $\pi_5$ and $\pi_{20}$ as
\beqs
\Sigma_6&\equiv&e^{\frac{2i \pi_5}{f_5}}\Omega=\Omega e^{\frac{2i \pi_5^{\mathrm{T}}}{f_5}}\,,
~~~~
\Sigma_{21}\,\equiv\,e^{\frac{2i \pi_{20}}{f_{20}}}\omega=\omega e^{\frac{2i \pi_{20}^{\mathrm{T}}}{f_{20}}}\,.
\label{Eq:Sigmas}
\eeqs
The decay constants  are denoted by $f_{5}$ and $f_{20}$.~\footnote{These  conventions are chosen 
so that, when applied to the QCD chiral Lagrangian, 
 the decay constant 
is $f_{\pi}\simeq 93$ MeV.} To write the 
EFT Lagrangian density, we further 
replace the mass terms with  (non-dynamical) spurion fields
$M_6 \equiv m^{f}\, \Omega$ and $M_{21} \equiv - m^{as} \,\omega$.
At the leading order in both the derivative expansion and the expansion in small masses, 
the Lagrangian density for the PNGBs of the $SU(4)/Sp(4)$ breaking takes the form
\beqs
\label{Eq:Lpi}
{\cal L}_6&=&\frac{f_{5}^2}{4}\Tr\left\{\frac{}{}\partial_{\mu}\Sigma_6 (\partial^{\mu}\Sigma_6 )^{\dagger}\frac{}{}\right\}
\,-\,\frac{v_6^3}{4}\Tr \left\{\frac{}{}M_6 \Sigma_6 \frac{}{}\right\}\,+\,{\rm h.c.}\\
&=&\Tr\left\{\frac{}{}\partial_{\mu}\pi_5\partial^{\mu}\pi_5\frac{}{}\right\}\,+
\,\frac{1}{3f_{5}^2}\Tr\left\{\frac{}{}
\left[\partial_{\mu}\pi_5\,,\,\pi_5\right]\left[\partial^{\mu}\pi_5\,,\,\pi_5\right]\frac{}{}\right\}\,+\,\cdots\nonumber\,+\\
&&\,+\,\frac{1}{2}\,m^{f} v_6^3\,\Tr (\Sigma_6\Sigma_6^{\dagger})  \,-\, \frac{m^{(f)} v_6^3}{f_5^2}\Tr \pi_5^2
 \,+\,\frac{m^{f} v_6^3}{3 f_5^4}\Tr \pi_5^4 \,+\,\cdots\,,
\eeqs
where $v_6$ parameterises the condensate.
The matrix of the five PNGBs in the $SU(4)/Sp(4)$ coset can be written as follows~\cite{Bennett:2019cxd}:
\beqs
\pi_5(x)\hspace{-2pt}=\hspace{-2pt}\frac{1}{2\sqrt{2}}
\hspace{-2pt}
\left(\hspace{-4pt}
\begin{array}{cccc}
 \pi^3(x) & \pi^1(x)-i \pi^2(x) & 0 & -i \pi^4(x)+\pi^5(x) \\
 \pi^1(x)+i \pi^2(x) & -\pi^3(x) & i \pi^4(x)-\pi^5(x) & 0 \\
 0 & -i \pi^4(x)-\pi^5(x) & \pi^3(x) & \pi^1(x)+i \pi^2(x) \\
 i \pi^4(x)+\pi^5(x) & 0 & \pi^1(x)-i \pi^2(x) & -\pi^3(x)
\end{array}
\hspace{-4pt}
\right)\hspace{-4pt}.
\label{Eq:pion}
\eeqs

The expansion for the $SU(6)/SO(6)$ PNGBs is
formally identical---thanks to the opposite signs we chose in the definition of the mass matrices,
ultimately deriving from the fact that  $\Omega^2=-\mathbb{1}_4$, while
$\omega^2=\mathbb{1}_6$---and one just replaces
$v_6\rightarrow v_{21}$, and analogous replacements  for other quantities.\footnote{The trace of the identity matrix
 may introduce numerical factors that differ in the two expansions. 
 In the $SU(4)/Sp(4)$ case 
$\Tr \Sigma_6\Sigma_6^{\dagger}=4$, 
while in the $SU(6)/SO(6)$ case
$\Tr \Sigma_{21}\Sigma_{21}^{\dagger}=6$.}
For instance, the matrix $\pi_{20}$ describing the PNGBs 
can be written as $\pi_{20}(x)=\sum_{B=1}^{20}\pi^B(x)t^B$,
where $t^B$ are the aforementioned broken generators of $SU(6)$.

As explained in detail in Ref.~\cite{Bennett:2019cxd}, one can extend 
 the  EFT description
 to include  the behaviour of the lightest 
vector and axial-vector states, besides the pNGBs,
by applying the principles of  Hidden Local Symmetry (HLS)~\cite{
Bando:1984ej,Casalbuoni:1985kq,Bando:1987br,Casalbuoni:1988xm,Harada:2003jx} 
(see also~\cite{Georgi:1989xy,Appelquist:1999dq,Piai:2004yb,Franzosi:2016aoo}).
There are well known limitations to the applicability of this type of EFT treatment, 
and  while we intend to come back to this topic in future publications,
we will not explore it further in this study. 

\subsubsection{Coupling to the Standard Model}

This paper studies the $Sp(4)$ gauge dynamics
coupled only to (f) and (as) fermions.
Nevertheless, to motivate it in terms of  composite Higgs and partial top compositeness, 
we recall briefly how the model can be (weakly) coupled to the SM gauge fields
of the $SU(3)_c \times SU(2)_L \times U(1)_Y$ gauge group---details can be found in 
Refs.~\cite{Barnard:2013zea,Ferretti:2013kya,Cacciapaglia:2019bqz,Bennett:2019cxd}.

The $SU(4)/Sp(4)$ coset is relevant to EWSB.
The $SU(2)_L \times SU(2)_R\sim SO(4)$ symmetry of the SM Higgs potential
is a subgroup of the unbroken $Sp(4)$. The
unbroken subgroup $SO(4)\sim SU(2)_L\times SU(2)_R$ 
has the following generators:
\beqs
\label{Eq:SU2L}
T^{1}_L&=&\frac{1}{2}\left(\begin{array}{cccc}
0 & 0 & 1 & 0\cr
0 & 0 & 0 & 0\cr
1 & 0 & 0 & 0\cr
0 & 0 & 0 & 0\cr
\end{array}\right)\,,\,\,
T^{2}_L\,=\,\frac{1}{2}\left(\begin{array}{cccc}
0 & 0 & -i & 0\cr
0 & 0 & 0 & 0\cr
i & 0 & 0 & 0\cr
0 & 0 & 0 & 0\cr
\end{array}\right)\,,\,\,
T^{3}_L\,=\,\frac{1}{2}\left(\begin{array}{cccc}
1 & 0 & 0 & 0\cr
0 & 0  & 0 & 0\cr
0 & 0 & -1 & 0\cr
0 & 0 & 0 & 0\cr
\end{array}\right)\,,\\
T^{1}_R&=&\frac{1}{2}\left(\begin{array}{cccc}
0 & 0 & 0 & 0\cr
0 & 0 & 0 & 1\cr
0 & 0 & 0 & 0\cr
0 & 1 & 0 & 0\cr
\end{array}\right)\,,\,\,
T^{2}_R\,=\,\frac{1}{2}\left(\begin{array}{cccc}
0 & 0 & 0 & 0\cr
0 & 0 & 0 & -i\cr
0 & 0 & 0 & 0\cr
0 & i & 0 & 0\cr
\end{array}\right)\,,\,\,
T^{3}_R\,=\,\frac{1}{2}\left(\begin{array}{cccc}
0 & 0 & 0 & 0\cr
0 &1  & 0 & 0\cr
0 & 0 & 0 & 0\cr
0 & 0 & 0 & -1\cr
\end{array}\right)\,.
\label{Eq:SU2R}
\eeqs
In decomposing $Sp(4)\rightarrow SO(4)$,
the PNGBs decompose as 
$5=1\oplus 4$, where the $4 \sim 2_{\mathbb{C}}$ is the Higgs doublet.
More explicitly, the real fields $\pi^1$, $\pi^2$, $\pi^4$, and $\pi^5$ combine into the $4$ of $SO(4)$.
The remaining  $\pi^3$ is a SM singlet.
The hypercharge assignments 
for the five PNGBs correspond to the 
action of the $T^3_R$ diagonal generator of $SU(2)_R$.

The $SU(6)/Sp(6)$ coset plays the important  part of introducing color $SU(3)_c$,
as the diagonal combination of the 
natural $SU(3)_L \times SU(3)_R$ subgroup of $SU(6)$.
The PNGBs decompose as $20 \sim 8 \oplus 6_{\mathbb{C}}$ under $SU(3)_c$.
An additional $U(1)_X$ subgroup of $SO(6)$ commutes with $SU(3)_L \times SU(3)_R$,
so that the SM hypercharge $U(1)_Y$
 is a linear combination of  $U(1)_X$ and the $U(1)$ group generated by the aforementioned $T^3_R$.

With these  assignments of quantum numbers, 
composite fermion operators emerge 
which combine two (f) fermions $Q$ (to make a $SU(2)_L$ doublet) and  
one (as) fermion $\Psi$ (a triplet of $SU(3)_c$). The resulting chimera baryon
has the same quantum numbers as a SM quark.
These are massive Dirac fermions. 
Elementary SM fermions, in particular the top and bottom quarks,
can couple to them.
This can be achieved  in two ways: either by coupling an SM bilinear
operator to a meson of the strong coupling theory---effectively reproducing in the low energy 
EFT a Yukawa coupling---or, alternatively, by coupling a (chiral) SM fermion to a chimera baryon.

The gauging of the SM gauge group introduces a new explicit source of breaking of the 
global symmetries (besides the mass terms). An analysis of the 1-loop effective
potential, along the lines of Ref.~\cite{Coleman:1973jx}, yields 
additional contributions to the masses of the PNGBs,
which are in general divergent, but controlled by the small, perturbative couplings of the SM gauge fields
circulating in the loops.
Furthermore, they 
 introduce an instability in the Higgs effective potential:
the negative sign of fermion loops 
ultimately  triggers EWSB. Because of the weakness of the
 couplings, these effects can be  arranged 
 to be small, and yield a value for $v_W$ 
 that is smaller than the decay constant of the PNGBs 
as it would emerge in isolation,
from the strong dynamics sector only. In the literature, the combination of these phenomena goes
under the name of vacuum misalignment.

 If the strongly-coupled regime of the underlying dynamics is 
 very different from that of a QCD-like theory---in particular if the theory 
 has enough fermions to be close to the conformal window---the emergence of large 
 anomalous dimensions may enhance 
the effective couplings at low energy, hence explaining why the top quark mass is large.
This is one motivation  for composite Higgs models  with partial top compositeness,
and this model provides the simplest template.
Similar ideas were put forward long time ago,
 in the  context of walking technicolor (see  for instance 
Refs.~\cite{Cohen:1988sq,Leung:1989hw}), top compositeness~\cite{Kaplan:1991dc}, and
warped extra-dimensions~\cite{Grossman:1999ra,Gherghetta:2000qt}.
It would go beyond our scope to review the rich literature on 
the subject, and we refer the interested reader to the discussion
in Ref.~\cite{Chacko:2012sy},
 in the context of dilaton-Higgs models, and 
to follow the references therein.

Lattice studies provide  non-perturbative information that is 
essential for the programme of phenomenological applications described in this subsection,
with potentially transformative reach.
As we shall demonstrate in the body 
of the paper, our research programme has reached the stage
at which we can compute the spectrum of masses and decay
 constants of the composite particles (mesons and chimera  baryons). 
In the future, we will further improve our numerical studies in order to
measure other quantities,  such as the size of the condensates, 
the scaling dimension of the operators in the non-perturbative regime, the magnitude of 
non-trivial matrix elements that feed into the effective potential for the PNGBs
and  scattering amplitudes of mesons.

\subsection{Of mesons and chimera baryons}
\label{Sec:mb}

While it is easier to discuss the symmetries of the system 
by writing the fermions in the 2-component notation, we
revert to 4-component spinors to prepare for the lattice numerical studies.
In switching to the 4-component spinor notation, it is useful to explicitly write the
charge-conjugated spinors as follows:
\beqs
\label{Eq:qdagger}
 Q^{i\,a}_{C}&\equiv& 
 \left(\begin{array}{c}q^{N_f+i\,a} \cr -\Omega^{ab}(\tilde{C} 
 q^{i\,{\ast}})_b\end{array}\right)\,,
 \\
 \Psi^{i\,ab}_{C}&\equiv&
  \left(\begin{array}{c}\psi^{n_f+i\,ab} \cr 
  -\Omega^{ac}\Omega^{bd}(\tilde{C} \psi^{i\,^{\ast}})_{cd}\end{array}\right)
 \,.
\label{Eq:psidagger}
 \eeqs 
The meson operators sourcing the five PNGBs  are the following: 
 \beqs
 {\cal O}_{{\rm PS}, 1}\nonumber
 &=&\left(\overline{Q^{1\,a}} \gamma^5 Q^{2\,a}+\overline{Q^{2\,a}} \gamma^5 Q^{1\,a}\right)\,,\\
{\cal O}_{{\rm PS}, 2}&=&
\nonumber
i\left(-\overline{Q^{1\,a}} \gamma^5 Q^{2\,a}+\overline{Q^{2\,a}} \gamma^5 Q^{1\,a}\right)\,,\\
\label{Eq:5}
{\cal O}_{{\rm PS}, 3} &=&\left(\overline{Q^{1\,a}}  \gamma^5 Q^{1\,a}-\overline{Q^{2\,a}}\gamma^5Q^{2\,a}\right)\,,\\
 {\cal O}_{{\rm PS}, 4}&=&\nonumber
-i\,\left(\overline{Q^{1\,a}}  Q^{2\,a}_{\,C}+\overline{Q_C^{2\,a}}  Q^{1\,a}\right)\,,\\
 {\cal O}_{{\rm PS}, 5}&=&\nonumber
i\,\left(-i\,\overline{Q^{1\,a}}  Q^{2\,a}_C+i\overline{Q^{2\,a}_C} Q^{1\,a}\right)\,.
 \eeqs
We expect the lightest states of the theory to appear in 2-point correlation functions of these operators.

The theory possesses also an 
anomalous, axial $U(1)_A$, which is both spontaneously and explicitly broken (by the mass term,
as well as  the anomaly).
Hence, there are  $U(1)_A$ partners to the meson operators,
sourcing the counterparts of the $a_0$ particles of QCD,
that can be obtained by  replacing $\mathbb{1}_4\rightarrow i\gamma^5$ inside the expressions
in Eqs.~(\ref{Eq:5}), to yield:
 \beqs
 {\cal O}^{\prime}_{{\rm PS}, 1}\nonumber
 &=&
 i\left(\overline{Q^{1\,a}}  Q^{2\,a}+\overline{Q^{2\,a}}  Q^{1\,a}\right) \,,\\
{\cal O}^{\prime}_{{\rm PS}, 2}&=&\nonumber
\left(\overline{Q^{1\,a}}  Q^{2\,a}-\,\overline{Q^{2\,a}}  Q^{1\,a}\right)\,,\\
\label{Eq:5prime}
 {\cal O}^{\prime}_{{\rm PS}, 3} &=&i\left(\overline{Q^{1\,a}}   Q^{1\,a}-\overline{Q^{2\,a}} Q^{2\,a}\right)\,,\\
 {\cal O}^{\prime}_{{\rm PS}, 4}&=&\nonumber
 \,\left(\overline{Q^{1\,a}}  \gamma^5 Q^{2\,a}_{\,C}+\overline{Q_C^{2\,a}}  \gamma^5 Q^{1\,a}\right)\,,\\
 {\cal O}^{\prime}_{{\rm PS}, 5}&=&\nonumber
 i\,\left(\,\overline{Q^{1\,a}} \gamma^5  Q^{2\,a}_C-\overline{Q^{2\,a}_C} \gamma^5 Q^{1\,a}\right)\,.
 \eeqs
Mesons made of $\Psi^{i\,ab}$ are built in a similar way, and we do not list them
explicitly---details can be found in Ref.~\cite{Bennett:2019cxd}.

The chimera baryons we are interested in
 must have the same quantum numbers as the SM quarks, which 
 transform as a $(2,2)$ of $SU(2)_L\times SU(2)_R$ in the standard model.
But they also carry $SU(3)_c$ color, and hence require inserting $\Psi^{i\,ab}$, with $i=1,\,2,\,3$
being identified with the QCD color index.
 A simple way to achieve this and build a $Sp(4)$ singlet is to rewrite, 
 in the first line of Eqs.~(\ref{Eq:5}), $Q^{2\,a}=Q^{2\,b}\delta^a_{\,\,\,b}$ and $Q^{1\,a}=Q^{1\,b}\delta^a_{\,\,\,b}$, 
 and then replace
 $\delta^a_{\,\,\,b}\rightarrow  P_{L,R}\Psi^{k\,ac} \Omega_{cb} $,
 where 
 \beqs
 P_{L,R}&\equiv&
 \frac{1}{2}\left(\frac{}{}\mathbb{1}_4 \pm \gamma_5\frac{}{}
 \right)\,.
 \label{eq:chiral_projector}
 \eeqs
 After performing the same substitution on all the mesons, we obtain a list of
chimera baryon operators ${\cal O}^{L,R}_{{\rm CB}}$:
 \beqs
 {\cal O}^{L,R}_{{\rm CB}, 1}\nonumber
 &=&
 \left(\overline{Q^{1\,a}} \gamma^5 Q^{2\,b}+\overline{Q^{2\,a}} \gamma^5 Q^{1\,b}\right)  \Omega_{bc} P_{L,R}\Psi^{k\,ca} \,,\\
{\cal O}^{L,R}_{{\rm CB}, 2}&=&\nonumber
i\left(-\overline{Q^{1\,a}} \gamma^5 Q^{2\,b}+\overline{Q^{2\,a}} \gamma^5 Q^{1\,b}\right)\Omega_{bc} P_{L,R}\Psi^{k\,ca}\,,\\
{\cal O}^{L,R}_{{\rm CB}, 3}&=&\label{Eq:top}
\left(\overline{Q^{1\,a}}  \gamma^5 Q^{1\,b}-\overline{Q^{2\,a}}\gamma^5Q^{2\,b}\right)\Omega_{bc} P_{L,R}\Psi^{k\,ca}\,,\\
 {\cal O}^{L,R}_{{\rm CB}, 4}&=&\nonumber
 -i\,\left(\overline{Q^{1\,a}}  Q^{2\,b}_{\,C}+\overline{Q_C^{2\,a}}  Q^{1\,b}\right)\Omega_{bc} P_{L,R}\Psi^{k\,ca}\,,\\
 {\cal O}^{L,R}_{{\rm CB}, 5}&=&\nonumber
 i\,\left(-i\,\overline{Q^{1\,a}}  Q^{2\,b}_C+i\overline{Q^{2\,a}_C} Q^{1\,b}\right)\Omega_{bc} P_{L,R}\Psi^{k\,ca}\,.
 \eeqs
Analogously, the $U(1)_A$ partners of the chimera baryons are the following:
  \beqs
 {\cal O}^{\prime\,L,R}_{{\rm CB}, 1}\nonumber
 &=&
 i\left(\overline{Q^{1\,a}}  Q^{2\,b}+\overline{Q^{2\,a}}  Q^{1\,b}\right) \Omega_{bc} P_{L,R}\Psi^{k\,ca}\,,\\
{\cal O}^{\prime\,L,R}_{{\rm CB}, 2}&=&\nonumber
\left(\overline{Q^{1\,a}}  Q^{2\,b}-\,\overline{Q^{2\,a}}  Q^{1\,b}\right)\Omega_{bc} P_{L,R}\Psi^{k\,ca}\,,\\
{\cal O}^{\prime\,L,R}_{{\rm CB}, 3}&=&\label{Eq:topH}
i\left(\overline{Q^{1\,a}}   Q^{1\,b}-\overline{Q^{2\,a}}Q^{2\,b}\right)\Omega_{bc} P_{L,R}\Psi^{k\,ca}\,,\\
 {\cal O}^{\prime\,L,R}_{{\rm CB}, 4}&=&\nonumber
 \,\left(\overline{Q^{1\,a}}  \gamma^5 Q^{2\,b}_{\,C}+\overline{Q_C^{2\,a}}  \gamma^5 Q^{1\,b}\right)\Omega_{bc} P_{L,R}\Psi^{k\,ca}\,,\\
 {\cal O}^{\prime\,L,R}_{{\rm CB}, 5}&=&\nonumber
 i\,\left(\,\overline{Q^{1\,a}} \gamma^5  Q^{2\,b}_C-\overline{Q^{2\,a}_C} \gamma^5 Q^{1\,b}\right)\Omega_{bc} P_{L,R}\Psi^{k\,ca}\,.
 \eeqs
The ${\cal O}^{\prime\,L,R}_{{\rm CB}}$ operators are expected to source heavier particles,
in respect to the ${\cal O}^{L,R}_{{\rm CB}}$.

\section{The lattice theory}
\label{Sec:latticesetup}

In this section, we describe in detail the 
lattice gauge theory of interest,
and the implementation of the numerical algorithms we adopt. 
Our software is based upon the HiRep code, originally developed 
in the BSM context and presented in Ref.~\cite{DelDebbio:2008zf}. 
In earlier studies of  $Sp(2N)$ lattice gauge theories~\cite{Bennett:2017kga,Bennett:2020qtj},
we both generalised the Cabibbo-Marinari
prescription~\cite{Cabibbo:1982zn},
and implemented an efficient resymplectization projection.
For the purpose of this study,
we further wrote original code to implement dynamical 
calculations in the presence of 
matter in multiple representations.
It is worth reminding the reader that most  
lattice code publicly available 
has been optimised for QCD and QCD-like theories, 
and only a handful of 
codes allowing to treat multiple representations exist
(see for instance~\cite{Ayyar:2017qdf,
Ayyar:2018zuk,Cossu:2019hse}, for $SU(4)$ gauge theories).
Hence, we describe our algorithm in some detail, and 
we  provide a number of tests, 
both in this as well as in the
subsequent sections, to demonstrate that our
implementation 
reproduces the expected results, in the appropriate limits.

\subsection{Lattice action}
\label{Sec:action}

We write the Euclidean action, discretised in four dimensions,
 of non-Abelian $Sp(2N)$ gauge theories coupled to fermionic matter
  as the sum of  the gauge $S_g$ and fermion $S_f$ actions, 
\beq
S = S_g + S_f.
\label{eq:lattice_action}
\eeq
The generic lattice site is denoted by $x$, while  $\hat{\mu}, \hat{\nu}$ are
 unit displacements in the space-time directions $\mu, \nu$, so that
 the first term of Eq.~(\ref{eq:lattice_action}),  the Wilson plaquette action, is 
\beq
S_g\equiv\beta \sum_x \sum_{\mu<\nu} \left(1-\frac{1}{2N} {\rm Re}\, {\rm Tr}\, 
U_\mu (x) U_\nu (x+\hat{\mu}) U_\mu^\dagger(x+\hat{\nu}) U_\nu^\dagger(x)\right),
\label{eq:gauge_action}
\eeq
where $U_{\mu}(x) \in Sp(2N)$ is the group 
variable living on the link $(x,\mu)$, and
$\beta\equiv \frac{4N}{g_0^2}$, with $g_0$ the gauge coupling.

The second term of Eq.~(\ref{eq:lattice_action}) is the massive Wilson-Dirac action:
\beq
S_f \equiv a^4 \sum_{j=1}^{N_f}\sum_x \overline{Q}^j(x) D^{(f)}_m Q^j(x)+
a^4 \sum_{j=1}^{n_f}\sum_x \overline{\Psi}^j(x) D^{(as)}_m \Psi^j(x),
\label{eq:fermion_action}
\eeq
where $a$ is the lattice spacing, $Q^{j}$ and $\Psi^j$ the fermions (flavour indices are explicitly shown, while
color and spinor indices are understood), and the Dirac operators $D^{(f)}_m$ for the fundamental
and $D^{(as)}_m$ for the 2-index antisymmetric representation will be defined shortly. 
Here and in the following, we  restrict the number of colors to $N_c=4$ (or $N=2$), and the number of Dirac flavours to $N_f=2$ and $n_f=3$ for the fundamental and antisymmetric representations, respectively. 
Nevertheless, where possible we leave explicit the dependence on arbitrary $N\geq 2$, 
as our construction can be applied to all $Sp(2N)$ gauge theories.

For the $(f)$ fermions, the link variable  appearing in the Dirac operator
coincides with $U_\mu(x)$ in \Eq{gauge_action}:
\beqs
U^{(f)}_\mu (x) = U_\mu (x) \in Sp(2N).
\eeqs
In the case of the $(as)$ fermions, we construct link variable $U^{(as)}_\mu (x)$, 
and thus the Dirac operator $D^{(as)}_m$, by following the prescription in Ref.~\cite{DelDebbio:2008zf}. 
We first define an orthonormal basis $e_{(as)}^{(ab)}$ (the multi-index $(ab)$ runs
 over ordered  pairs with   $1 \leq a < b \leq 2N$)
for the appropriate vector space of 
$2N\times 2N$ antisymmetric (and $\Omega$-traceless) matrices.
There are  $N(2N-1)-1$ such matrices. 
For $b=N+a$ and $2\leq a\leq N$, they have the following non-vanishing entries:
\beqs
(e_{(as)}^{(ab)})_{c,N+c}\equiv -(e_{(as)}^{(ab)})_{N+c,c}\equiv 
\left\{\begin{matrix}
&\frac{1}{\sqrt{2\,a\,(a-1)}},~~~\textrm{for}~c<a,\\
&\frac{-(a-1)}{\sqrt{2\,a\,(a-1)}},~~~\textrm{for}~c=a,\\
\end{matrix}\right.
\label{eq:eas_diag}
\eeqs
and for $b\neq N+a$ 
\beqs
(e_{(as)}^{(ab)})_{cd}\equiv \frac{1}{\sqrt{2}}(
\delta_{ad}\delta_{bc}
-\delta_{ac}\delta_{bd}
)\,. 
\label{eq:eas}
\eeqs
The $\Omega$-traceless
condition can be rewritten explicitly as $ \Omega^{dc} \left(e^{(ab)}_{(as)}\right)_{cd}=0$.
Specialising to  the $Sp(4)$ case,  the matrix $e_{(as)}^{(13)}$ vanishes by construction, 
and one can verify that the remaining five non-vanishing matrices satisfy the orthonormalisation 
condition $\Tr e_{(as)}^{(ab)}e_{(as)}^{(cd)}=-\delta^{(ab)(cd)}$. The ordering
 of pairs $(ab)$ in our convention
 is $(12)$, $(23)$, $(14)$, $(24)$ and $(34)$. We show their explicit forms  in~\App{A3}.
The  link variables $U^{(as)}_\mu(x)$ descend from the fundamental
 link variables $U_\mu(x)$ and take the form of
\beqs
\left(U^{(as)}_\mu\right)_{(ab)(cd)}(x)\equiv
{\rm Tr}\left[
(e_{(as)}^{(ab)})^\dagger U_\mu(x) e_{(as)}^{(cd)} U^{\mathrm{T}}_\mu(x)\right],~~~{\rm with}~a<b,~c<d.
\label{eq:U_AS}
\eeqs

With all of the above, the massive Wilson-Dirac operators are defined by
\beqs
D^{(f)}_m Q_j(x) &\equiv& (4/a+m^{f}_0) Q_j(x)\label{eq:DiracF} \\
&&-\frac{1}{2a}\sum_\mu \nonumber
\left\{(1-\gamma_\mu)U^{(f)}_\mu(x)Q_j(x+\hat{\mu})
+(1+\gamma_\mu)U^{(f),\,\dagger}_\mu(x-\hat{\mu})Q_j(x-\hat{\mu})\frac{}{}\right\}\,,
\eeqs
for the fundamental representation, and
\beqs
D^{(as)}_m \Psi_k(x) &\equiv& (4/a+m^{as}_0) \Psi_k(x) \label{eq:DiracAS}\\
&&-\frac{1}{2a}\sum_\mu \nonumber
\left\{(1-\gamma_\mu)U^{(as)}_\mu(x)\Psi_k(x+\hat{\mu})
+(1+\gamma_\mu)U^{(as),\,\dagger}_\mu(x-\hat{\mu})\Psi_k(x-\hat{\mu})\frac{}{}\right\}\,,
\eeqs
for the 2-index antisymmetric representation. 
$m_0^{f}$ and $m_0^{as}$ are the (degenerate) bare masses of $Q$ and $\Psi$, respectively.

\subsection{Numerical implementation}
\label{Sec:hmc}

We have extended the 
HiRep code~\cite{DelDebbio:2008zf},\footnote{ The code is publicly available, and can be 
accessed at https://github.com/claudiopica/HiRep 
for the main $SU(N_c)$ version, and at https://github.com/sa2c/HiRep for the $Sp(2N)$ fork.} to 
adapt it to treat  $Sp(2N)$  (rather than $SU(N_c)$) 
gauge theories and couple them to fermions in multiple representations of the group.
Ensembles with dynamical fermions can be produced by combining  the hybrid Monte Carlo (HMC) algorithm, 
and its extension with rational approximations for the Dirac matrix with fractional powers---the rational 
hybrid Monte Carlo (RHMC).  
The standard (R)HMC algorithm consists of the following three main steps.
\begin{itemize}
\item Generation of new pseudofermion fields from a heat-bath distribution.
\item Molecular dynamics (MD) evolution---dynamical evolution of the gauge
 field configuration with a fictitious Hamiltonian.
\item Metropolis test at the end of each MD trajectory to correct for errors in the
numerical integration of the equations of motion. 
\end{itemize}
Let us provide some more technical details about these three steps.

As anticipated, the implementation of HMC/RHMC algorithms for fermions in arbitrary representations 
of $SU(N_c)$ gauge groups is extensively discussed in Ref.~\cite{DelDebbio:2008zf}, 
and its generalisation to the fundamental representation of $Sp(2N)$ in Ref.~\cite{Bennett:2019cxd}. 
We pause here to discuss in further depth the case of multiple representations, given
the limited extent of the literature on the subject~\cite{Ayyar:2017qdf,
Ayyar:2018zuk,Cossu:2019hse}. In the rest of this subsection,
we follow closely the discussion in Ref.~\cite{DelDebbio:2008zf}, and 
refer the reader to this publication for details, while we highlight  the differences
required in our implementation.

The fermion action in \Eq{fermion_action} is quadratic in the fermion fields.
It can be explicitly integrated when we compute the partition function of the theory,
a process that results in the fermion determinant $\textrm{det}(D_m)$. 
If we suppress  spin and color indices, for convenience, and consider a generic number of flavours $n$,
we can  replace this determinant by introducing complex bosonic fields $\phi$ and $\phi^{\dagger}$, 
called pseudofermions, with the generic definition:
\beq
\label{Eq:determinant}
(\textrm{det}(D_m))^{n} \equiv (\textrm{det}(Q_m))^{n} =
 \int \mathcal{D}\phi\mathcal{D}\phi^\dagger e^{-a^4\sum_x \phi^\dagger(x) (Q_m^2)^{-n/2}\phi(x)}\,.
\eeq
The  Dirac operator 
$
Q_m \equiv \gamma_5 D_m
$ is hermitian.
The square of $Q_m$ is positive definite. 
In the rest of this section, we set the lattice spacing $a=1$, for notational convenience.

 As explained in Ref.~\cite{DelDebbio:2008zf}, 
one defines the MD evolution in fictitious time $\tau$ to be governed by a Hamiltonian 
which receives contributions $H_g$ from  gauge fields,  and $H_f^R$ from each 
species of fermions  in  representation $R$ of the group---see
Eqs.~(15)-(18) in Ref.~\cite{DelDebbio:2008zf}.
If we want to  describe  $n^R$ degenerate (Dirac) fermions in a given representation $R$,
we need to be more precise in the definition of the pseudofermions
and how they enter the exponent in Eq.~(\ref{Eq:determinant}), and the Hamiltonian $H_f^R$.
We introduce $N_{\rm pf}$ pseudofermions $\phi_k^{R}$ and $\phi_k^{R\,\dagger}$,
and their Hamiltonian is determined by the Dirac operator in the representation $R$:
\beq
H_f^R=\sum_{k=1}^{N_{\rm pf}}\sum_x\phi^{R,\dagger}_k(x) \left(\left(Q_m^{R}\right)^2\right)^{-l_k}\phi^R_k(x)\,,
\label{eq:hamiltonian_f}
\eeq
subject to the constraint $\sum_{k=1}^{N_{\rm pf}} l_k=n^R/2$.
If the number  $n^R$ of species of type $R$ is even, then
we can set $l_k=1$ for all $k$ and $N_{\rm pf}=n^R/2$, 
because  the inverse of $\left(Q_m^{R}\right)^2$ can be computed, $Q_m$ being
 hermitian. 

  In the case of odd $n^R$, on the other hand, it is 
  possible to set $N_{\rm pf}=n$, and $l_k=1/2$, by
applying the rational
 approximation~\cite{Clark:2003na} to  $H_f^R$---see Sect.~IIIB of Ref.~\cite{DelDebbio:2008zf},
 also for the definition of the numerical  coefficients appearing in the RHMC approximation.

In the calculation we perform for
 this paper, we use an admixture of the above.
 For the $N_f=2=n^{(f)}$ Dirac fermions in the fundamental representation, 
 we set $N_{\rm pf}=1$, and  adopt the HMC evolution.
 As for the $n_f=3=n^{(as)}$ Dirac fermions in the antisymmetric representation, we further split them
 into $n^{(as)}-1=2$, which requires $N_{\rm pf}=1$ pseudofermions in the HMC evolution,
and a third degenerate $(as)$ fermion, which we describe by one additional pseudofermion, for which the
 evolution is ruled by the RHMC algorithm---$l_1=1/2$ in its Hamiltonian in Eq.~(\ref{eq:hamiltonian_f}).\footnote{
We made this choice so that the rational approximation is applied only to one of the pseudofermions. 
We checked numerically that, for the range of masses relevant to this paper, had we treated all 
 three $(as)$ fermions with the RHMC algorithm, 
with  $l_k=1/2$ for $k=1,\,2,\,3$, we would have obtained consistent results.} 

\begin{figure}
\begin{center}
\includegraphics[width=.59\textwidth]{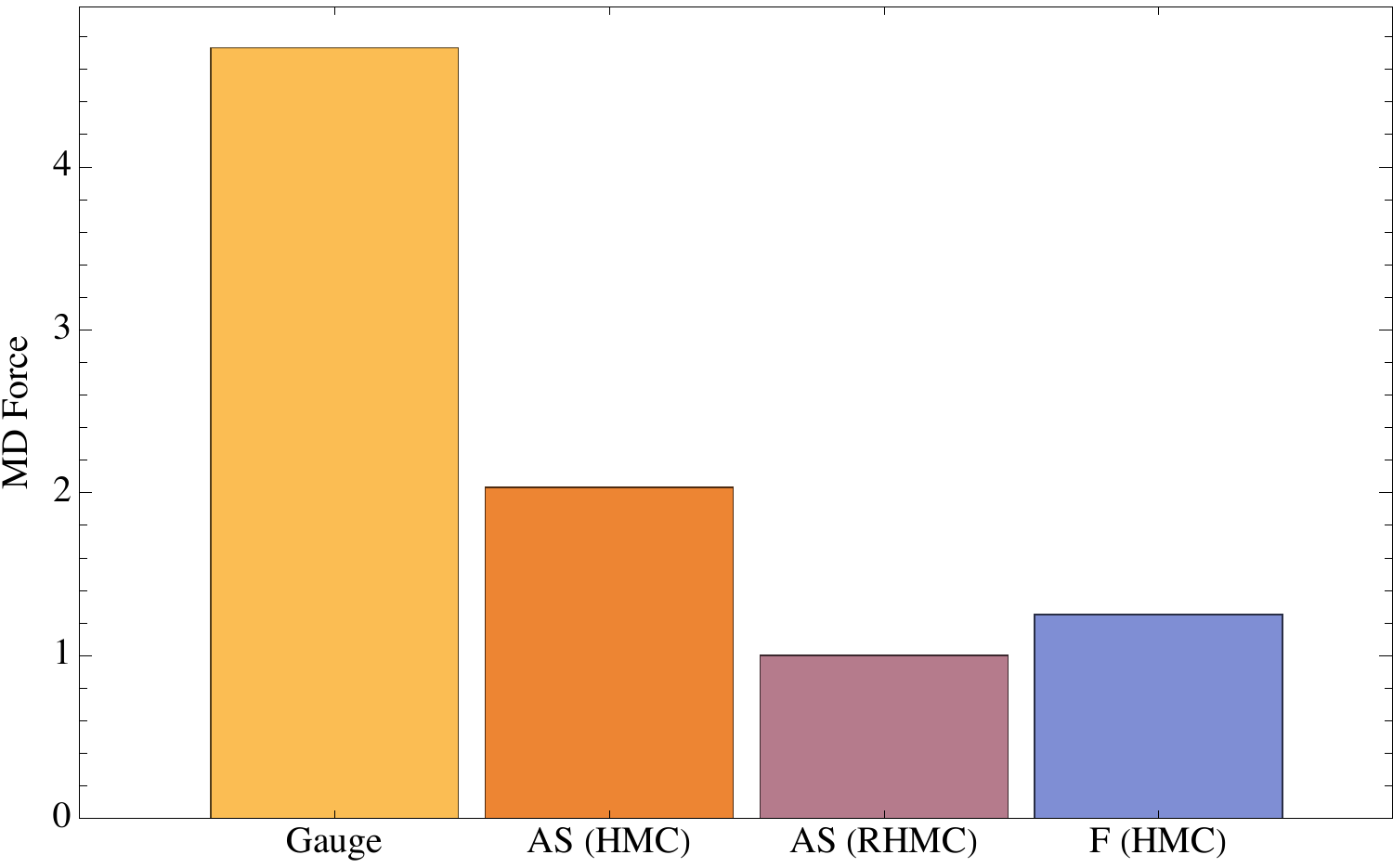}
\caption{%
\label{fig:force_chart}%
The relative contribution of gauge and fermion fields to the 
molecular dynamics force, averaged over the ensemble
with $\beta=6.5$, $am_0^f=-0.7$ and $am_0^{as}=-0.9$, on lattice of size $8^4$,
chosen for illustration purposes. 
The fermion force receives three separate contributions, one for each
of the pseudofermion fields: 
one HMC pseudofermion for two fundamental flavours, denoted F (HMC), 
and one each for the antisymmetric AS (HMC) and AS (RHMC) 
pseudofermions making up the
 three antisymmetric flavours. The forces are normalised to the one due to
  the antisymmetric fermion with the RHMC implementation.
}
\end{center}
\end{figure}

We hence have four  contributions to the MD evolution: 
 the gauge contribution is supplemented by those coming from the
 HMC treatment of 
 the $(f)$ pseudofermion, from the HMC treatment of one $(as)$ pseudofermion,
 and from the RHMC application to the third, $(as)$ pseudofermion.
We illustrate the size of each, by showing  in \Fig{force_chart}
their contribution to the force as it enters the Hamiltonian evolution---see the 
Hamilton equations governing the MD evolution, in Eqs.~(19) and~(20) of Ref.~\cite{DelDebbio:2008zf}---averaged 
over one of the ensembles.
The acceptance rate is in the range of $75-85\%$. To accelerate the
(computationally demanding) inversion of the
 Dirac operator $Q^2_m$, 
  we use the second order Omelyan integrator~\cite{Takaishi:2005tz} 
  in the MD evolution 
  and the even-odd preconditioning of the fermion matrix~\cite{DeGrand:1990dk},
applied to the (R)HMC algorithm  as discussed in Ref.~\cite{DelDebbio:2008zf}.

\subsection{Symmetry properties of the Dirac operator}
\label{Sec:symmetry}

The Wilson-Dirac formulation for mass-degenerate Dirac fermions in \Eq{fermion_action} 
explicitly breaks the global $SU(4)\times SU(6)$ symmetry to its 
$Sp(4)\times SO(6)$ subgroup, as in the continuum theory
 discussed in \Sec{model}.  This is accompanied by the formation of a non-zero fermion condensate,
 which in the massless limit would result in the 
 spontaneous breaking of the symmetry.
This is reflected in the spectrum of the Dirac operator~\cite{Banks:1979yr}:
 universal features in this spectrum can be modelled
by  chiral random matrix theory (chRMT)~\cite{Verbaarschot:1994qf}---see 
Ref.~\cite{Verbaarschot:2000dy} for a comprehensive review. 
In this subsection, following the discussion in Ref.~\cite{Cossu:2019hse},
we summarise the chRMT
analytical predictions.
In subsequent sections we will present our numerical results,
obtained by computing explicitly the spectrum of  Dirac eigenvalues,  and 
 compare them to chRMT predictions, hence providing a non-trivial test of the accuracy of the
numerical algorithms.

An antiunitary transformation is an antilinear map between two complex Hilbert spaces $\mathcal{H}_1$ and $\mathcal{H}_2$
\beq
{\cal A} : \mathcal{H}_1\rightarrow \mathcal{H}_2, 
\eeq
with ${\cal A}(a x + b y) = a^* {\cal A}(x) + b^* {\cal A}(y)$ such that
\beq
\langle {\cal A}(x),{\cal A}(y) \rangle = \langle x,y \rangle^*,
\eeq
where $\langle \cdot , \cdot \rangle$ denotes the inner products in the two spaces. 
$x$, $y$  are elements of $\mathcal{H}_1$, while  $a$, $b$ are complex numbers.
 If $\mathcal{H}_1=\mathcal{H}_2$ or, equivalently, the map is invertible, we call ${\cal A}$ an antiunitary operator. 
Any antiunitary operator can  be written as 
\beq
{\cal A}=\mathcal{V} K,
\label{eq:antiunitaryO}
\eeq 
where $\mathcal{V}$ is a unitary operator and $K$ is the complex-conjugation operator.

Let us now consider the discretised Dirac operator $D_m^R$---generalising
Eqs.~(\ref{eq:DiracF}) and~(\ref{eq:DiracAS}) to arbitrary representations $R$.
If we find an  antiunitary operator ${\cal A}^R$ that obeys the relation
\beq
[{\cal A}^R,\gamma^5 D_m^R]=0,
\label{eq:A_commute_D}
\eeq
then we can use this property to characterise the degeneracies in the  spectrum of the Dirac operator. 
There are actually three possibilities~\cite{Verbaarschot:2000dy}, precisely related to the Dyson index as follows: 
\begin{itemize}
\item  $({\cal A}^{R})^2=\mathbb{1}$, in which case the Dyson index is $\bar{\beta}=1$,
and there exists a basis in which the Dirac operator is real,
\item  there exists no such ${\cal A}^R$, in which case the Dyson index is $\bar{\beta}=2$,
and the Dirac operator is complex,
\item $({\cal A}^{R})^2=-\mathbb{1}$, in which case the Dyson index is $\bar{\beta}=4$,
and there exists a basis in which the Dirac operator is real quaternionic (pseudo-real).
\end{itemize}

In the context of chiral random matrix theory (chRMT), this classification parallels that of the ensembles,
itself  reflected in the chiral symmetry breaking pattern for the $N_f$ Dirac fermions of the theory, as follows.
\begin{itemize}
\item $\bar{\beta}=1$: the chRMT ensemble is called chiral Gaussian Orthogonal Ensemble (chGOE), 
because of the real matrix elements, and the  breaking pattern is $SU(2N_f)\rightarrow Sp(2N_f)$.
\item $\bar{\beta}=2$: the chRMT ensemble is called chiral Gaussian Unitary Ensemble (chGUE), 
because of the complex  elements, and the  breaking pattern is $SU(N_f)\times SU(N_f)\rightarrow SU(N_f)$.
\item $\bar{\beta}=4$: the chRMT ensemble is called chiral Gaussian Symplectic Ensemble (chGSE), 
because of the quaternionic  elements, and the  breaking pattern is $SU(2N_f)\rightarrow SO(2N_f)$.
\end{itemize}

Let us first consider the case of fermions in the fundamental representation of $Sp(4)$. As implied by 
Eqs.~(\ref{eq:antiunitaryO}) and~(\ref{eq:A_commute_D}), and using the facts that 
$\Omega^{-1} T_{(f)}^A \Omega = -T_{(f)}^{A\,{ T}}= -T_{(f)}^{A\,\ast}$, that $C$ 
commutes with $\gamma_5$ and that $C^2=-\mathbb{1}=-\gamma_5^2$,
we see that
\beq
{\cal A}^{(f)} = \Omega C \gamma^5 K\,,
\label{eq:A_F}
\eeq
commutes with $\gamma_5 D_{m}^{(f)}$, and that
(${\cal A}^{(f)})^2=\mathbb{1}$ and thus belongs to the class of $\bar{\beta}=1$. 
Indeed, the  $SU(4)$ global symmetry acting on the $(f)$ fermions  
breaks to  its $Sp(4)$ subgroup.

In the case of fermions in the antisymmetric representation of $Sp(4)$, the construction of the 
antiunitary operator requires first to generalise the generators to this representation. 
We first recall that the color indices of the link variables $\left(U^{(as)}_{\mu}\right)_{(ab)(cd)}(x)$ in \Eq{U_AS} 
are denoted by the multi-indices $(12),~(23),~(14),~(24),$ and $(34)$. 
Using this ordering convention, we find the following $5\times 5$ matrix
\beq
W=\left(
\begin{array}{ccccc}
0 & 0 & 0 & 0 & 1 \\
0 & 0 & 1 & 0 & 0 \\
0 & 1 & 0 & 0 & 0 \\
0 & 0 & 0 & 1 & 0 \\
1 & 0 & 0 & 0 & 0
\end{array}
\right),
\eeq
which is real, symmetric and unitary, and satisfies
\beq
W^{-1} T^A_{(as)} W = -T^{A*}_{(as)}\,,
\eeq
where a basis of $T_{(as)}$ is shown explicitly in Appendix~\ref{Sec:A3}.
In analogy with \Eq{A_F}, we find the antiunitary operator
\beq
{\cal A}^{(as)}=W C \gamma^5 K\,,
\eeq
to commute with $\gamma_5 D_m^{(as)}$.
The square of $W$ is the identity matrix, hence we conclude that  $({\cal A}^{(as)})^2=-\mathbb{1}$, and 
$\bar{\beta}=4$.
The $SU(6)$ symmetry acting on the $(as)$ fermions is broken to its $SO(6)$ subgroup.

A noticeable consequence of the fact that $({\cal A}^{(as)})^2 = -\mathbb{1}$ is that  the determinant of $D_m^{(as)}$ is real and positive~(see, e.g.,~\cite{Hands:2000ei}). Therefore, numerical simulations of Sp(4) gauge theories involving an odd number of antisymmetric Dirac flavours are not plagued by the sign problem. This enables us to have controlled numerical results for our systems using standard Monte Carlo methods.

One of the interesting predictions of chRMT is that
the distribution of the unfolded density of spacings $s$ between subsequent eigenvalues of $\gamma_5 D_m^{R}$
assumes  the following  functional dependence (the Wigner surmise)
\beq
P(s) = N_{\bar{\beta}} s^{\bar{\beta}} e^{-c_{\bar{\beta}} s^2},~~~~{\textrm{with}}~~N_{\bar{\beta}}=2\frac{\Gamma^{{\bar{\beta}}+1}\left(\frac{{\bar{\beta}}}{2}+1\right)}{\Gamma^{{\bar{\beta}}+2}
\left(\frac{{\bar{\beta}}+1}{2}\right)},
~~c_{\bar{\beta}}=\frac{\Gamma^2\left(\frac{{\bar{\beta}}}{2}+1\right)}{\Gamma^2\left(\frac{{\bar{\beta}}+1}{2}\right)},
\label{eq:dist_f_chrmt}
\eeq
where $\Gamma$ is the Euler gamma function. 
This prediction can be tested numerically, as we shall see later in the paper
(see also Ref.~\cite{Cossu:2019hse}).

\section{Lattice Observables}
\label{Sec:observables}

This section is devoted to defining and discussing the lattice observables of interest in the numerical study.
We start from the spectrum of the Dirac operator, which as explained in Sect.~\ref{Sec:symmetry} 
is closely related to the breaking of the global symmetry.
We then provide details about the lattice implementation of meson and (chimera) baryon operators, and
refresh  for the reader some standard material about the extraction of masses and (renormalised)
decay constants from the appropriate
2-point functions.

Before proceeding, we pause to make two comments of a technical nature. In what follows, we 
express the masses 
and decay constants of composite states in  units of the lattice spacing $a$. 
The reader might, with some reason, think that it
 would be best practice to introduce a non-perturbative scale-setting procedure that 
allows to take the continuum $a\rightarrow 0$ limit without ambiguities. And indeed,
in previous publications 
our collaboration elected to 
adopt to this purpose the Wilson flow~\cite{Luscher:2010iy,Luscher:2013vga}.
Yet, as in this work we do not attempt the continuum limit extrapolation, but rather only extract
lattice measurements in a small number of ensembles, this is not necessary.
Furthermore,  in this theory the fermions have
non negligible dynamical effects---see for example \Fig{force_chart}---and hence the Wilson flow observables
are expected to be quite sensitive to the choice of fermion mass, making a future, dedicated study necessary.
We plan to do so when we will have enough numerical ensembles to perform 
the continuum and chiral limit extrapolations.

The second comment is even more dreary. Throughout this work we use $Z_2\times Z_2$ single time slice stochastic sources~\cite{Boyle:2008rh} in the studies of 2-point correlation functions for mesons, while we use simple point sources for the chimera baryon. However, it is a well known fact among lattice practitioners that extracting the masses of heavy composite states, particularly
in the case of fermionic operators such as the chimera baryon, is complicated by
heavy state contamination and numerical noise~\cite{Lepage:1989hd}.
And it is a known fact that such shortfallings can be addressed by combining 
(Wuppertal) smeared
source and sink operators~\cite{Gusken:1989qx}, 
by (APE) smearing of the gauge links~\cite{APE:1987ehd}
and by adopting
variational methods in treating the eigenvalue problems~\cite{Luscher:1990ck,Blossier:2009kd}.
Again, applying these techniques to our current ensembles would bring us unnecessarily beyond the scopes of this paper.
And yet, as anticipated in Ref.~\cite{Bennett:2021mbw}, at the time of editing this manuscript we have developed most 
of the necessary processes for 
our model, and some of us have been extensively testing them
 on a simpler theory: the partially quenched model in which only
the $(as)$ fermions are included in the MD evolution, while the $(f)$ fermions are treated as external probes.
We will report on this process elsewhere~\cite{AS}, and apply such techniques to 
 the multi-representation theory of interest  in future precision 
studies.

\subsection{Eigenvalues of the lattice Dirac operator}
\label{Sec:dirac}

For the tests described in this subsection, we use ensembles obtained in the quenched approximation.
We denote as $\lambda$ each eigenvalue of the hermitian Dirac operator $Q_m$, 
defined after Eq.~(\ref{Eq:determinant}). 
We compute such eigenvalues via matrix diagonalisation, using the Jacobi algorithm, 
 which is accurate enough to yield all the eigenvalues of the Dirac matrix with dimension up
 to $\sim 5000$.\footnote{
If we restrict ourselves to the computation of the low-lying eigenvalues, 
we can use several techniques for acceleration, such as the subspace iteration 
with Chebyshev acceleration and eigenvalue locking
 (e.g. see the Appendix of Ref.~\cite{DelDebbio:2005qa}), as implemented in the HiRep code.} 
We then sum 
  the eigenvalues of $Q_m^2$, and find
\beq
\Tr Q_m^2 \equiv \sum_{\lambda=\lambda_{\rm min}}^{\lambda_{\rm max}} {\lambda^2}\,,
\label{eq:trq2}
\eeq
which we can  compare to the analytical expression
\beq
\Tr  Q_m^2 = 4 \times d_R \times N_T \times N_S^3 \times (4+(a\,m_0+4)^2),
\label{trqmsquared}
\eeq
where the trace is over  color and spinor indices, while $d_R$ is the dimension 
of the representation $R$, and $N_T$ and $N_S$ are 
the extents of the lattice in the temporal
and spactial directions, respectively.
 As a first test of the numerical processes,
 we calculate the difference between Eqs.~(\ref{eq:trq2}) and (\ref{trqmsquared}), 
 denoted as $\Delta \Tr Q^2_m$, in \Tab{traceQ2}. In the table we report the
 result of our exercise, for several 
 gauge groups and matter representations.  
As can be seen $\Delta \Tr Q_m^2/\Tr Q_m^2 \sim O(10^{-14})$ for all the cases we considered.

\begin{table}
\caption{%
\label{tab:traceQ2}%
The numerical error in the calculation of  $\Tr Q^2_m$ on a lattice of dimensions $4^4$, for
the values of lattice parameters indicated, and for 
five different combinations of quenched theory and fermion representation.
}
\begin{center}
\begin{tabular}{|c|c|c|c|c|c|}
\hline
~~Gauge group~~ & ~~Representation~~ & $~~~\beta~~~$ & $~~~am_0~~~~$ & ~~$\Tr Q^2_m$ from Eq.~(\ref{trqmsquared})~~ & $~~~\Delta \Tr Q^2_m~~~$\\
\hline
$SU(2)$ & $(f)$ & 1.8 & $-1.0$ & 26624 & $8.7 \times 10^{-11}$ \\
$SU(4)$ & $(f)$ & 10.0 & $-0.2$ & 75530.24 & $5.8 \times 10^{-10}$ \\
$SU(4)$ & $(as)$ & 10.0 & $-0.2$ & 113295.36 & $1.5 \times 10^{-9}$ \\
$Sp(4)$ & $(f)$ & 8.0 & $-0.2$ & 75530.24 & $1.9 \times 10^{-9}$ \\
$Sp(4)$ & $(as)$ & 8.0 & $-0.2$ & 94412.8 & $3.5 \times 10^{-9}$ \\
\hline
\end{tabular}
\end{center}
\end{table}

In order to make a comparison with the chRMT prediction in \Eq{dist_f_chrmt},
we need to implement an unfolding procedure 
which consists of  rescaling of the spacing between adjacent eigenvalues by the local spectral density. 
Because the functional form of the density is not known a priori, 
in practice we replace it by the density over many lattice configurations.   
To do so, following the prescription of Ref.~\cite{Cossu:2019hse},
we first compute the eigenvalues of $Q_m$
for a set of $N_{\rm{conf}}$ different configurations.
Each such calculation, for $c=1,\,\cdots,\,N_{\rm{conf}}$, 
yields eigenvalues $\lambda_i^{(c)}$, which we 
list in increasing order, discarding degeneracies.
We then combine all the eigenvalues thus computed in
one, increasingly ordered long list.
And for each $c=1,\,\cdots,\,N_{\rm{conf}}$ we produce a new list,
in which instead of $\lambda_i^{(c)}$ we include $n_i^{(c)}$, defined as the positive integer position
of the eigenvalue $\lambda_i^{(c)}$ in the long list.
The density of spacing, $s$,  is then replaced by the
sequence of $s^{(c)}_i$ given by
\beq
 s^{(c)}_i\equiv\frac{n_{i+1}^{(c)} - n_i^{(c)}}{{\cal N}}\,.
\label{eq:spacing}
\eeq
The constant ${\cal N}$  is defined in such a way that $\langle s \rangle= 1$,
after  averaging $s^{(c)}_i$ over the whole ensemble.
We then define the unfolded density of spacings $P(s)$ as the limiting case
of the  normalised (and discretised) 
distribution function obtained
by binning our numerical results for $s^{(c)}_i$.
We will return in Sect.~\ref{Sec:results} to the explicit comparison of the 
numerical results with the analytical predictions of chRMT.

\subsection{Of mesons on the lattice}
\label{Sec:mesons}

We have  already discussed how the interpolating operators sourcing mesons are defined in the (Minkowski)
continuum theory, 
in particular for pseudoscalars, in Sect.~\ref{Sec:mb}. 
We come now to the (Euclidean) lattice formulation.
Gauge-invariant operators  associated with mesonic states are generically denoted by
\beq
\mathcal{O}_{M}^R(x) = \overline{\chi}(x) \Gamma_M \chi^{\prime}(x),
\label{eq:meson_ops}
\eeq
where $\chi,\chi^{\prime}=Q$ or $\chi,\chi^{\prime}=\Psi$, for 
fermions in representation
$R=(f)$ or $R=(as)$, respectively. 
We suppress here color, flavour, and spinor indexes, for notational simplicity, but we will make them manifest 
when useful.
Adopting Euclidean signature, and specialising to rest-frame (zero-momentum) observables,
the Dirac structures of interest are\footnote{As on the lattice one measures 
correlation functions for zero momentum, 
it is convenient to use $\gamma^0\gamma^{\mu}$ and $\gamma^5\gamma^0\gamma^{\mu}$,
instead of $\sigma^{\mu\nu}$ and $\gamma^5\sigma^{\mu\nu}$, respectively.}
\beq
\Gamma_M =\gamma^5,\,\mathbb{1}, \,\gamma^\mu,\,\gamma^5\gamma^\mu,\,
\gamma^0\gamma^{\mu},\,\gamma^5\gamma^0\gamma^{\mu}
\label{eq:meson_channels}
\eeq
which we label by PS, S, V, AV, T, and AT, 
corresponding to the pseudoscalar, scalar, vector, axial-vector, tensor, and axial-tensor mesons.\footnote{
In the continuum limit, after chiral symmetry breaking, correlation functions involving 
 tensor operator $T$ and vector operator $V$ mix. 
 Also, we anticipate here that will face numerical difficulties in extracting masses
  for the axial-tensor states---for comparison, these states are
 called $b_1$ in two-flavor QCD. 
} 
We restrict our attention to flavoured meson states with $\chi\neq \chi^{\prime}$, so that contributions from
 disconnected diagrams to 2-point functions are absent. 
 As explicitly shown in Eq.~(\ref{Eq:5}), mesons and diquarks combine together to form irreducible
  representations of  $Sp(4)$. For example,  masses and decay 
  constants
  of the five PNGBs are degenerate (see also Ref.~\cite{Bennett:2019jzz}).

The 2-point correlation function for mesons can be written as follows:
\beqs
\langle O_M^R (x) O_{M'}^{R\dagger}(y)\rangle
&=& \langle \overline{\chi}(x) \Gamma_M \chi^{\prime}(x) \overline{\chi^{\prime}}(y) \bar{\Gamma}_{M'} \chi(y)\rangle\nn \\
&=&-\Tr\left[\Gamma_M S^{R\,\prime} (x,y) \bar{\Gamma}_{M'} S^R(y,x)\right]\nn \\
&=&-\Tr\left[\gamma^5 \Gamma_M S^{R\,\prime} (x,y) \bar{\Gamma}_{M'} \gamma^5 S^{R\,\dagger} (x,y)
\right], 
\label{eq:meson_2p_corr}
\eeqs
where $\bar{\Gamma}=\gamma^0 \Gamma^\dagger \gamma^0$.
The fermion propagators are defined by
\beq
S^{\,i\,a}_{Q\,\,\,\,b\,\alpha\beta}(x,y) = 
\langle Q^{i\,a}_{\,\,\,\,\,\alpha}(x) \overline{Q^{i\,b}}_{\beta}(y) \rangle
~{\textrm{and}}~S^{\,k\,ab}_{\Psi\,\,\,\,\,\,\,\,cd\,\alpha\beta}(x,y) =
 \langle \Psi^{k\,ab}_{\,\,\,\,\,\,\,\,\,\,\alpha}(x) 
\overline{\Psi^{k\,cd}}_{\beta}(y) \rangle\,,
\label{eq:fermion_prop}
\eeq
where $a,\,b,\,c,\,d$ are color indices, 
$i,\,k$ are flavor indices, and $\alpha,\,\beta$ are spinor 
indices. 
We also use the $\gamma^5$-Hermiticity property, $S_R(x,y)^\dagger = \gamma^5 
S_R(y,x) \gamma^5$ (see the \App{A2}), in the last line of \Eq{meson_2p_corr}.
With the notation $x\equiv (t,\vec{x})$ and $y\equiv (t_0,\vec{y})$, the zero-momentum correlation function is
\beqs
\label{eq:meson_corr}
C_{\mathcal{O}_{M\,M'}^R}(t-t_0) &=& \sum_{\vec{x}\vec{y}} \langle O_M^R (x) O_{M'}^{R\dagger}(y)\rangle\\
&=&-\sum_{\vec{x}\vec{y}}\Tr\left[\gamma^5 \Gamma_M S^{R\,\prime} (x,y) \bar{\Gamma}_{M'} \gamma^5 S_{R}^{\,\,\,\dagger} (x,y)
\right]. 
\eeqs

At large Euclidean time $t$, the correlation function in \Eq{meson_corr}
 for $M=M'$ has the following asymptotic form
\beq
C_{\mathcal{O}_{M\, M}^R}(t) \longrightarrow
|\langle 0| \mathcal{O}_M^R |M  \rangle|^2
\frac{1}{2 m_M^R} \left[e^{-m_M^R t} +e^{-m_M^R (T-t)}
\right],
\eeq
where $T$ is the temporal extent of the lattice and $m^R_M$ is the mass of the
 ground state meson $|M\rangle$ of type $M$,
 composed of fermions in  representation $R$. 
 The overlap of the interpolating operator $\mathcal{O}_M$ with 
 the  $|PS\rangle$, $|V\rangle$ and $|AV\rangle$ states 
 can be parametrised by
\beq
\langle 0| \mathcal{O}_{\rm AV}^R | {\rm PS} \rangle = \sqrt{2} f_{\rm PS}^R p^\mu,~
\langle 0| \mathcal{O}_{\rm V}^R | {\rm V} \rangle = \sqrt{2} f_{\rm V}^R m_{\rm V} \epsilon^\mu,~{\rm and}~
\langle 0| \mathcal{O}_{\rm AV}^R | {\rm AV} \rangle = \sqrt{2} f_{\rm AV}^R m_{\rm AV} \epsilon^\mu,
\eeq
where $f^R_M$ are the decay constants of the 
corresponding three (ground-state) mesons.\footnote{ 
 The normalisations of the matrix elements are consistent with those
that for 2-flavour QCD yield the pion decay constant $f_\pi \simeq 93$ MeV. }
The polarisation four-vector $\epsilon^\mu$ obeys the two defining relations
 $p_\mu \epsilon^\mu=0$ and $\epsilon_\mu^\ast \epsilon^\mu =1$. 
 To extract the pseudoscalar decay constant, besides
  $C_{\mathcal{O}_{{\rm PS}\,{\rm PS}}^R}(t)$, we need to extract the additional 
  correlation function with $M=AV$ and $M'=PS$:
\beq
C_{\mathcal{O}_{{\rm AV}\, {\rm PS}}^R}(t) \longrightarrow
\frac{1}{\sqrt{2}}
f_{\rm PS}^R \langle 0| \mathcal{O}_{\rm PS}^R |PS  \rangle^\ast
\left[e^{-m_{\rm PS}^R t} -e^{-m_{\rm PS}^R (T-t)}
\right].
\eeq

The decay constants receive multiplicative renormalisation. We computed
 the renormalisation factors in  lattice perturbation theory for Wilson fermions 
 at the one-loop level, with tadpole improvement, 
 following the prescriptions dictated by Refs.~\cite{Martinelli:1982mw,Lepage:1992xa}. 
 The tadpole-improved gauge coupling is defined as $\tilde{g}^2=g^2/\langle P\rangle$,
 with $\langle P\rangle$ the average plaquette.
With the definitions
\beqs
f_{PS}^{R\,ren}&\equiv& Z_A f_{PS}^{R}\,,
\,\,\,\,\,\,\,\,\,\,
f_{V}^{R\,ren}\,\equiv\, Z_V f_{V}^{R}\,,
\,\,\,\,\,\,\,\,\,\,
f_{AV}^{R\,ren}\,\equiv\, Z_A f_{AV}^{R}\,,
\eeqs
and
\beqs
Z_{A,V}&=&1+C^R\left(\Delta_{\Sigma_1}+\Delta_{\Gamma}\right) \frac{\tilde{g}^2}{16\pi^2}\,,
\eeqs
one finds the numerical coefficients required by replacing
 $C^{(f)}=5/4$, $C^{(as)}=2$, $\Delta_{\Sigma_1}=-12.82$, $\Delta_{V}=-7.75$,
 and $\Delta_{AV}=-3.0$~\cite{Bennett:2019cxd}.

\subsection{Of chimera baryons on the lattice}
\label{Sec:chimera}

As discussed in \Sec{model},  chimera baryons are composed of two fermions
 in the fundamental $(f)$ and one in the antisymmetric $(as)$ representations of $Sp(4)$. 
The operators which interpolate the would-be 
 top partners (and their $U(1)_A$ counterparts) in a phenomenologically realistic model
  are displayed in Eqs.~(\ref{Eq:top}) and~(\ref{Eq:topH})---for the purposes of this paper, 
  we can ignore the chiral projection with $P_{L,R}$ in Eqs.~(\ref{Eq:top}) and ~(\ref{Eq:topH}).
  The operators in Eq.~(\ref{Eq:top}) are similar to the non-flavour singlet spin-$1/2$ $\Lambda$ baryon
 operators considered in lattice QCD calculations.  
In general  the interpolating operators of the chimera baryon are
\beq
\mathcal{O}_{\rm CB (CC)}^{\alpha} (x)= D^{\alpha\beta\gamma\delta}
\Omega_{ac} \Omega_{bd} 
Q^{i\,a}_{(C)\,\beta}(x) Q^{j\,b}_{(C)\,\gamma}(x) \Psi^{k\,
cd}_{\,\,\,\,\,\,\,\,\,\,\delta}(x), 
\label{eq:generic_ocb}
\eeq
where $a,\,b,\,c,\,d$ are color indices,  $i,\,j,\,k$ are flavor indices, and $\alpha,\,\beta,\,\gamma,\,\delta$ are spinor 
indices.\footnote{The subscript $Q_C$ denotes the charge-conjugate of the four-component spinor $Q$: 
 because of the pseudo-real nature of the two $(f)$ fermions, the global symmetry acting of them
  is enhanced from $SU(2)_L\times SU(2)_R\times U(1)_A\times U(1)_B$ to $SU(4)\times U(1)_A$,
  and hence the irreducible representations of the global symmetry contain what one would naively
  associate with states with different $U(1)_B$.}    
The tensor (in spinor space) $D^{\alpha\beta\gamma\delta} $ 
can be written as a combination of gamma matrices, which projects onto the desired spin state. 

We restrict our attention to spin-0 combinations of the two $(f)$ fermions, introduce the notation
$(\Gamma^1,\,\Gamma^2)\equiv (C\gamma^5,\,\mathbb{1})$, and restrict
$D^{\alpha\beta\gamma\delta} $ in \Eq{generic_ocb} to be made of combinations of $\Gamma^1$ and 
$\Gamma^2$.\footnote{Extending the basis 
 to include  other gamma structures goes beyond our current purposes.
Nevertheless, allowing for redundancies  in defining the
variational basis might  improve the numerical signal in a precision study.}
For instance, the linear combination 
$\frac{1}{2}\left(i\mathcal{O}_{{\textrm{CB},\,4}}
-\mathcal{O}_{\textrm{CB},\,5}\right)$  can be written as follows:
\beq
\overline{Q^{2\,a}_C} Q^{1\,b} \Omega_{bc} \Psi^{k\,ca} 
=-\Omega_{da}\Omega_{bc}  (
Q^{2\,d\,T} \Gamma^1 Q^{1\,b})
\Gamma^2 \Psi^{k\,ca},
\label{eq:ocb4}
\eeq
where the Dirac adjoint of $Q$ and its charge conjugate $Q_C$ are given by
\beqs
\overline{Q^a} &=& (Q^{a})^\dagger \gamma^0 = -Q_C^{b\,T} \Omega_{ba} (C\gamma^5), \\
\overline{Q^a_C} &=& -(Q^{a}_C)^\dagger \gamma^0 = -Q^{b\,T} \Omega_{ba} (C\gamma^5). 
\eeqs
In our numerical studies for the spin-$1/2$ chimera baryon, 
we find it convenient to use the operator in \Eq{ocb4}, rewritten as follows: 
\beq
\mathcal{O}^{k\,\gamma}_{\rm CB}(x) = (\Gamma^{1})^{\,\alpha\beta} (\Gamma^{2})^{\,\gamma\delta} \Omega_{da} \Omega_{bc} {Q^{2\,a}_{\,\,\,\,\,\,\alpha}(x)}  Q^{1\,b}_{\,\,\,\,\,\,\beta}(x) \Psi^{k\,\, cd}_{\,\,\,\,\,\,\,\,\,\,\,\,\delta}(x)\,.
\label{eq:cb_ops}
\eeq
Its Dirac conjugate operator is
\beq
\overline{\mathcal{O}^{k}_{\rm CB}}^{\,\gamma}(x) = 
\left(\Gamma^1\right)^{\alpha\beta} \left(\Gamma^2\right)^{\delta\gamma}
 \Omega^{da} \Omega^{bc}
\overline{\Psi^{k\,cd}}_{\,\delta}(x)  \overline{Q^{2\,a}}_\alpha(x) \overline{Q^{1\,b}}_\beta(x). 
\label{eq:cb_ops_conjugate}
\eeq
After Wick contractions, the propagator for the chimera baryon with flavour $k$ reads
\beqs
\langle {\mathcal{O}^{k}_{\rm CB}}^{\gamma}(x) 
\overline{\mathcal{O}^{k}_{{\rm CB}}}^{\,\gamma'}(y) \rangle
&=& 
\Omega_{da} \Omega_{bc} \Omega^{d'a'} \Omega^{b'c'} (\Gamma^1)^{\alpha\beta} (\Gamma^1)^{\alpha'\beta'} 
(\Gamma^2)^{\gamma\delta} (\Gamma^2)^{\delta^{\prime}\gamma^{\prime}}\times
\nn\\
&& ~~~
\times S^{\,k\,cd}_{\Psi\,\,\,\,\,\,c'd'\,\delta\delta^{\prime}}(x,y) 
S^{2\,a}_{Q\,\,\,\,a'\,\alpha\alpha'}(x,y) 
S^{1\,b}_{Q\,\,\,\,b^{\prime}\,\beta\beta'}(x,y),
\label{eq:cb_corr}
\eeqs
with the fermion propagators  in \Eq{fermion_prop}.

If we define, for convenience, 
\beqs
\tilde{S}_{\Psi}^k\equiv\Gamma^2 S_{\Psi}^k \Gamma^{2\,T}\,,~
S^U\equiv\Omega S_Q^2 \Omega^T~{\rm and}~
\tilde{S}^D\equiv\Gamma^1\left(\Omega^T S_Q^1 \Omega
\right)\Gamma^{1\,T}\,,
\eeqs
with $k$ the flavour index, and color indexes understood
(but notice that $S^U$ and $S^D$ have lower first 
and upper second  color index,
thanks to the action of $\Omega_{ab}$ on the left and $\Omega^{ab}$ on the right),
then the correlation function in \Eq{cb_corr}, evaluated at positive Euclidean time $t-t_0>0$ 
and zero momentum $\vec{p}=0$, 
for $\gamma=\gamma^{\prime}$, 
takes the more compact form
\beqs
C_{\rm CB}^k(t-t_0) 
=\sum_{\vec{x}\vec{y}}
\Tr_{\hspace{-2pt} s} \tilde{S}^{k\,cd}_{\Psi\,\,\,\,\,\,\,c'd'}(x,y) \Tr_{\hspace{-2pt} s}\left[
S^{U\,\,\,\,d'}_{\,\,\,\,\,d}(x,y) 
\left(\tilde{S}^{D\,\,\,c^{\prime}}_{\,\,\,\,\,c}(x,y)\right)^T
\right],
\label{Eq:cb_corr_compact}
\eeqs
with $\Tr_{\hspace{-2pt} s}$ the trace over spinor indexes, and the transposition only acts on the spinorial indexes. 
Because of the antisymmetric properties of the $\tilde{S}_{\Psi}$ indexes, 
we can rewrite the color contractions by  antisymmetrising over the color indices 
of the fundamental propagators $S^U$ and $S^D$, by defining a
new object:
\beq
S_{DQ\,A}^{\,\,\,\,\,\,\,\,\,\,\,\,\,\,\,B}(x,y)\equiv \Tr_{\hspace{-2pt} c}\left[
(e_{AS}^A)^\dagger S^U(x,y) (e_{AS}^B) \left(S^D(x,y)\right)^T
\right], 
\label{eq:sdq}
\eeq
where $\Tr_{\hspace{-2pt} c}$ is a trace over color, while
 $A,B=1,\,\cdots,\,5$ denote the ordered pairs of  color indices $(ab)$, 
 with the convention  introduced in \Sec{action}---see  Eqs.~(\ref{eq:eas_diag}) and (\ref{eq:eas}).  
Using $S_{DQ}$ in \Eq{sdq}, we arrive at 
\beq
C_{\rm CB}^k(t-t_0) = 
\sum_{\vec{x}\vec{y}}
\Tr_{\hspace{-2pt} s} S_{\Psi\,\,\,\,B}^{k\,A}(x,y) \Tr_{\hspace{-2pt} s} S_{DQ\,A}^{\,\,\,\,\,\,\,\,\,\,\,\,\,\,B}(x,y)\,.
\label{eq:cb_corr_final}
\eeq
While we have considered the chimera baryon propagators built
 out of $\mathcal{O}_{{\rm CB},4(5)}$ in the above discussion, in \App{B} 
 we explicitly show that those built out of $\mathcal{O}_{{\rm CB},1(2)}$
  are identical to $C_{\rm CB}(t-t_0)$ in \Eq{cb_corr_final}. 

As in the case of mesons, at large Euclidean time the (zero-momentum) 2-point correlation functions
involving chimera baryon are dominated by the contributions of the lowest states
 in the given channel.  
 Without loss of generality, we localise the source at the origin $\overrightarrow{y}=\overrightarrow{0}$. 
 As $t\rightarrow \infty$, the asymptotic behaviour of correlator is a textbook example~\cite{Montvay:1994cy}:
\beqs
C_{\rm CB}(t) &\equiv& \sum_{\vec{x}} \langle  \mathcal{O}_{\rm CB} (x) \overline{\mathcal{O}}_{\rm CB}(0) \rangle \nn \\
&\longrightarrow& \mathcal{P}_+ \left[ c_{{\rm CB}}^+ e^{-m_{\rm CB}^+ t}
 + c_{\rm CB}^- e^{-m_{\rm CB}^-(T-t)}\right] - \mathcal{P}_- 
 \left[ c_{{\rm CB}}^- e^{-m_{\rm CB}^- t} + c_{\rm CB}^+ e^{-m_{\rm CB}^+(T-t)}\right],\nn \\
\label{eq:cb_prop_asymptotic}
\eeqs
where the prefactor $\mathcal{P}_\pm\equiv(1\pm\gamma^0)/2$ arises from the sum over spin at zero 
momentum, which is nothing but the parity projector in the nonrelativistic limit. 
(Note that we impose antisymmetric boundary condition for fermions in the temporal extent.)
The coefficients $c_{\rm CB}^\pm$ denote the overlap of the interpolating operator
 $\mathcal{O}_{\rm CB}$ with positive and negative parity states.
 Indeed, in the infinite volume lattice ($T\rightarrow \infty$), the second terms in the 
 brackets in \Eq{cb_prop_asymptotic}---the backward propagators---vanish.

In order to extract the masses of both parity even and odd chimera baryon states we
 isolate those states as yielded by \Eq{cb_prop_asymptotic}. 
In the nonrelativistic limit, the operator which interpolates the chimera baryon 
with definite spin and parity $\frac{1}{2}^\pm$ is defined by
\beq
\mathcal{O}_{\rm CB}^\pm(x) \equiv \mathcal{P}_\pm \mathcal{O}_{\rm CB}(x),
\label{eq:cb_ops_parity}
\eeq
where the interpolating operator $\mathcal{O}_{\rm CB}$ is defined in \Eq{cb_ops}. Accordingly, we define the $2$-point correlation function for $\mathcal{O}_{\rm CB}^\pm$ at zero momentum as
\beqs
C_{\rm CB}^{\pm}(t) &\equiv& \sum_{\vec{x}} \langle  \mathcal{O}_{\rm CB}^\pm (x) \overline{\mathcal{O}}_{\rm CB}^{\,\pm}(0) \rangle, \nn \\
&=& 
\sum_{\vec{x}}
\Tr_{\hspace{-2pt}s} \left[\Gamma^2 \mathcal{P}_\pm \Gamma^2 S_{\Psi}^{A,B}(x,0)\right]\, \Tr_{\hspace{-2pt}s} 
 S_{DQ}^{A,B}(x,0),
\eeqs
where $S_{DQ}$ is defined in \Eq{sdq}. At large Euclidean time, the asymptotic behaviour of $C_{\rm CB}^\pm$ can be written by
\beq
C_{\rm CB}^{\pm}(t) \longrightarrow c_{{\rm CB}}^\pm e^{-m_{\rm CB}^\pm t} + c_{\rm CB}^\mp e^{-m_{\rm CB}^\mp(T-t)}.
\label{eq:cb_corr_parity}
\eeq
The forward and backward propagators for the parity even state decay with the masses of 
$m_{\rm CB}^+$ and $m_{\rm CB}^-$, and conversely for the parity odd state. 

Without parity projection, and at finite $T$
but for $t$ large enough to see the asymptotic 
behaviours of $C_{\rm CB}(t)$,
  the correlation function in \Eq{cb_prop_asymptotic} 
  is eventually 
   dominated by the lightest state,
and the forward and backward contributions have 
the same coefficients up to opposite sign. 
As will be discussed in the next section, it turns 
out that the lightest state is parity-even. After 
taking the trace over the spin, hence, we find 
$C_{\rm CB}(t)$ at large Euclidean time as
\beq
C_{\rm CB}(t)\longrightarrow c_{{\rm CB}}^+ \left(e^{-m_{\rm CB}^+ t} - e^{-m_{\rm CB}^+(T-t)}\right).
\label{eq:cb_corr_no_parity}
\eeq

\section{Numerical results}
\label{Sec:results}

\begin{figure}
\begin{center}
\includegraphics[width=.59\textwidth]{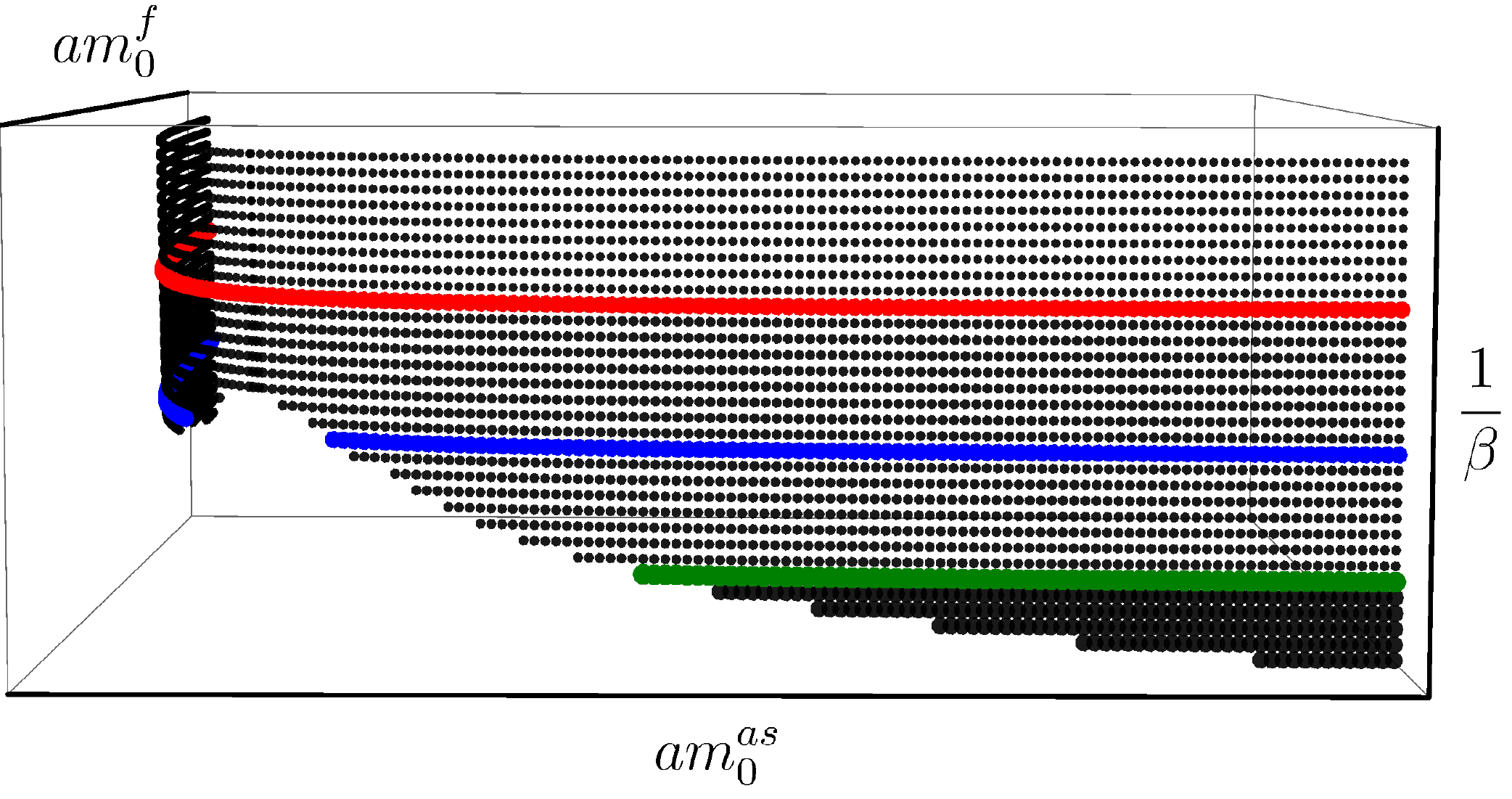}
\caption{%
\label{fig:mr_bulk_phase}%
A schematic representation of the phase diagram in the  space of bare parameters
of  the $Sp(4)$ gauge theory
coupled to $N_f=2$ fundamental and $n_f=3$ two-index antisymmetric Wilson-Dirac fermions. 
The three bare parameters are the lattice gauge coupling, $\beta$, and 
the bare fermions masses, $am^{\rm f}_0$ and $am^{\rm as}_0$,
 for the fundamental and antisymmetric representations, respectively. 
The black-dotted surface denotes the location of  first-order bulk phase transitions. 
On this surface with boundary, we identify three lines of phase-transitions at fixed coupling  $\beta$: 
red, blue and green denote choices of decreasing coupling.
The red line is continuous, while the blue and green lines are interrupted,
as they cross the boundary of the surface.
A critical line of second-order phase transitions is met at the end of first-order lines,
followed by intervals in the numerical values of the masses for which a
smooth  cross-over takes place.
}
\end{center}
\end{figure}

In this section we present our main numerical results for the $Sp(4)$ theory of interest.
We study the phase space of the lattice theory, the spectrum of the Dirac operator (quenched), 
the spectrum of mesons, 
and some  important features of the chimera baryon correlation functions.
We also assess the size of finite-volume effects.
Our results are available in machine-readable form
 in Ref.~\cite{datapackage}. The software workflow used to analyse the data 
 and prepare the plots and tables are made available in Ref.~\cite{analysiscode}.

\subsection{Phase structure of the lattice theory}
\label{Sec:pmspace}

In the limit of infinite volume, the lattice action in \Eq{lattice_action} has three tunable parameters:
 the lattice coupling $\beta$ and the two bare masses $am_0^f$ and $am_0^{as}$ of the $(f)$ and $(as)$
 (Wilson-Dirac)  fermions, respectively.
 The continuum theory is expected to be recovered at the quantum critical point of the lattice theory, 
 which is connected to the (appropriately defined) limit of large $\beta$ and small lattice spacing.
 In practical numerical studies, we work with finite lattice parameters, and therefore it is important
  to choose the lattice parameters  in a way that can be smoothly connected and extrapolated to
  the desired continuum theory.
To do so, in this subsection we explore the parameter space of the lattice theory,
 identify the phase boundary between its strong- and weak-coupling regimes, 
 and investigate the properties of the phase transitions.

Firstly, we  recall that the bulk phase structures of $Sp(4)$ with and without (Wilson) fermions,
 either in the fundamental or the antisymmetric representations,
  have already been studied numerically on the lattice. 
  In the Yang-Mills case, the study of the standard plaquette action shows
   that there is no bulk phase transition~\cite{Holland:2003kg}.
In the presence of fermionic matter,  first order bulk phase transitions have been found,
for both choices of $(f)$ and $(as)$ (Wilson-Dirac) fermions~\cite{Bennett:2017kga,Lee:2018ztv}. 
Interestingly, by comparing the results for the theory with 
$N_f=2$ fundamental fermions, against the theory with  $n_f=3$ antisymmetric fermions,
one finds that the weak coupling regime extends to different values of $\beta$,
reaching to smaller values  in the case of $(as)$ fermions---the critical values of $\beta$, demarcating
 strong and weak coupling regimes, are $\beta_{\rm cr}^f\sim 6.7$~\cite{Bennett:2017kga}
 and~$\beta_{\rm cr}^{as}\sim 6.5$~\cite{Lee:2018ztv}.

\begin{figure}
\begin{center}
\includegraphics[width=.59\textwidth]{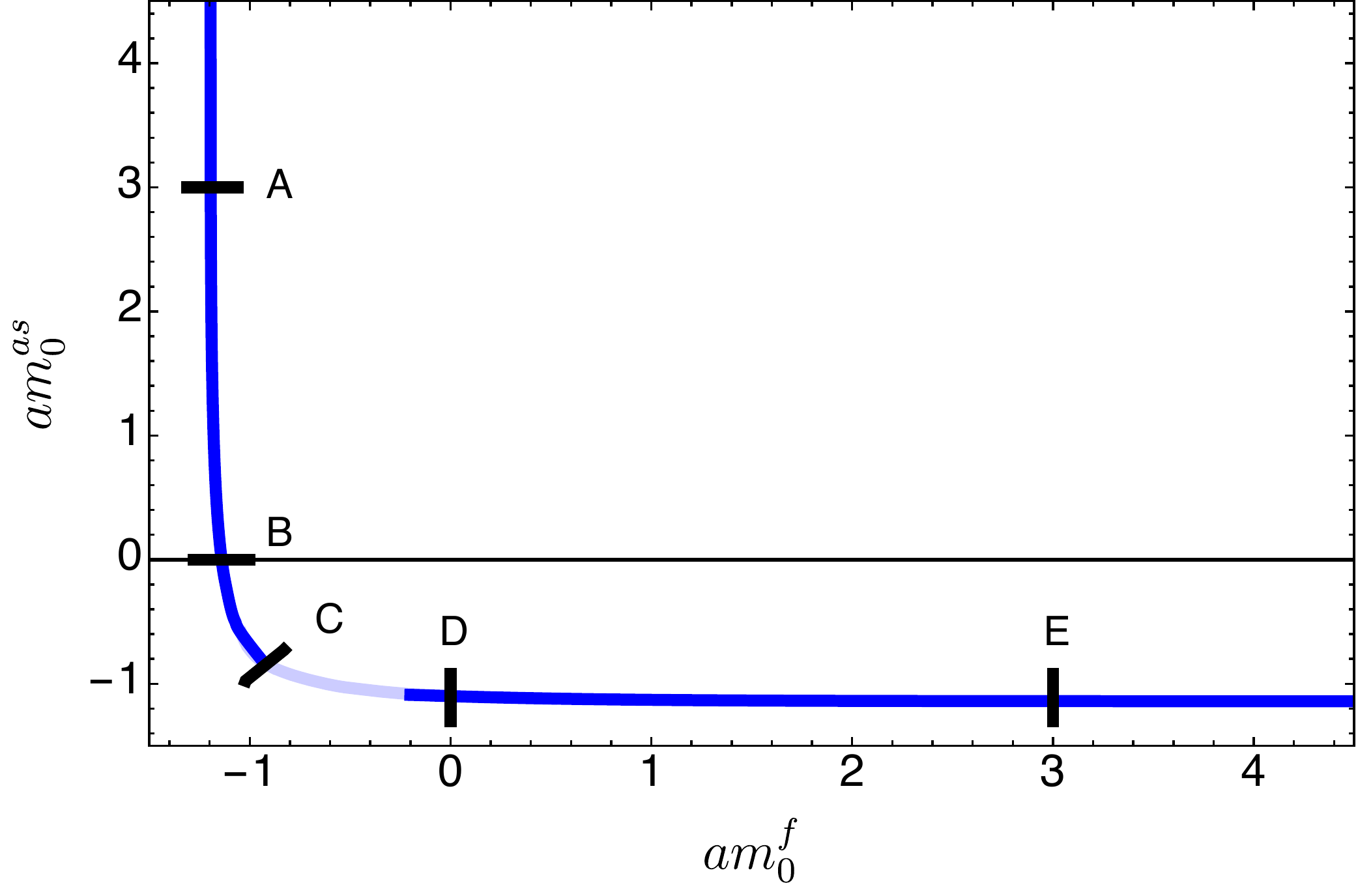}
\caption{%
\label{fig:bulk_phase_b6.4}%
A schematic representation of the  phase structure in the 
$Sp(4)$ gauge theory with $\beta=6.4$, concomitantly
coupled to $N_f=2$ fundamental and $n_f=3$ two-index antisymmetric fermions,
as a function of the two bare masses $a m_0^f$ and $a m_0^{as}$.
The blue solid line is the same that appears  in \Fig{mr_bulk_phase},
and it consists of first-order bulk phase transitions, 
while along the light blue a smooth crossover takes place. 
}
\end{center}
\end{figure}

\begin{figure}
\begin{center}
\includegraphics[width=.49\textwidth]{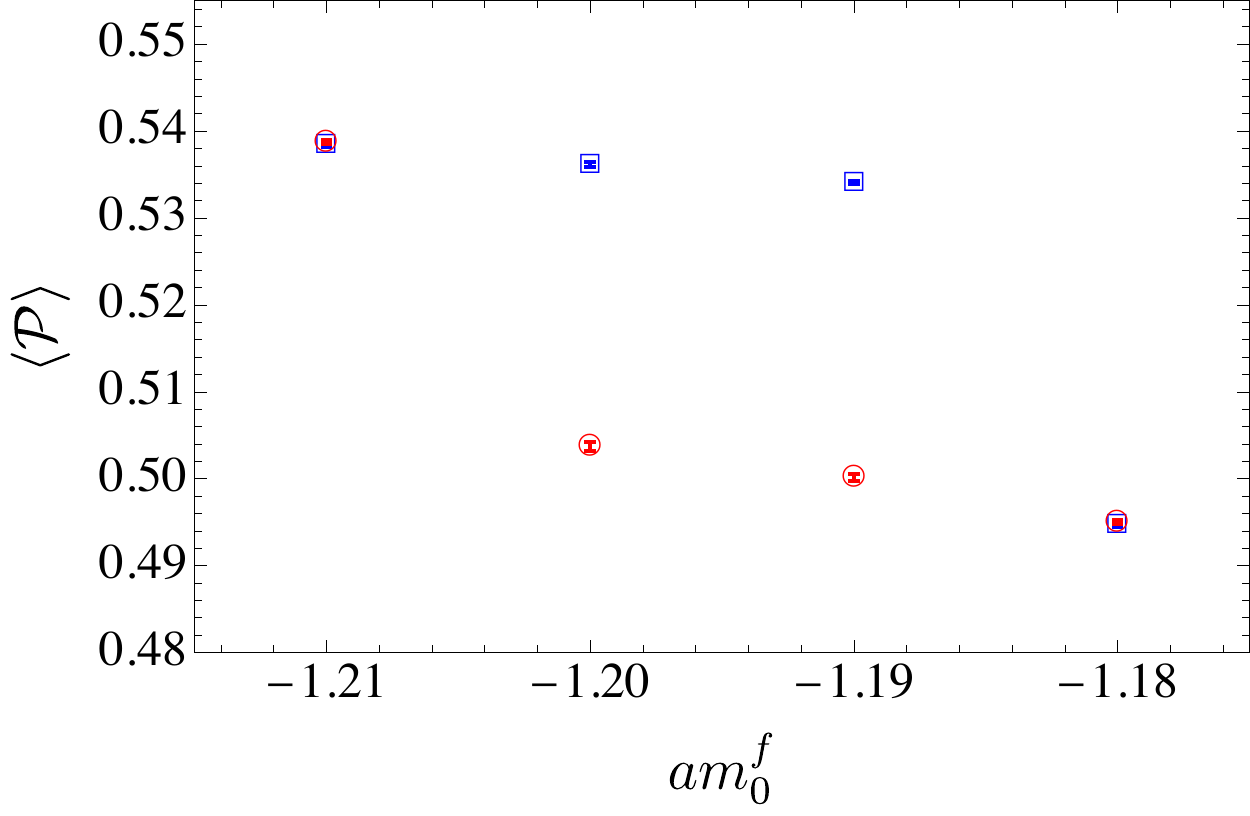}
\includegraphics[width=.49\textwidth]{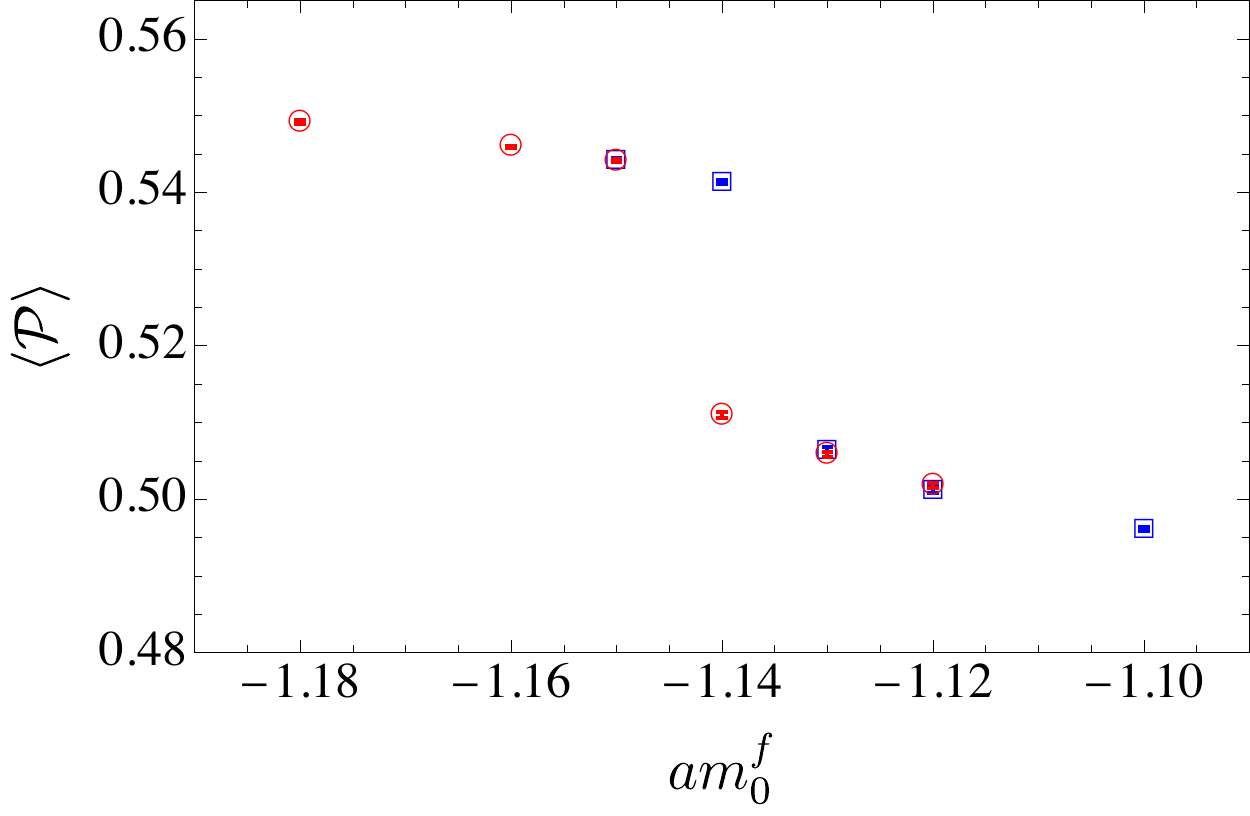}
\includegraphics[width=.49\textwidth]{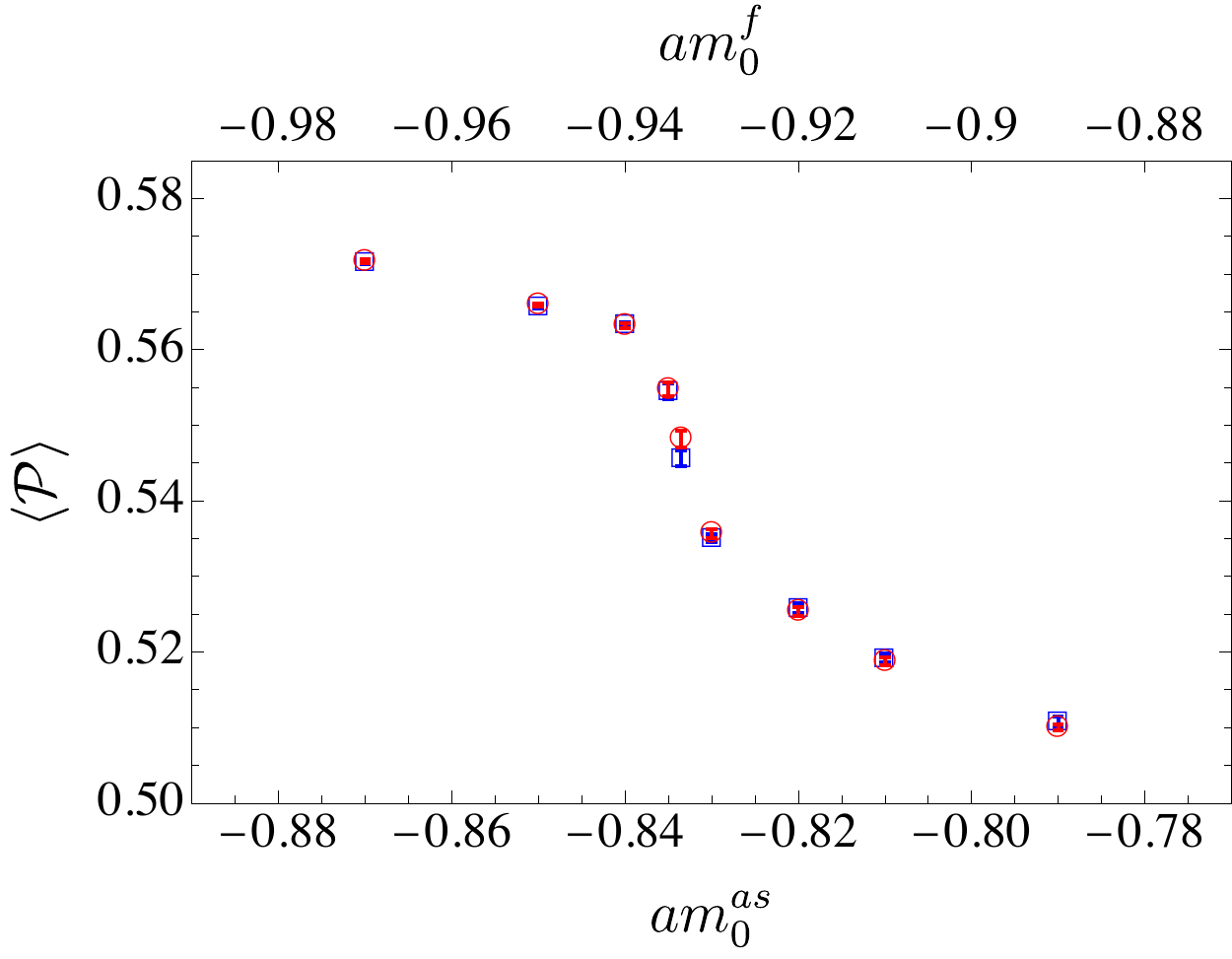}
\includegraphics[width=.49\textwidth]{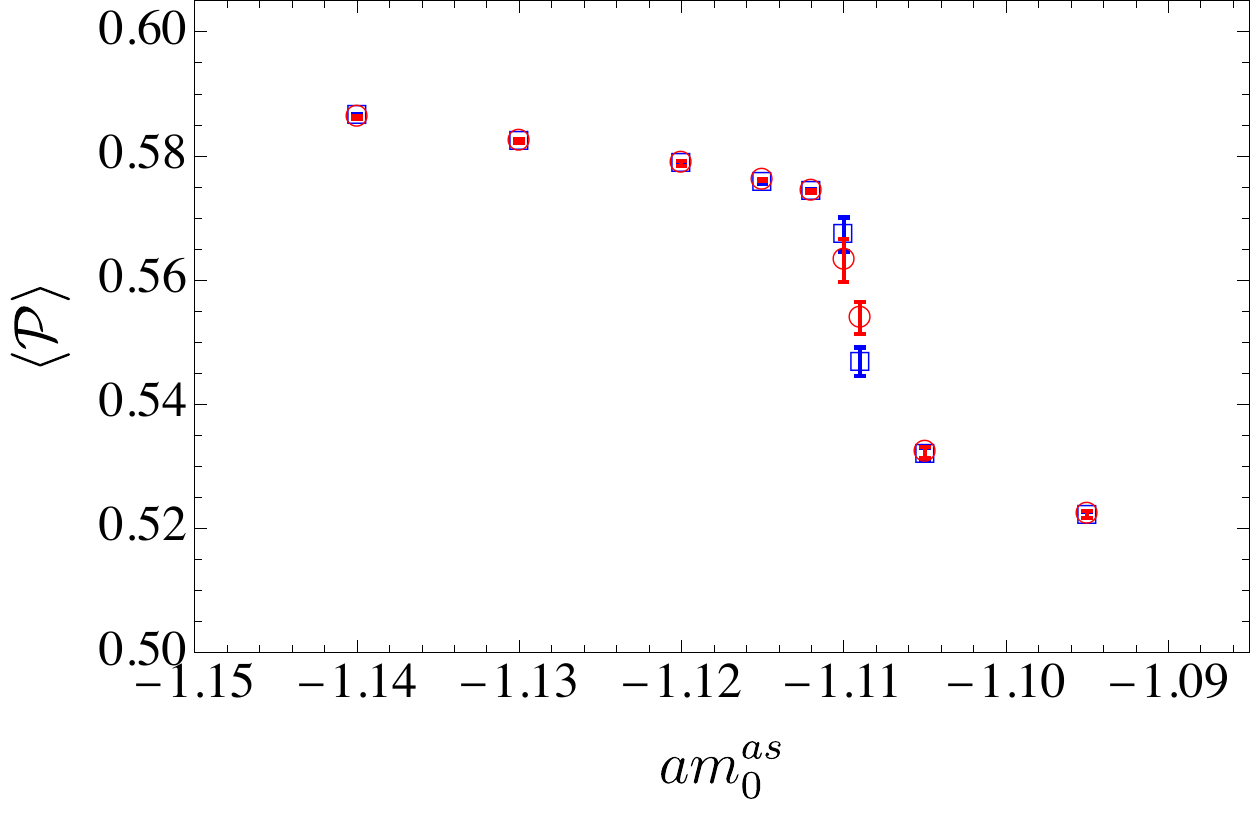}
\includegraphics[width=.49\textwidth]{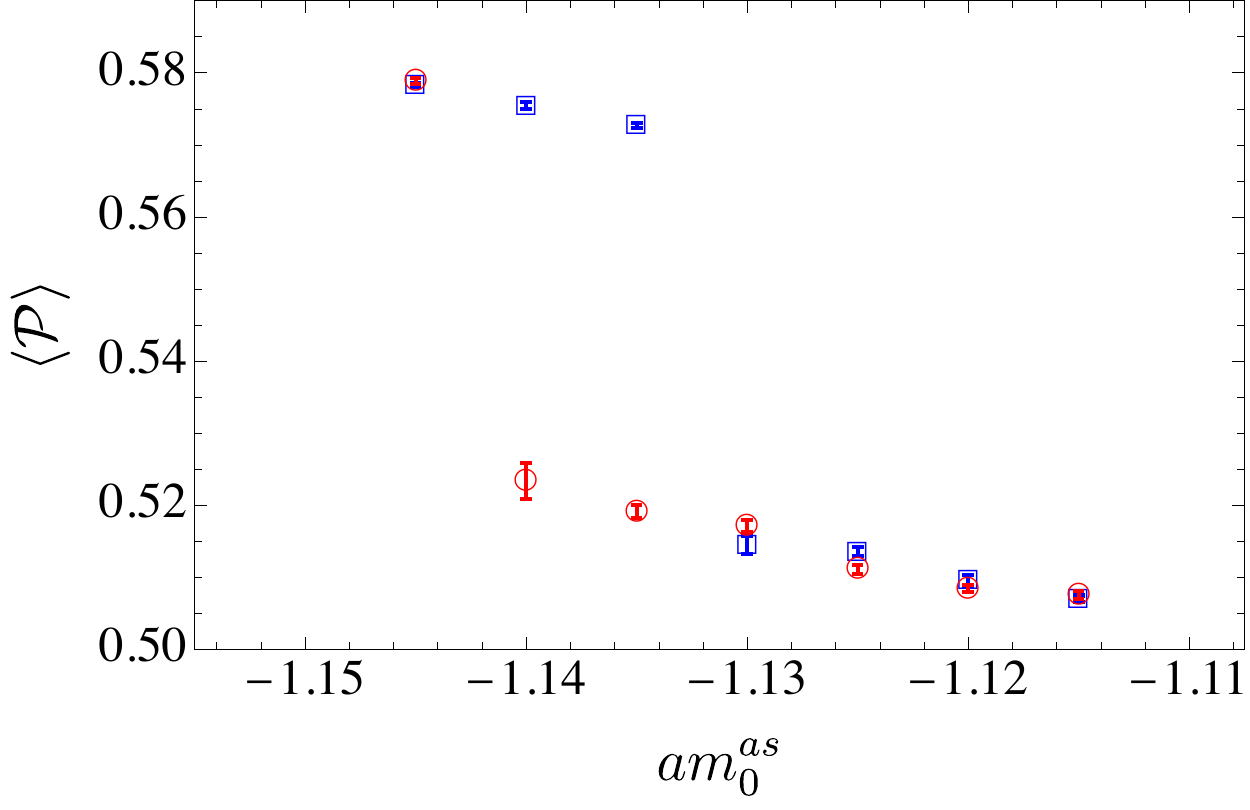}
\caption{%
\label{fig:plaq_b6.4}%
Average value $\langle {\cal P}\rangle$ of the plaquette, computed with
ensembles that start even  from a cold 
(unit, blue squares) and hot (random, red circles) configurations, 
for lattice with dimension  $8^4$. 
From top-left  to bottom panel, the lattice parameters correspond to
the segments denoted by  A, B, C, D, E in \Fig{bulk_phase_b6.4}. 
}
\end{center}
\end{figure}

Starting from these observations, we sketch in \Fig{mr_bulk_phase} the putative bulk phase diagram 
of the  $Sp(4)$ gauge theory coupled to mixed representation fermions. 
The black-dotted  surface represents a surface with boundary of first-order bulk phase transitions. 
For illustrative purposes, we also display
 three colored lines indicating 
 the first order phase transitions for fixed choices of  $\beta=6.2$ (red), $6.4$ (blue) and $6.6$ (green). 
 The red line illustrates how, for small values of $\beta$,
  we expect that a first order phase transition always occurs when we
  perform a mass scan in the 2-dimensional space of $am_0^f$ and $am_0^{as}$. 
  With moderate $\beta$, exemplified by the blue line, 
  the first order lines disappear in some central region of parameter space,
 in which both species of fermions have small masses. 
 We expect  the first-order surface to be asymmetric with respect to the exchange of  $am_0^{\rm as}$ 
 and $am_0^{\rm f}$, as suggested  by the different
 critical values of $\beta$. 
 If we further increase $\beta$,  one of the two lines disappears, and
 the line of first-order transitions only exists for heavy $(as)$ fermions,
  regardless of the treatment of the $(f)$ fermions.
Eventually, we expect even this line to disappear at larger values of $\beta$.

\begin{figure}
\begin{center}
\includegraphics[width=.49\textwidth]{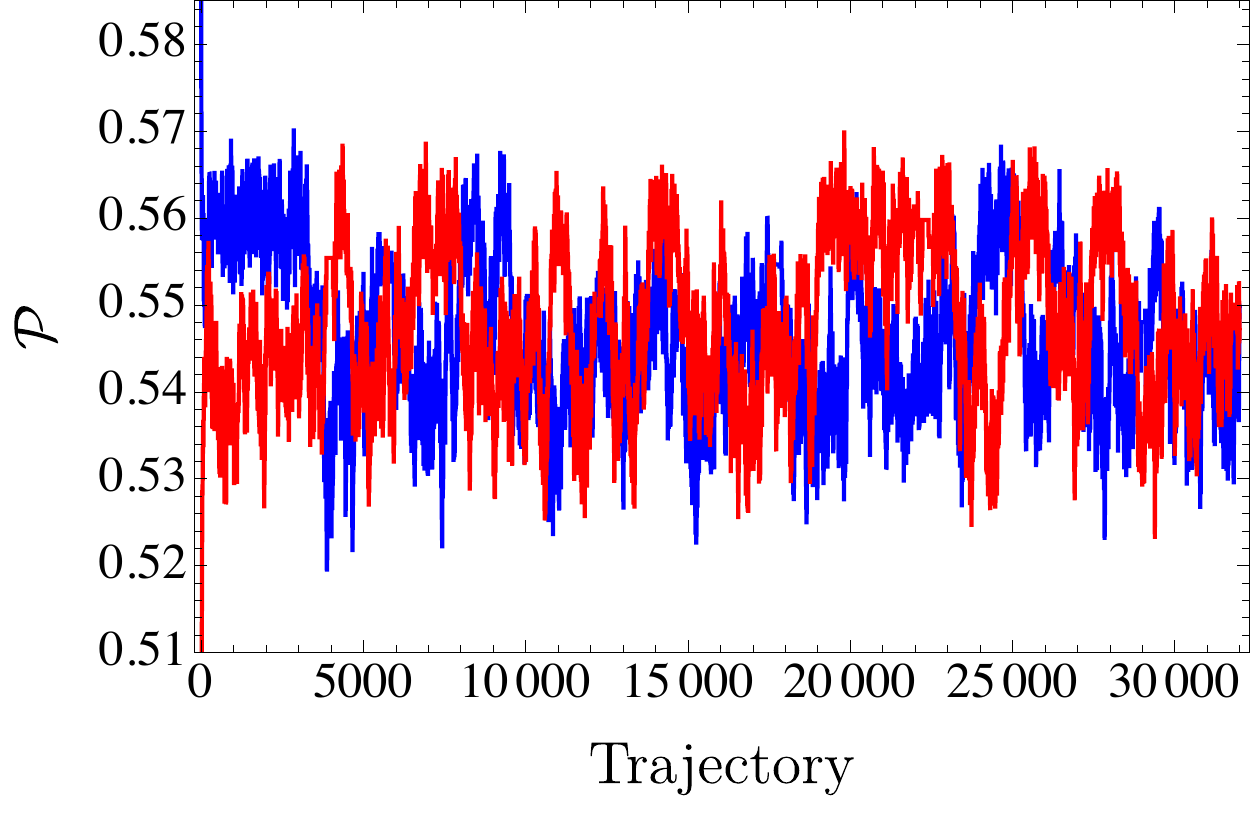}
\includegraphics[width=.49\textwidth]{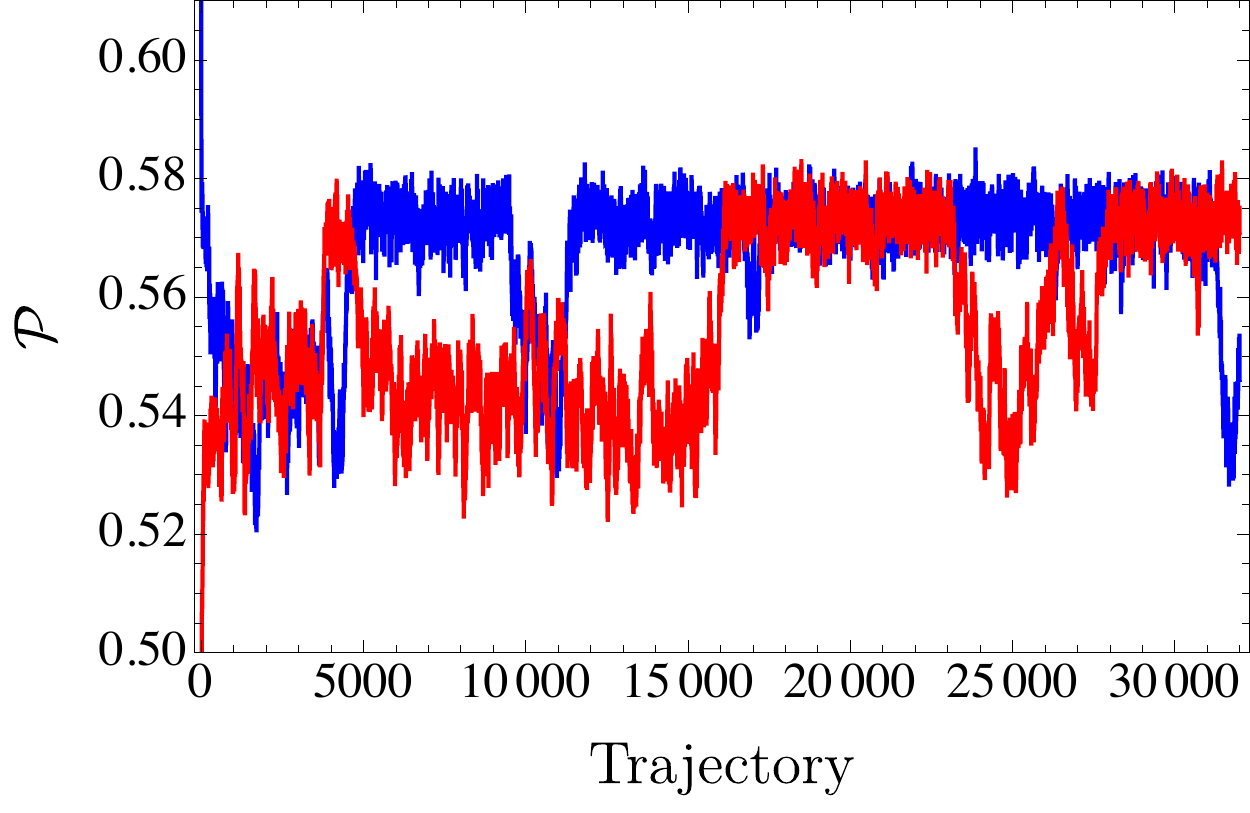}
\caption{%
\label{fig:plaq_traj}%
Illustrative examples of Monte Carlo trajectories of  
the plaquette ${\cal P}$. 
The lattice parameters are 
 $(\beta,\,am_0^f,\,am_0^{as})=(6.4,\,-0.8335,\,-0.9335)$ (left panel) 
and $(6.4,\,0.0,\,-1.11)$ (right panel). These correspond to two sets of parameters 
belonging to segments C and D, respectively, in \Fig{bulk_phase_b6.4}. We use the lattice with size $8^4$. 
The diagrams are obtained with ensembles generated from hot (red) and cold (blue) start.
 Small, but persistent hysteresis effects are clearly visible, with long self-correlation appearing.
}
\end{center}
\end{figure}

\begin{figure}
\begin{center}
\includegraphics[width=.49\textwidth]{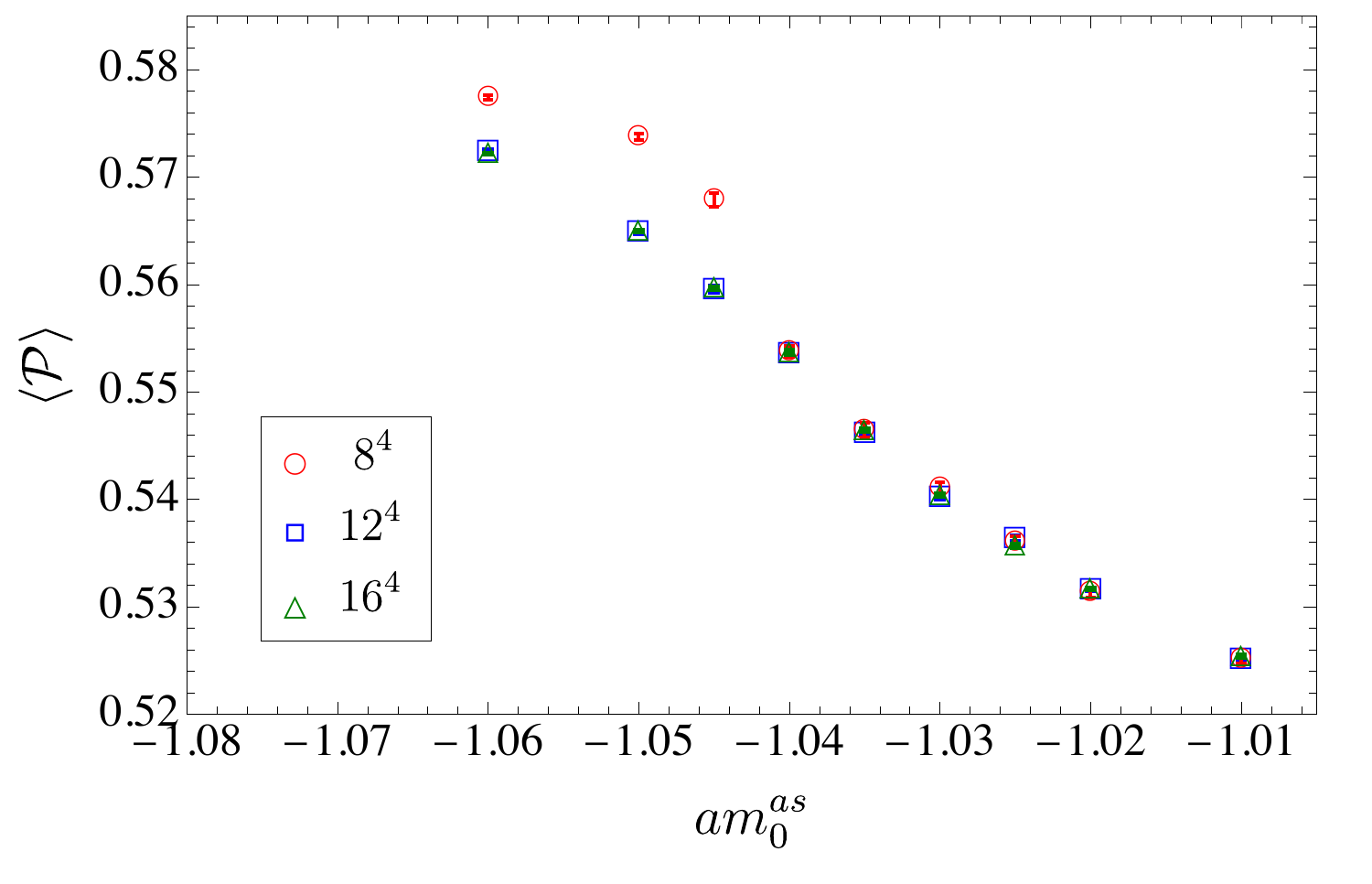}
\includegraphics[width=.49\textwidth]{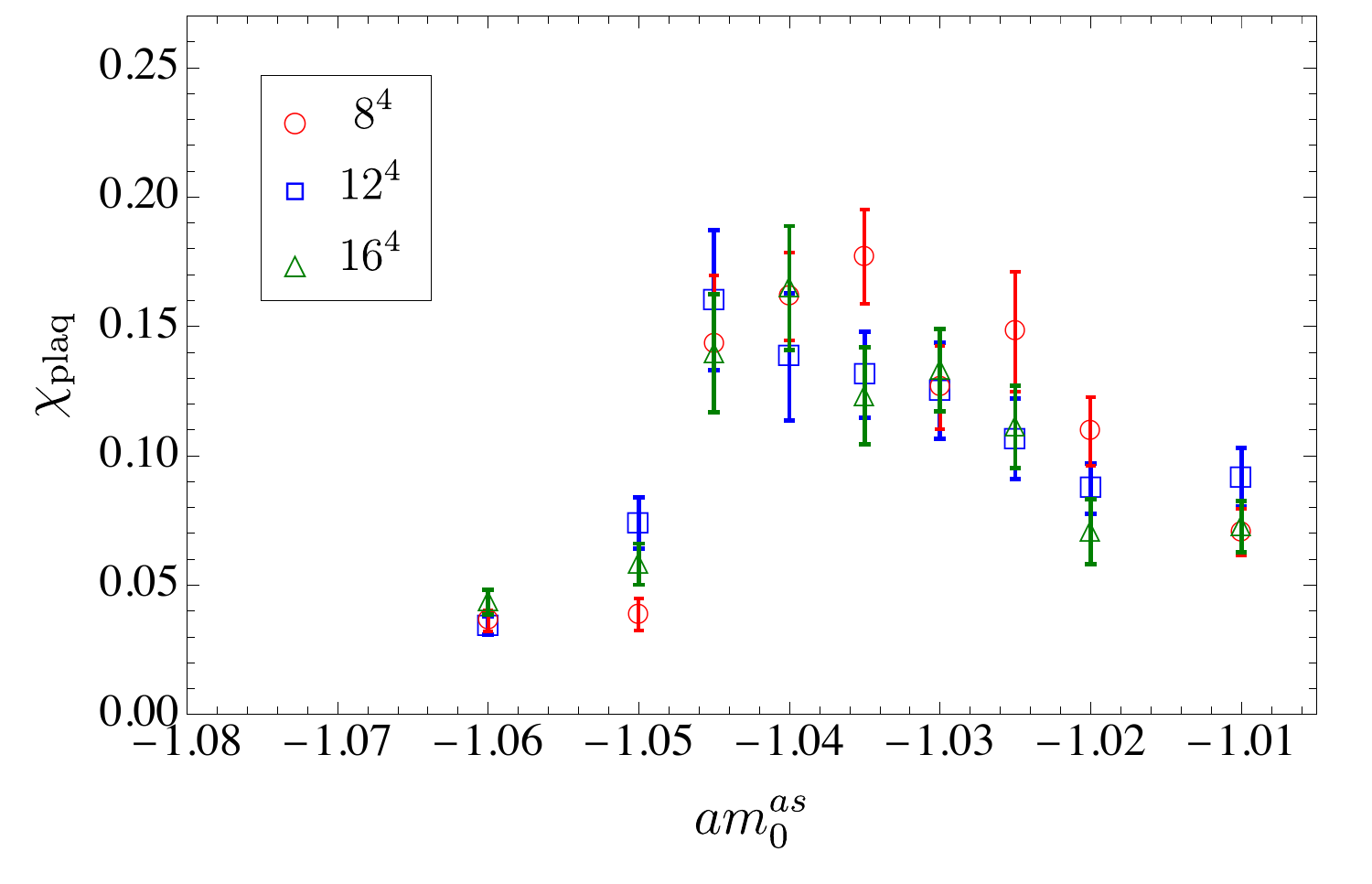}
\caption{%
\label{fig:plaq_sus_b6.4}%
Average plaquette $\langle {\cal P}\rangle$ (top panel) and its susceptibility $\chi_{\rm plaq}$  (bottom panel),
for three  choices of lattice volume, as indicated in the legends. 
The lattice parameters $\beta=6.4$ and $am_0^f=-0.6$ are fixed, and
display the dependence on $a m_0^{as}$.  
}
\end{center}
\end{figure}

To provide numerical support for the conjectured phase diagram in \Fig{mr_bulk_phase}, 
we start by performing mass scans for several representative sections of the
parameter space at fixed $\beta=6.4$, chosen to cut across the phase 
boundary---the blue line in the figure. 
Figure~\ref{fig:bulk_phase_b6.4} depicts
 the regions of parameter space of interest.
Numerical results in the five segments of parameter space denoted by
 A, B, C, D, and  E are shown in some detail  in \Fig{plaq_b6.4}. 
 We compute the average plaquette values using ensembles generated 
 on lattice of size  $8^4$, with an initial configuration 
 of either unit (cold) 
 or random (hot) link variables.

 We find strong evidence of hysteresis
 in cases A, B, and E, indicating the existence of a first-order phase transition,
in correspondence to 
the thick blue lines in \Fig{bulk_phase_b6.4}. 
 By comparing the behavior in the segments A and B, the wider mass range over which hysteresis exists
 in the former case seems to indicate that the strength of the phase transition grows as $am_0^f$ increases,
so that we expect that the first order lines persist all the way to the infinite mass case,
 for which either the fundamental or antisymmetric fermions are non-dynamical (quenched). 
In the heavy mass limits, this is consistent with recovering earlier results in the literature~\cite{Bennett:2017kga,Lee:2018ztv}.

 In cases  C and D,   \Fig{plaq_b6.4} no longer shows
clear evidence of  strong hysteresis. Yet, 
in proximity of the points with  steepest slope, we find that the fluctuations between two preferred  plaquette 
values in the Monte Carlo trajectories display  long autocorrelation time. 
Illustrative examples for the two cases are shown in \Fig{plaq_traj}. 
The combination of weaker transition and longer correlation length are  typical behaviors expected in proximity to the
end of first order lines, that reach critical points, before giving way to
a crossover region.  We illustrate this behaviour with the light blue line in \Fig{bulk_phase_b6.4}. 

To further substantiate these claims, we carry out a finite volume analysis of the plaquette 
susceptibilities at a fixed value of $\beta$ and of the mass of the fundamental fermions $am_0^f=-0.6$.
These choices identify a region  lying between  C and D in \Fig{bulk_phase_b6.4}. 
The results of this analysis are shown in \Fig{plaq_sus_b6.4}: 
in the upper and lower panels we plot the average value of the 
plaquette $\langle \mathcal{P}\rangle$ and the susceptibilities $\chi_{\rm plaq}$, respectively, 
measured in three different volumes and for various choices of the mass of the antisymmetric fermions. 
The value of the plaquette interpolates between two values typical of the two phases of the theory.
But we find that the height of the peak of $\chi_{\rm plaq}$ is independent of the volume, 
which is a typical signature of a smooth crossover.

\begin{figure}
\begin{center}
\includegraphics[width=.59\textwidth]{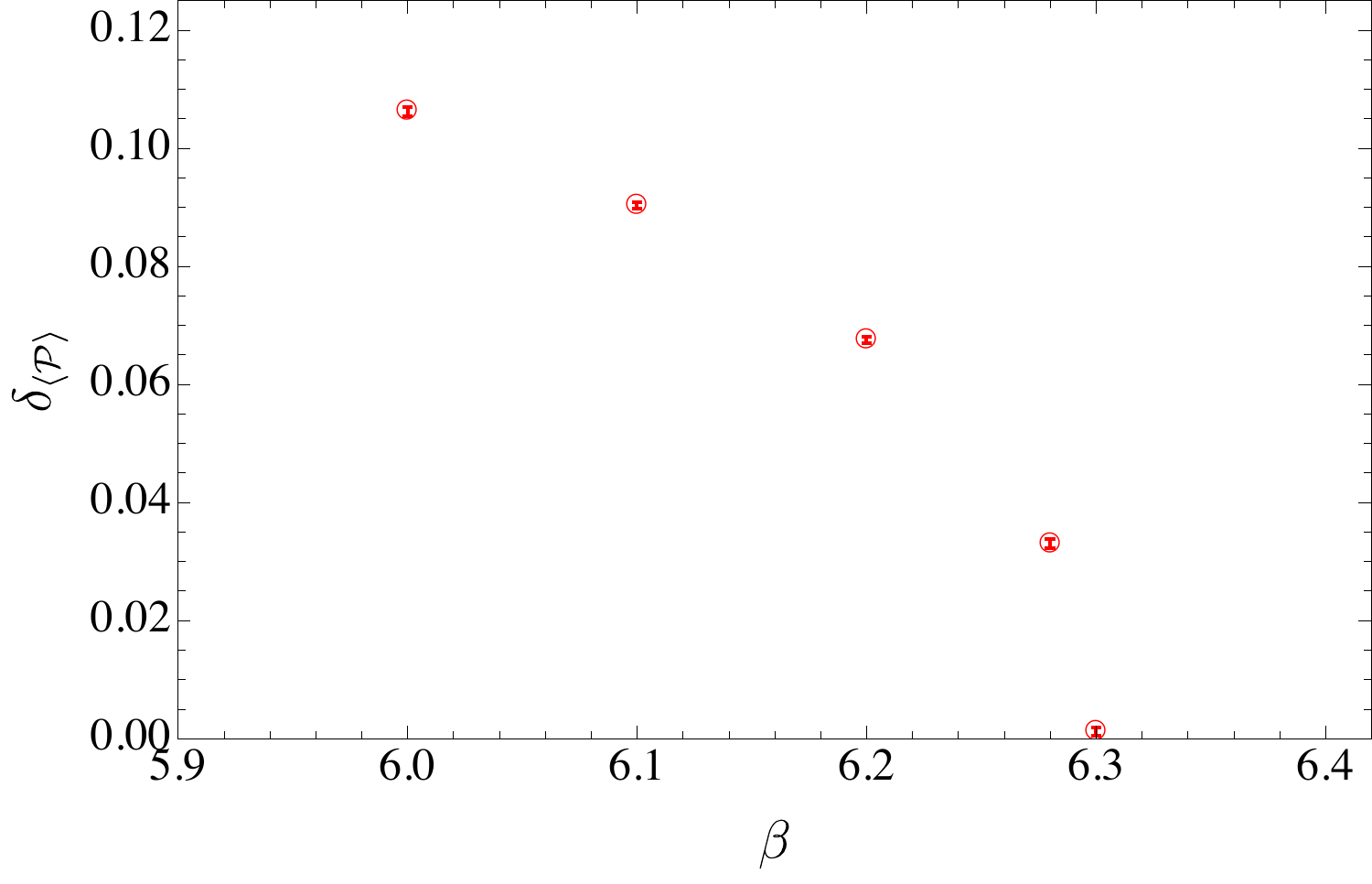}
\caption{%
\label{fig:1st_order_line}%
Illustrative example of the difference $\delta_{\langle {\cal P}\rangle}$ between the average plaquette values obtained 
from ensembles generated with cold (unit) and hot (random) initial configurations. 
The mass of the fundamental fermions $a m_0^f=-0.6$ is held fixed, and for each value of $\beta$ we
vary the bare mass of the antisymmetric fermions $a m_0^{as}$, until reaching 
the proximity of the transition---at which
  $\delta_{\langle {\cal P}\rangle}$ is maximised.
   The lattice size is $8^4$ for the three small $\beta$ values and $12^4$ for the rest. 
}
\end{center}
\end{figure}

\begin{figure}[t]
\begin{center}
\includegraphics[width=.49\textwidth]{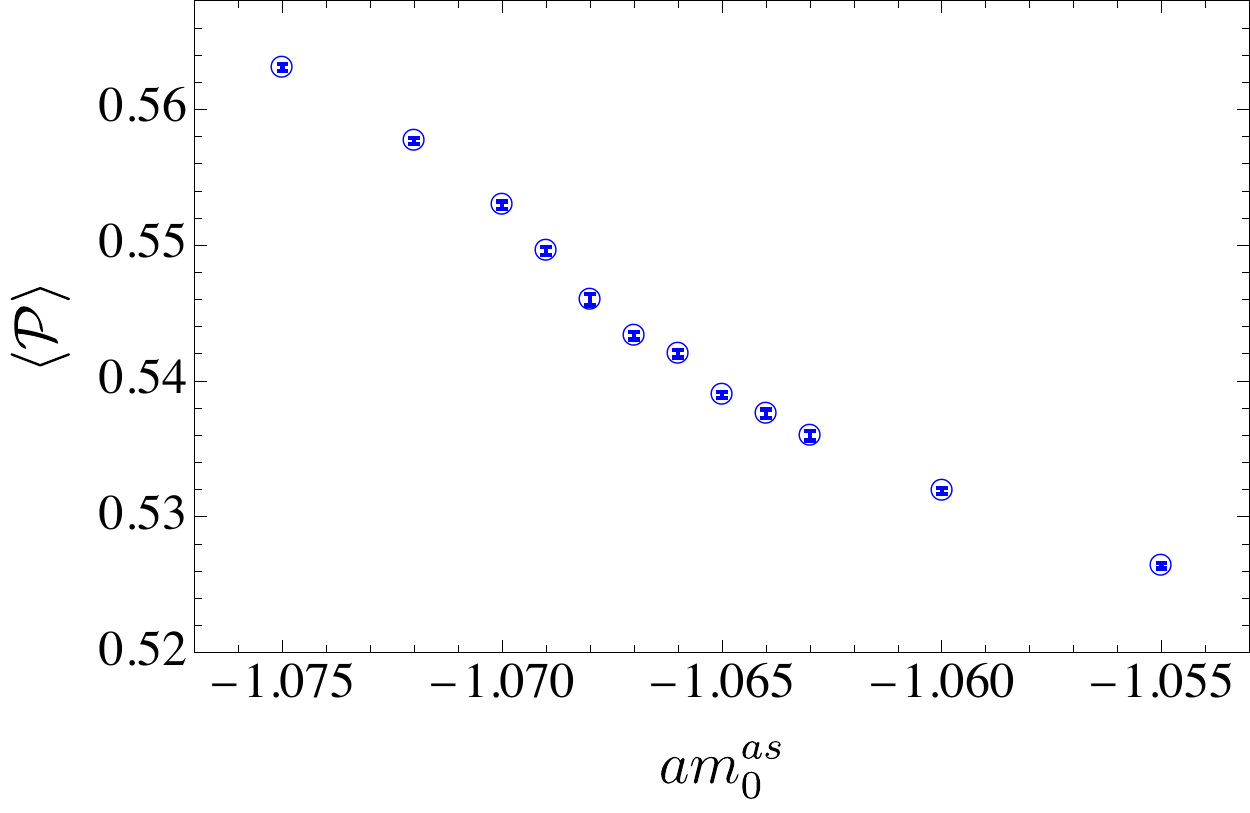}
\includegraphics[width=.49\textwidth]{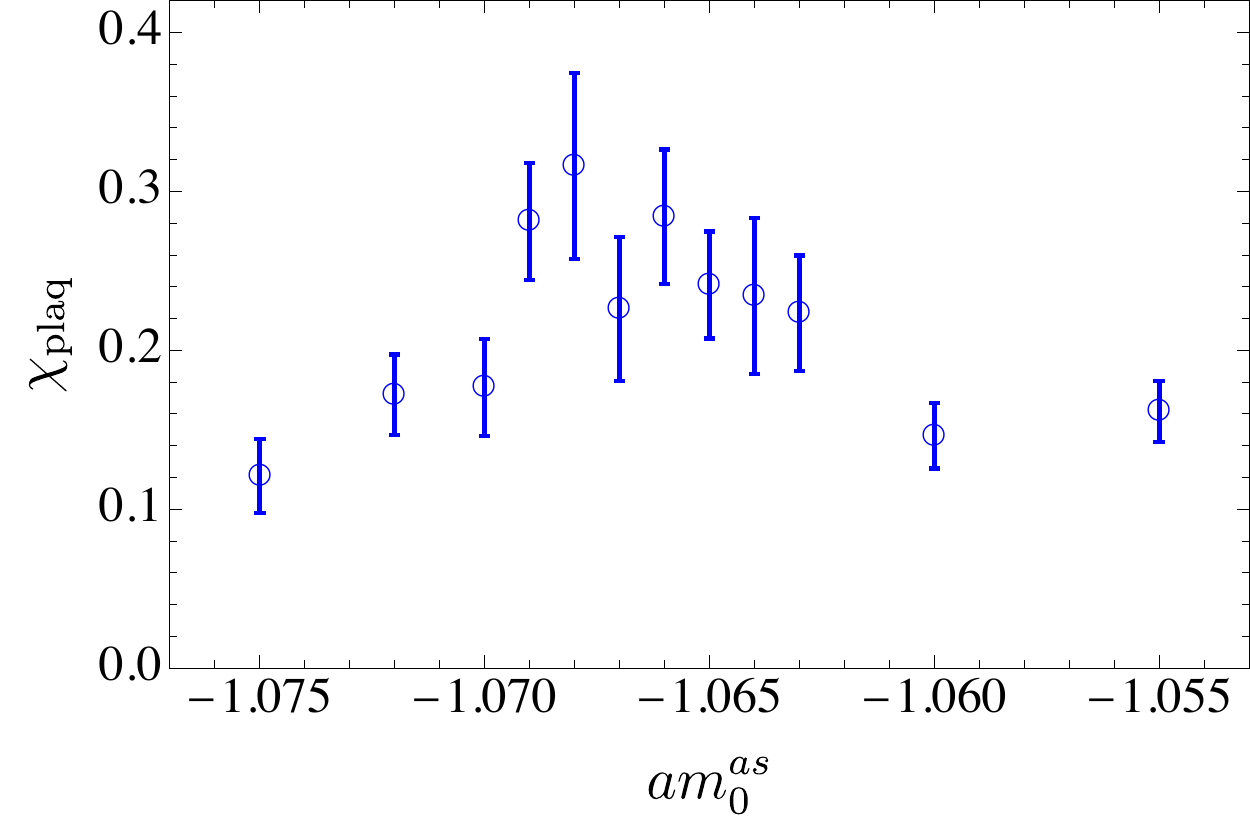}
\caption{%
\label{fig:plaq_b6p35}%
Average plaquette value $\langle {\cal P}\rangle$ (left panel) and its susceptibility $\chi_{\rm plaq}$ (right panel),
as a function of the mass $a m_0^{as}$, having fixed the other lattice parameter to
be $\beta=6.35$ and $am_0^f=-0.6$. The lattice volume is $24\times 12^3$. 
}
\end{center}
\end{figure}

We next would like to measure the critical coupling $\beta_{\rm cr}^{mr}$,
at the boundary of the surface of first-order phase transitions. 
We are particularly interested to determine  the
values of $\beta$ that are large enough that there is no phase transition,
for finite masses for the both types of fermions.
To exemplify the process, we start by fixing the fundamental fermion mass  $am_0^f=-0.6$. 
We consider a range of values of  $\beta$ smaller than $6.4$, 
 adjust the value of $a m_0^{as}$ in proximity of the phase transition,
and  calculate $\delta_{\langle \mathcal{P}\rangle}=|\langle \mathcal{P} 
\rangle_{\rm cold}-\langle \mathcal{P} \rangle_{\rm hot}|$,
 the difference between the average plaquette value measured in ensembles with cold and hot initial configurations. 
The results are shown in \Fig{1st_order_line}. 
The strong and weak coupling regimes are separated by the existence of
 a first-order phase transition for $\beta$ smaller than  the critical coupling 
 $\beta_{\rm cr}^{mr}\simeq 6.3$. Conversely, for larger values there are regions of parameter 
 space with $\delta_{\langle \mathcal{P}\rangle}=0$, signaling a cross-over. 
The final result of this analysis is that as long as our lattice calculations are performed with values of
$\beta \gtrsim 6.3$, for appropriate choices of fermion masses
 the theory is in the weak-coupling phase,
and  the results extrapolate smoothly to the continuum theory.
We notice that this numerical result is smaller than the aforementioned
cases where one of the fermion species is infinitely heavy.

\begin{figure}
\begin{center}
\includegraphics[width=.49\textwidth]{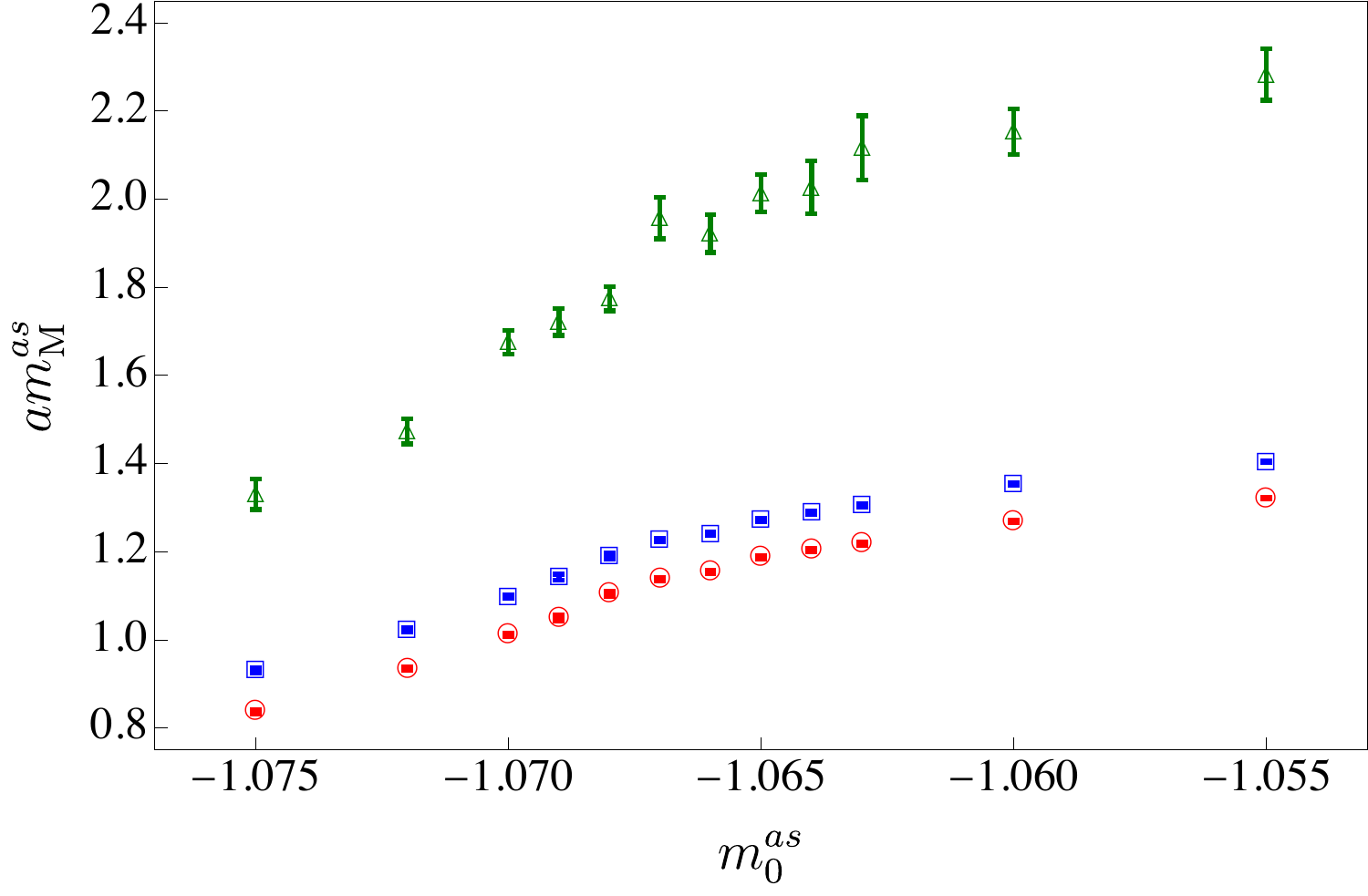}
\includegraphics[width=.49\textwidth]{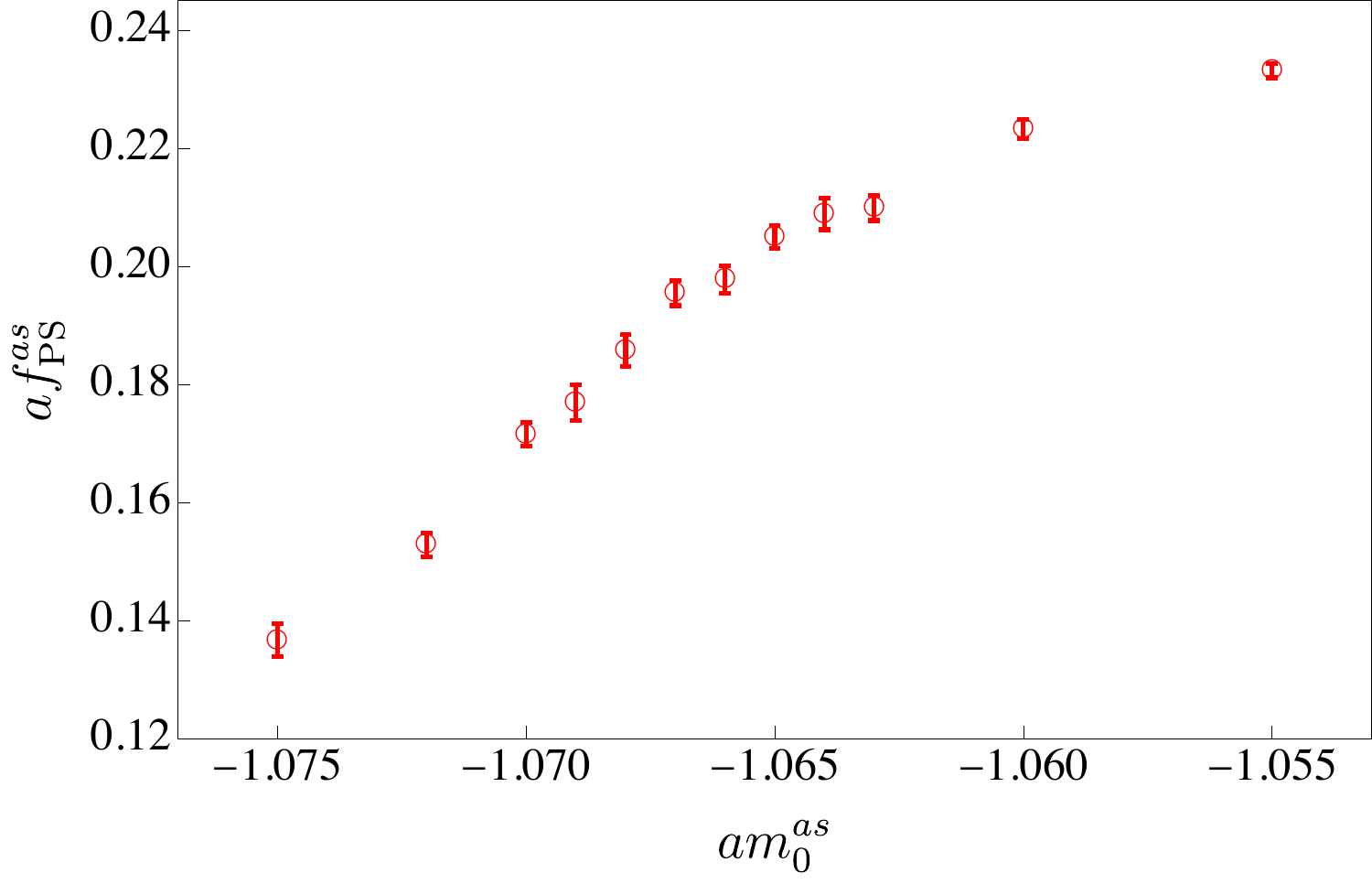}
\caption{%
\label{fig:mv_mps_fps}%
Left panel: masses, in lattice units,  of the pseudoscalar (red circles), vector (blue squares) and scalar (green triangles) flavoured mesons 
composed of fermions in the antisymmetric representation, as a function of $am_0^{as}$.
Right panel: decay constant, in lattice units, of the pseudoscalar meson composed of $(as)$ fermions,
 as a function of the bare mass $am_0^{as}$.
The other lattice parameters are fixed 
by $\beta=6.35$ and $am_0^f=-0.6$. The lattice volume is $24\times 12^3$. 
}
\end{center}
\end{figure}

\begin{figure}
\begin{center}
\includegraphics[width=.49\textwidth]{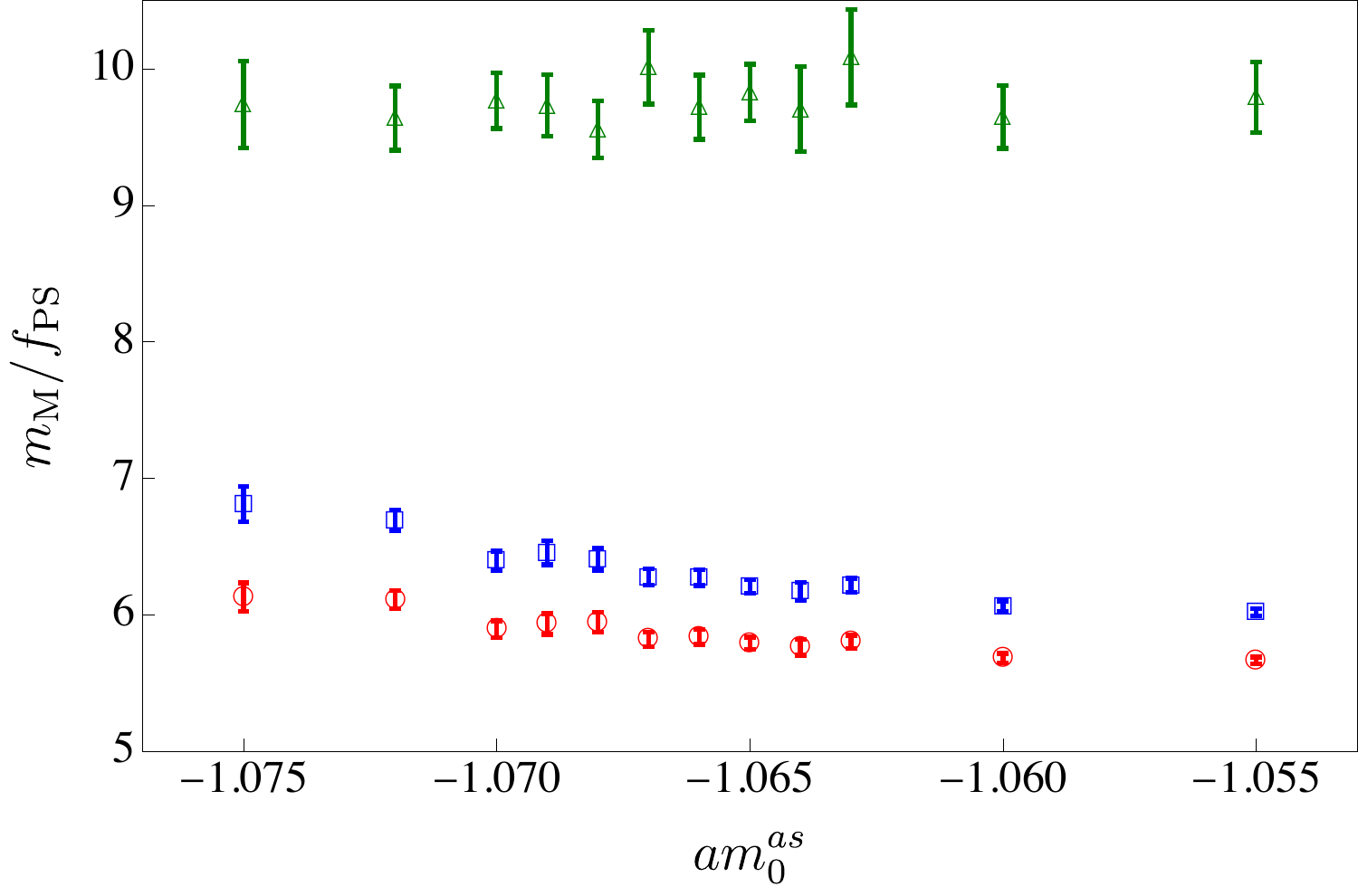}
\includegraphics[width=.49\textwidth]{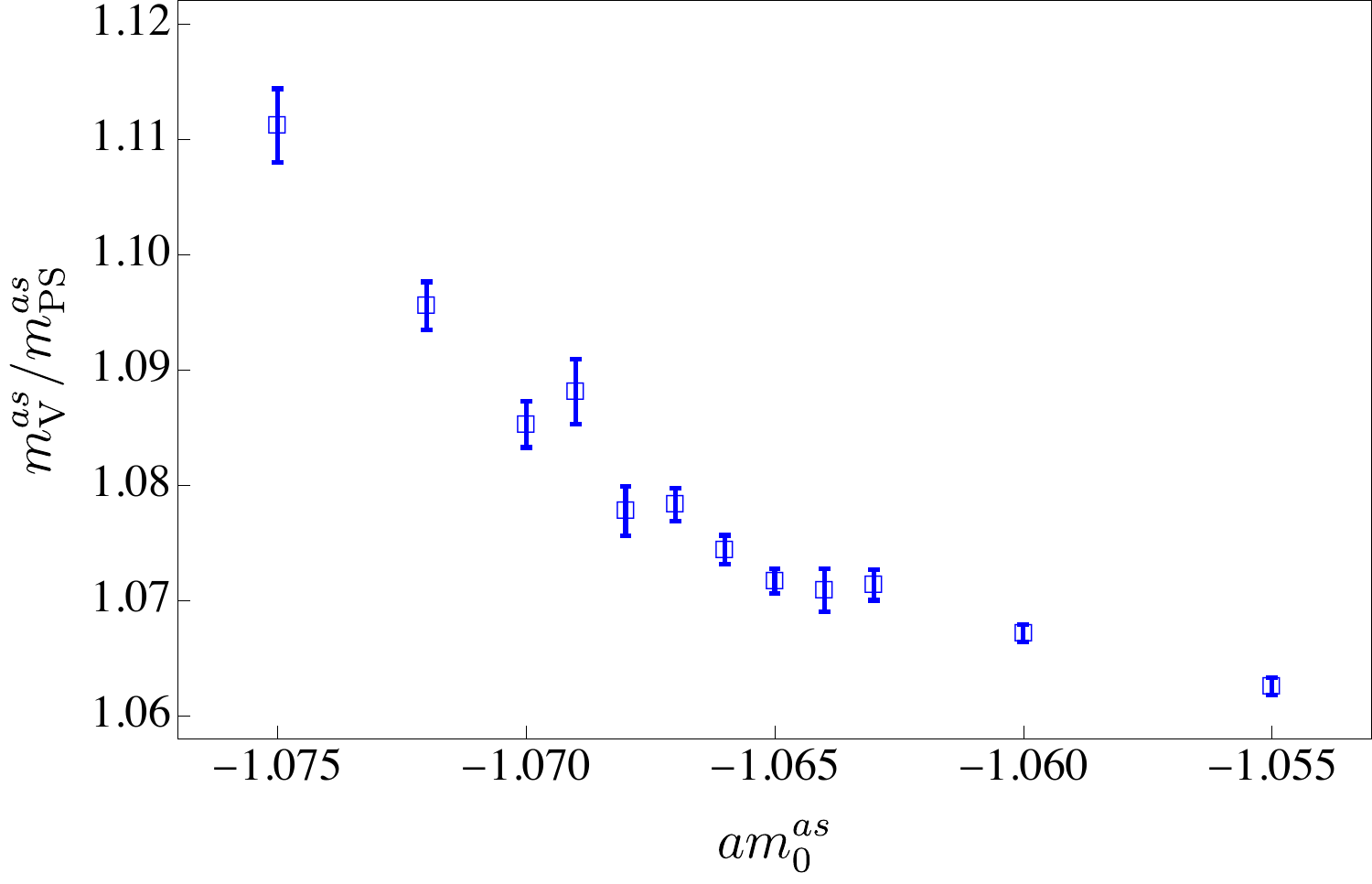}
\caption{%
\label{fig:ratio_mv_mps_fps}%
Left panel: masses, in units of the pseudoscalar decay constant,  
 of the pseudoscalar (red circles), vector (blue squares) and scalar (green triangles) flavoured mesons 
composed of fermions in the antisymmetric representation, as a function of  $am_0^{as}$.
Right panel: ratio of the masses of the vector and  pseudoscalar meson composed of $(as)$ fermions.
The other lattice parameters are fixed 
by $\beta=6.35$ and $am_0^f=-0.6$. The lattice volume is $24\times 12^3$. 
}
\end{center}
\end{figure}

The useful fixed point of our lattice theory, bringing it in contact with the desired continuum theory, is 
reached in proximity of  $a m_0^f=a m_0^{as}=\beta^{-1}=0$. 
Our investigation of the phase structure revealed the existence of a boundary to the surface 
 of first-order phase transitions, as shown in \Fig{mr_bulk_phase}.
Along this boundary,  we collected indications compatible with  the phase transition 
being of second order.
  Although these are bulk properties of the lattice theory, it is worth analysing
  the physical features associated with such second-order transitions, as these
 fixed points might be used to define the continuum limit
 to alternative theories. We want to understand whether such theories might be interesting 
 in themselves.

To  this purpose, we carry out  an exploratory study in proximity of the
second-order phase transitions.
We fix the lattice coupling slightly above its critical value, $\beta=6.35$, 
such that the theory  displays a  crossover region. 
We hold fixed also the mass of the fundamental fermions $am_0^f=-0.6$. 
We then perform a scan over values of $am_0^{as}$,
 to identify the crossover region. 
 In \Fig{plaq_b6p35}, we show the results of the average plaquette $\langle {\cal P}\rangle$,
  and its susceptibility $\chi_{\rm plaq}$,  adopting a lattice with size $24\times 12^3$.
The critical mass is  $am_{0,\,{\rm cr}}^{as}\simeq -1.068$. 
  
With the same ensembles, we then measure the masses of pseudoscalar, 
  vector and scalar mesons, as well as the decay constant of the pseudoscalar meson, 
  focusing on bound states with constituents  $(as)$ fermions. 
  As shown in \Fig{mv_mps_fps}, we find no non-trivial behaviours in these quantities. 
  In \Fig{ratio_mv_mps_fps}, we also present the masses in units of $f_{\rm PS}^{as}$,
  and the mass ratio between vector and pseudoscalar mesons. 
  Again, we do not find any interesting features associated with the fixed 
  points in the meson spectrum.
  Our findings are compatible with interpreting
   the theories living at the second-order fixed points
   along the critical boundary in terms of a non-interacting
   scalar field theory. 
   A dedicated, systematic, high-precision study of the theory in proximity of the critical values of the lattice
   parameters would be needed to ascertain whether this is the case, but we do not find any 
   alluring evidence
   to the contrary, at the level of precision of this study.

\subsection{Spectrum of the Dirac operator}
\label{Sec:diracspectrum}

As discussed in \Sec{symmetry}, the $SU(4)\times SU(6)$ 
global symmetry is expected to break to its $Sp(4)\times SO(6)$ subgroup.
The symmetry breaking pattern can be tested through a
comparison with the chRMT 
predictions, as was done for example in Ref.~\cite{Cossu:2019hse} for a $SU(4)$
theory with mixed fermion representations.
As a preliminary exercise, which we do not report here, we checked that
we could produce the expected results for the $SU(2)$ and $SU(4)$ theories with  
(quenched) fundamental fermions.
We discuss in the following the tests we carried out for the $Sp(4)$ 
gauge theory of interest to this paper.

\begin{figure}[t]
\centering
\includegraphics[width=0.59\textwidth]{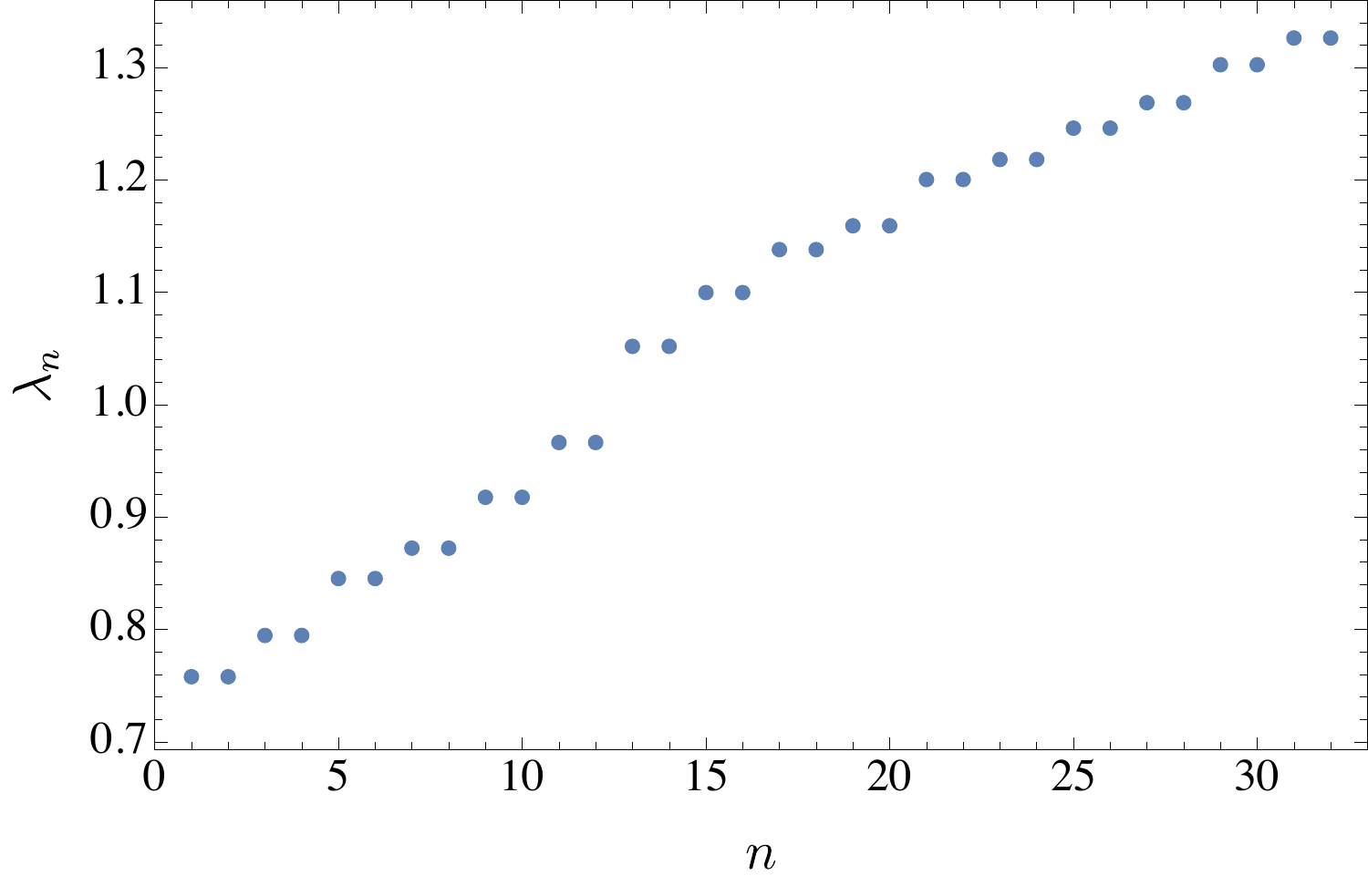}
\caption{%
Numerical results for the smallest  (positive) eigenvalues of the Dirac operators for $(as)$
fermions, measured in the quenched $Sp(4)$ ensemble with $\beta=8.0$ 
and the lattice size of $4^4$. The mass of the fermion in 
the antisymmetric representation is $a m_0^{as}=-0.2$.
}
\label{fig:degen_sp4as}
\end{figure}

Following the procedure illustrated in \Sec{dirac}, we compute the eigenvalues of the hermitian Wilson-Dirac operators, which are real regardless of the fermion representation, for fermions in the fundamental and antisymmetric. We then extract the distribution $P(s)$ of the unfolded density of spacings of the eigenvalues, with the discretised definition of $s$ in \Eq{spacing}.   
For this exercise, we use quenched ensembles with
 coupling $\beta=8.0$ and lattice size  $4^4$. 
 We fix the  masses of the fermions to be $am_0^{f}=am_0^{as}=-0.2$. 
 We recall that, in the case of $(as)$ fermions, the eigenvalues of the hermitian Wilson-Dirac operator are expected to have degeneracy $2$. This property follows from the fact that the fermionic determinant is positive definite, as discussed in \Sec{symmetry}. As an illustration, we show in \Fig{degen_sp4as} the sequence of the smallest positive eigenvalues of this operator for $(as)$ fermions for our choice of lattice parameters, which provides support for the expected double-degeneracy. 
 The presence of a largish  mass gap below the lowest eigenvalue
in our measurements is due to the comparatively large value
of the fermion mass. 

\begin{figure}[h!]
\centering
\includegraphics[width=0.59\textwidth]{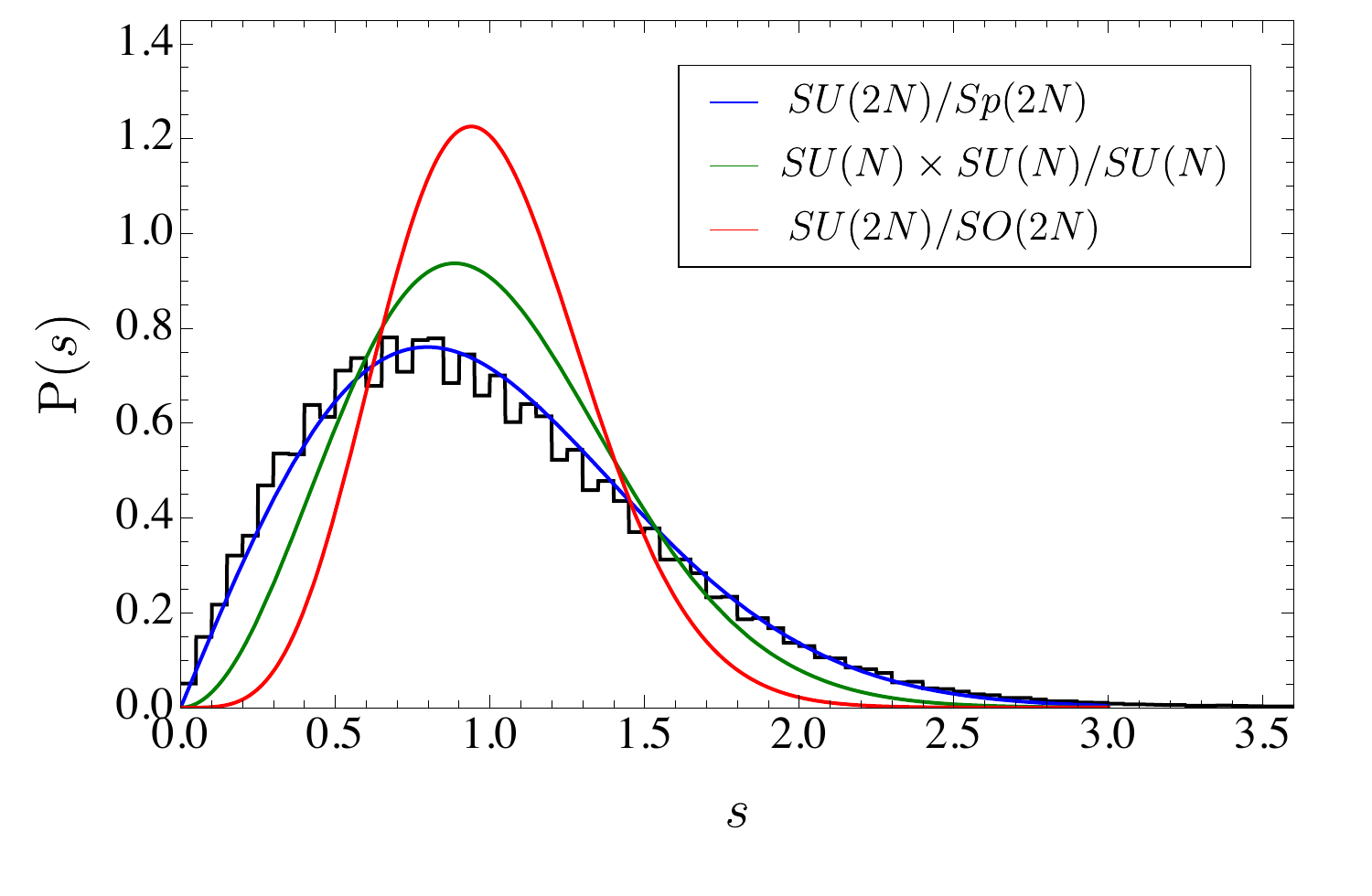}
\includegraphics[width=0.59\textwidth]{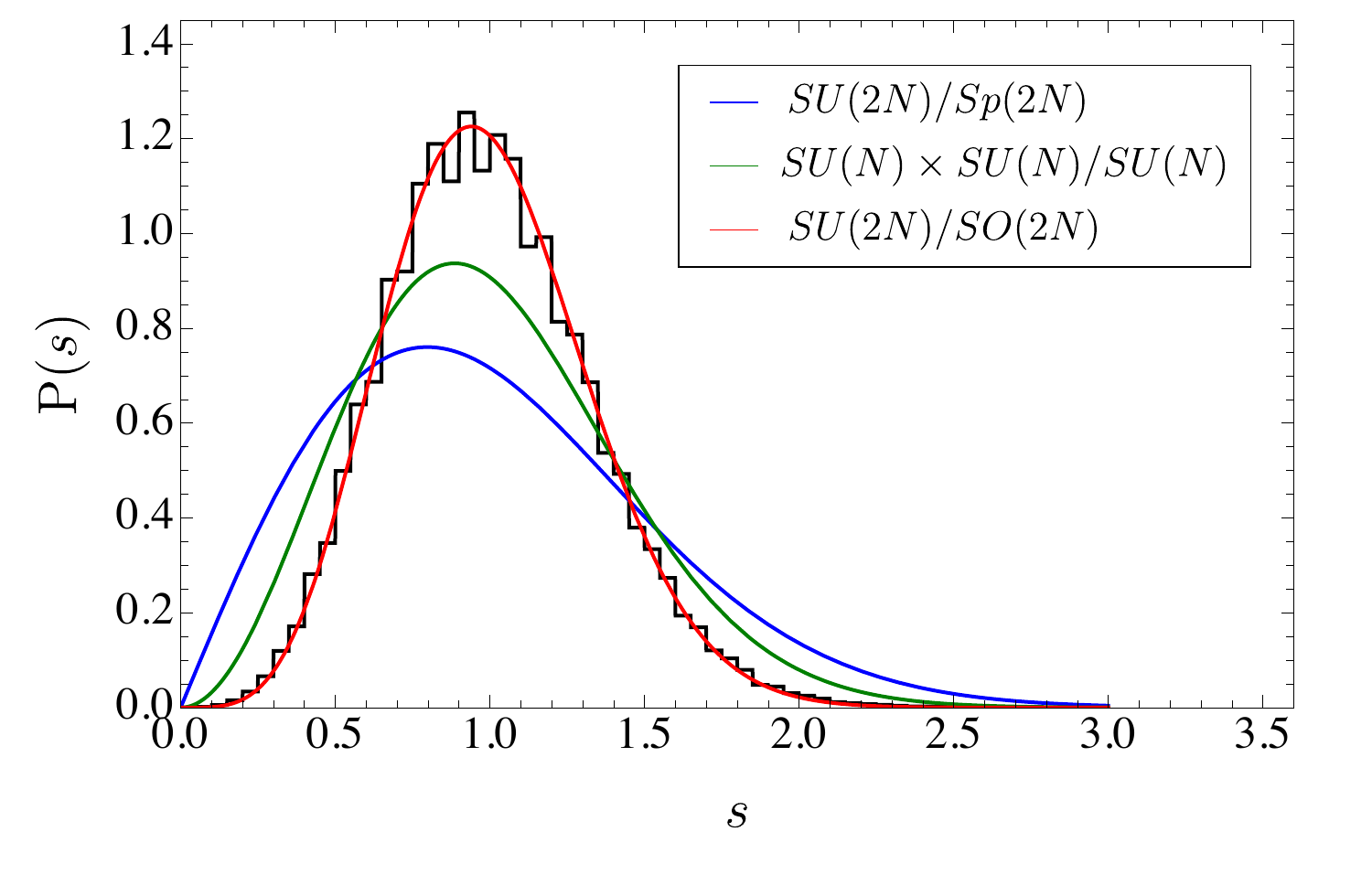}
\caption{%
Histogram (black solid lines)
of the distribution of unfolded density of spacing $P(s)$ between subsequent 
Dirac eigenvalues in the $Sp(4)$ lattice gauge theory in the quenched approximation, with
coupling $\beta=8.0$, fermion  masses $am_0^{f}=am_0^{as}=-0.2$, and lattice of size $4^4$.
The number of configurations is $192$, while the number of eigenvalues in 
each configuration used for the $(f)$ fermions (top panel) is $4096$,
while for the $(as)$ fermions (bottom panel) it is $5120$.
The curves depict, for different symmetry breaking patterns,
the predictions from matrix theory,
Eq.~(\ref{eq:dist_f_chrmt}).
}
\label{fig:pdf_sp4}
\end{figure}

\begin{figure}[h!]
\centering
\includegraphics[width=0.59\textwidth]{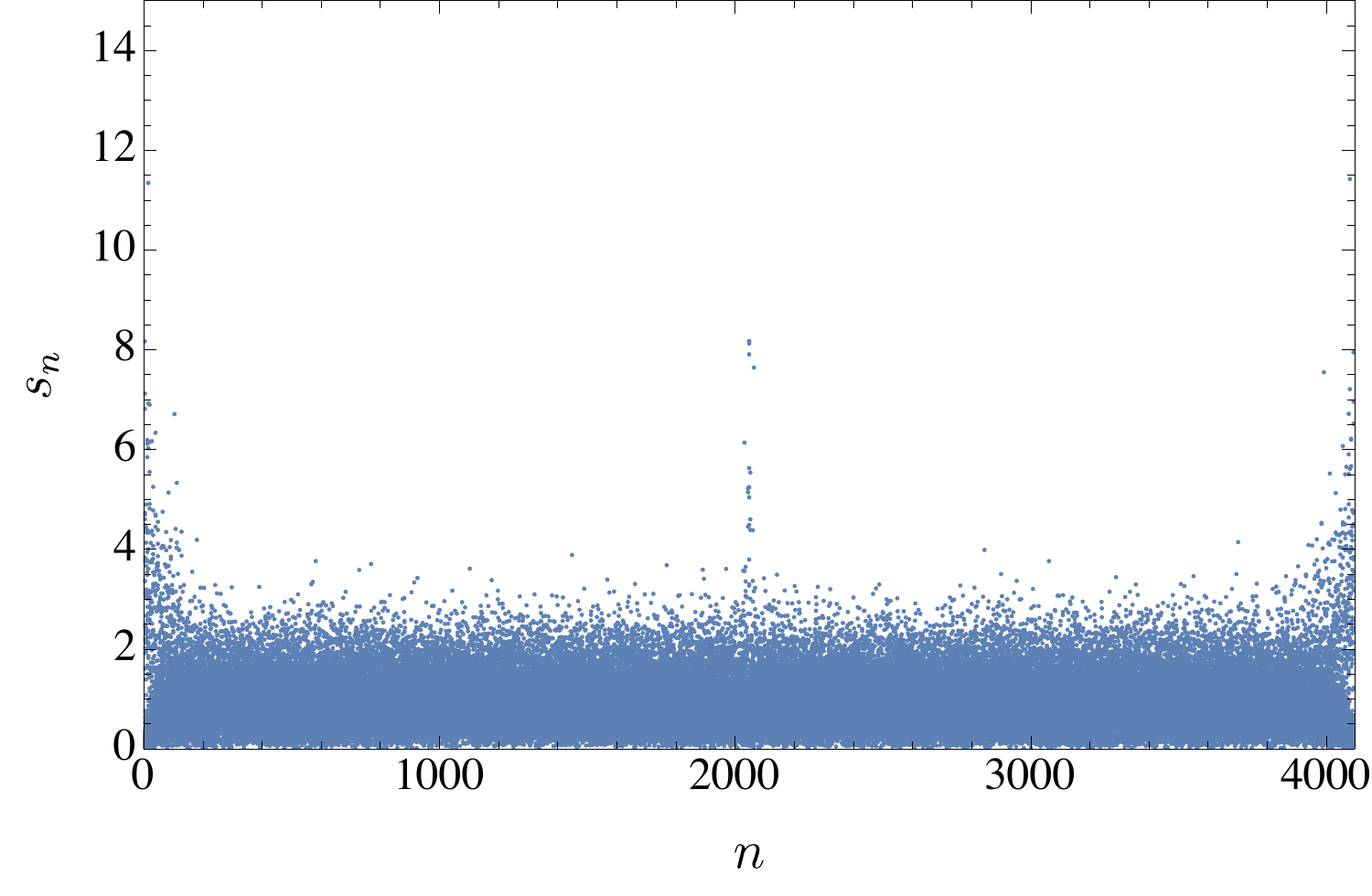}
\caption{%
The unfolded density of  spacing between subsequent Dirac eigenvalues $s_n$, 
as defined in Eq.~(\ref{eq:spacing}),
at the position of the $n$th eigenvalue for the $Sp(4)$ gauge theory
 in the quenched approximation, with $\beta=8.0$, for fermions in the fundamental representation. 
with  bare mass of $am_0^{f}=-0.2$. For this illustrative plot, we
 randomly chose $20$ out 
of the $192$ configurations.  
}
\label{fig:dist_sp4}
\end{figure}

In the upper and lower panels of \Fig{pdf_sp4}, we show histograms of
the unfolded density of the eigenvalue spacings 
for fermions in the fundamental 
and antisymmetric representations, respectively. The numerical results are 
compared to the chiral RMT predictions for chGOE, chGUE, and chGSE ensembles, 
defined in \Eq{dist_f_chrmt} with $\bar{\beta}=1,\,2,\,4$---for convenience,  
 in the legend we label the predictions by the associated 
 symmetry-breaking pattern.
As shown in the figures, we find that the distributions are in good 
agreement with the chRMT predictions.

\begin{figure}[h!]
\centering
\includegraphics[width=0.59\textwidth]{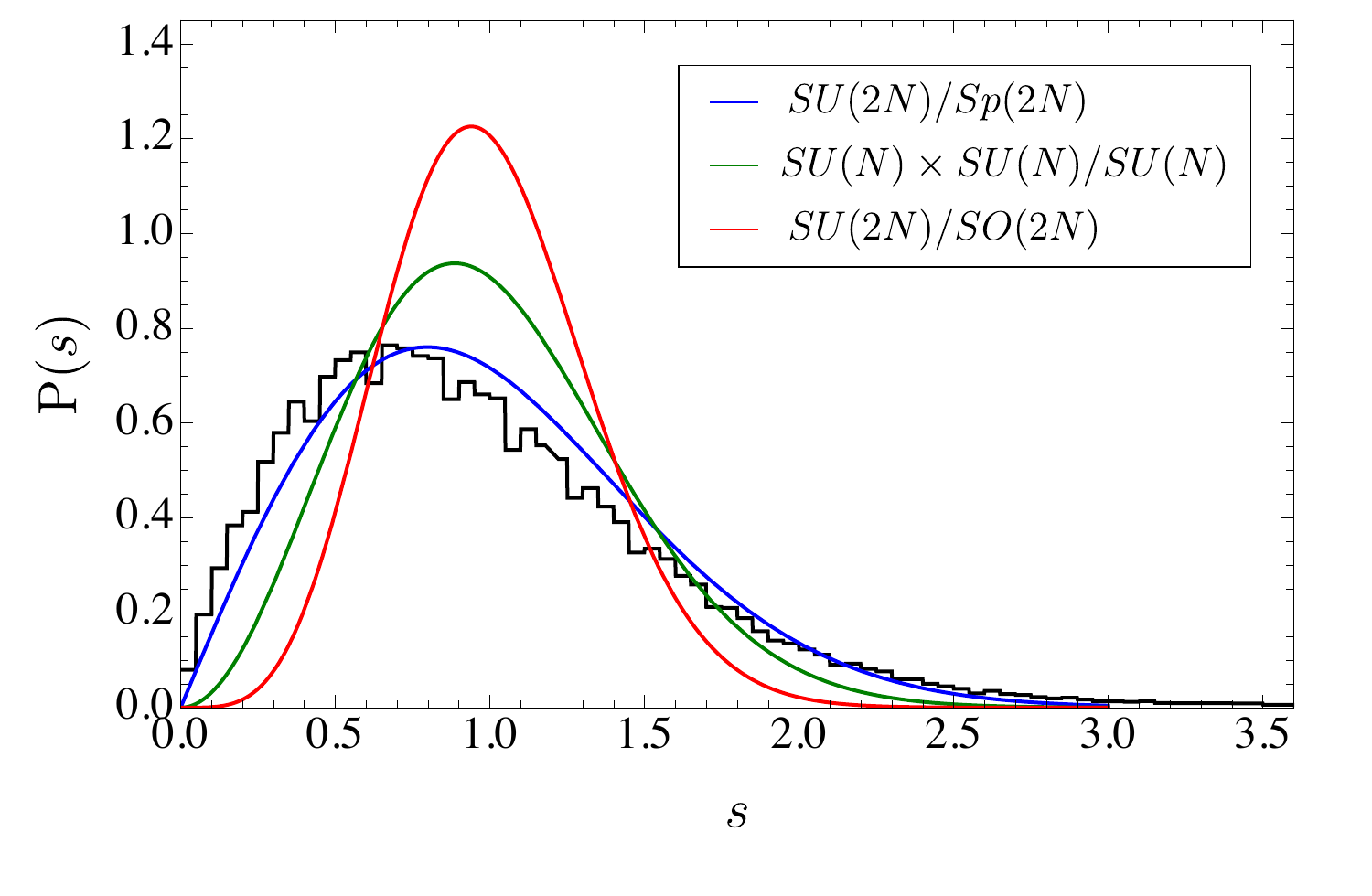}
\caption{%
Histogram of the distribution of unfolded density of the spacing 
between subsequent Dirac eigenvalues for fermions transforming in the fundamental representation of
 $Sp(4)$, in the quenched approximation,
with $\beta=8.0$, mass of the $(f)$ fermion  $am_0^{f}=-0.2$,
and on a lattice with size $3^4$.
The number of configurations is $196$, while the number of eigenvalues in each configuration is $1296$.
}
\label{fig:pdf_sp4_small_V}
\end{figure}

While the agreement is very convincing for $(as)$ fermions (bottom panel), 
one can detect a slight mismatch between the chRMT prediction and the numerical results in the 
case of $(f)$ fermions (top panel). 
By inspecting the details provided in \Fig{dist_sp4},
one sees that such a discrepancy is associated with some abnormally large spacings for  the
smallest and largest
 eigenvalues. 
 We interpret this as an artefact due to the finiteness of the size of the matrices.
 We hence expect the distortion of the distribution to
 becomes less pronounced as the size of Dirac matrix increases, i.e. by going towards larger $N$, 
 larger lattices,  and higher representations $R$. 
 For instance, the results of the same calculations for the $(f)$ fermions, but  on a smaller lattice volume of 
  $3^4$, is shown in \Fig{pdf_sp4_small_V}. The deviations with the chRMT predictions are larger,
   compared to the $4^4$ lattice. 
   Notice in particular that the total numbers of eigenvalues are 
   $4096$ and $5120$ for the fundamental and antisymmetric representations of $Sp(4)$
    with the $4^4$ lattice, respectively, while for the $(f)$ fermions  with 
    lattice volume of $3^4$  such number  is $1296$. 

\begin{figure}[h!]
\centering
\includegraphics[width=0.59\textwidth]{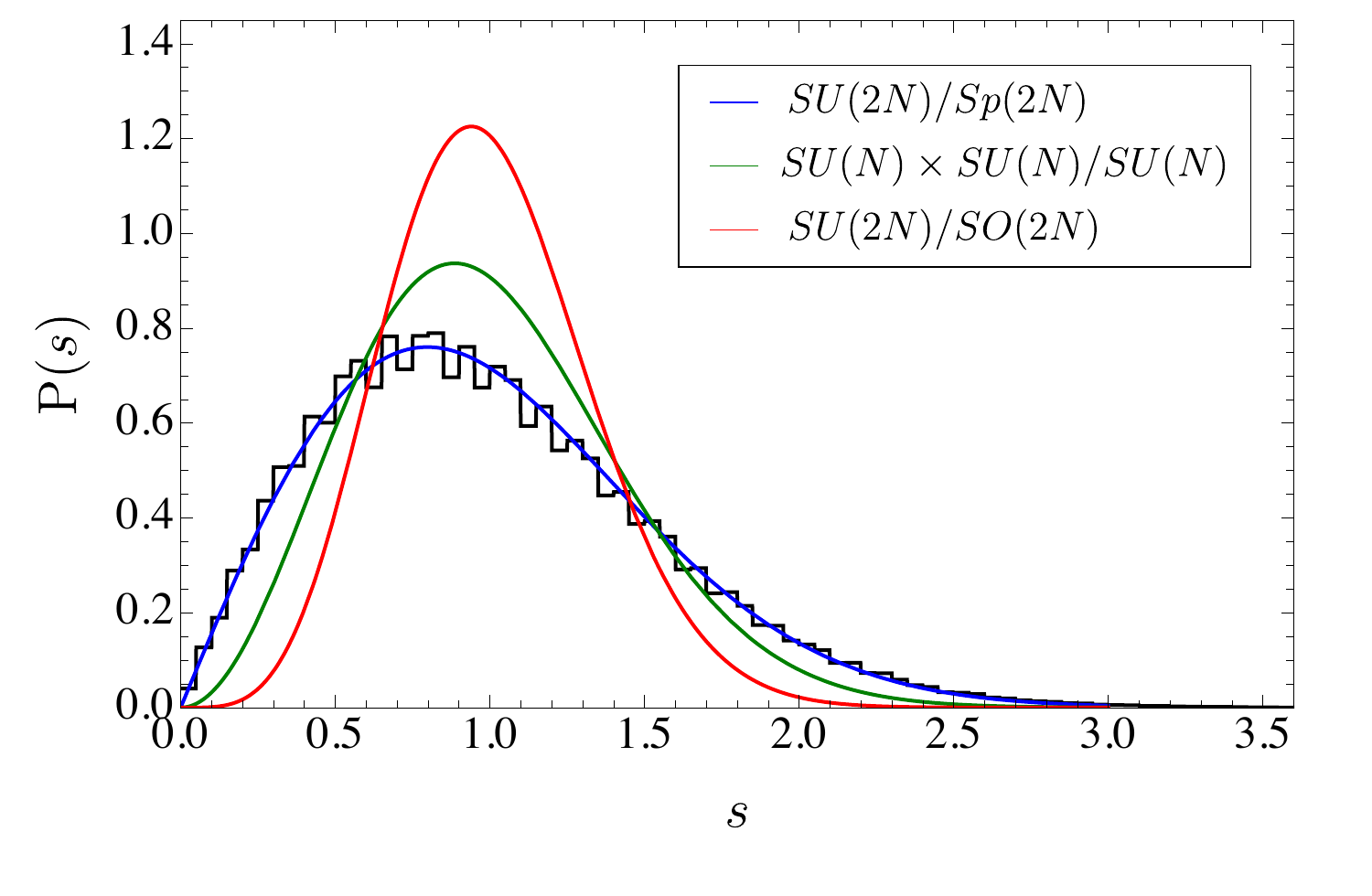}
\includegraphics[width=0.59\textwidth]{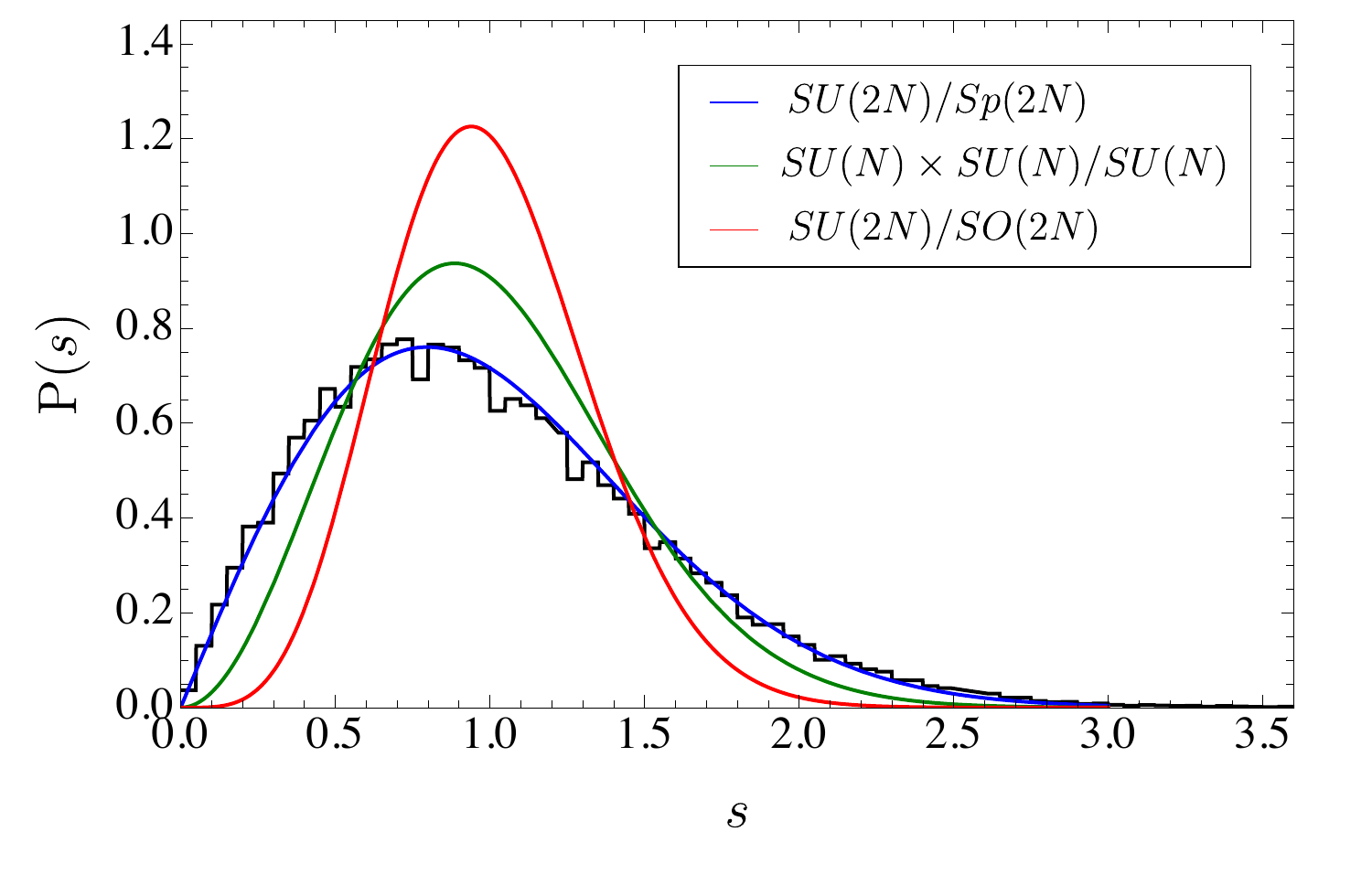}
\caption{%
Histograms of the unfolded density of spacing between subsequent Dirac 
eigenvalues for the $Sp(4)$ gauge theory in the quenched approximation, with
coupling $\beta=8.0$ and with $(f)$  fermions with mass
$am_0^{f}=-0.2$, on lattices of size $4^4$ (top panel) and $3^4$ (bottom panel).
The calculated eigenvaluesl  are the same used in Figs.~\ref{fig:pdf_sp4} and \ref{fig:pdf_sp4_small_V} 
with the notable exception that a few
hundred spacings at the smallest and largest eigenvalues have been discarded.
}
\label{fig:pdf_sp4_cut}
\end{figure}

To further support this interpretation,
we recalculate the unfolded density for the same theories,
but excluding small and large eigenvalues. By doing so we aim at demonstrating
that our action and algorithms yield a theory that 
reproduces the expected symmetry breaking patterns. 

We find that, to do so, it suffices to exclude a few hundred eigenvalues at the extrema of the spectrum. 
The resulting density distributions for fermions in the fundamental representation measured on lattices of sizes
$4^4$ and $3^4$  are shown in the upper and lower panels of \Fig{pdf_sp4_cut}, respectively. 
As expected, in this case the difference between the numerical results and chRMT predictions 
is no longer visible to the naked eye.
We remind the reader that these are quite small lattices, compared to what one normally considers for 
dynamical lattice calculations.
We can hence conclude that the HiRep code correctly implements also  Dirac fermions transforming 
in the fundamental and antisymmetric representations of the $Sp(4)$
 gauge group.

\subsection{Finite volume effects}
\label{Sec:fv}

\begin{figure}
\begin{center}
\includegraphics[width=.59\textwidth]{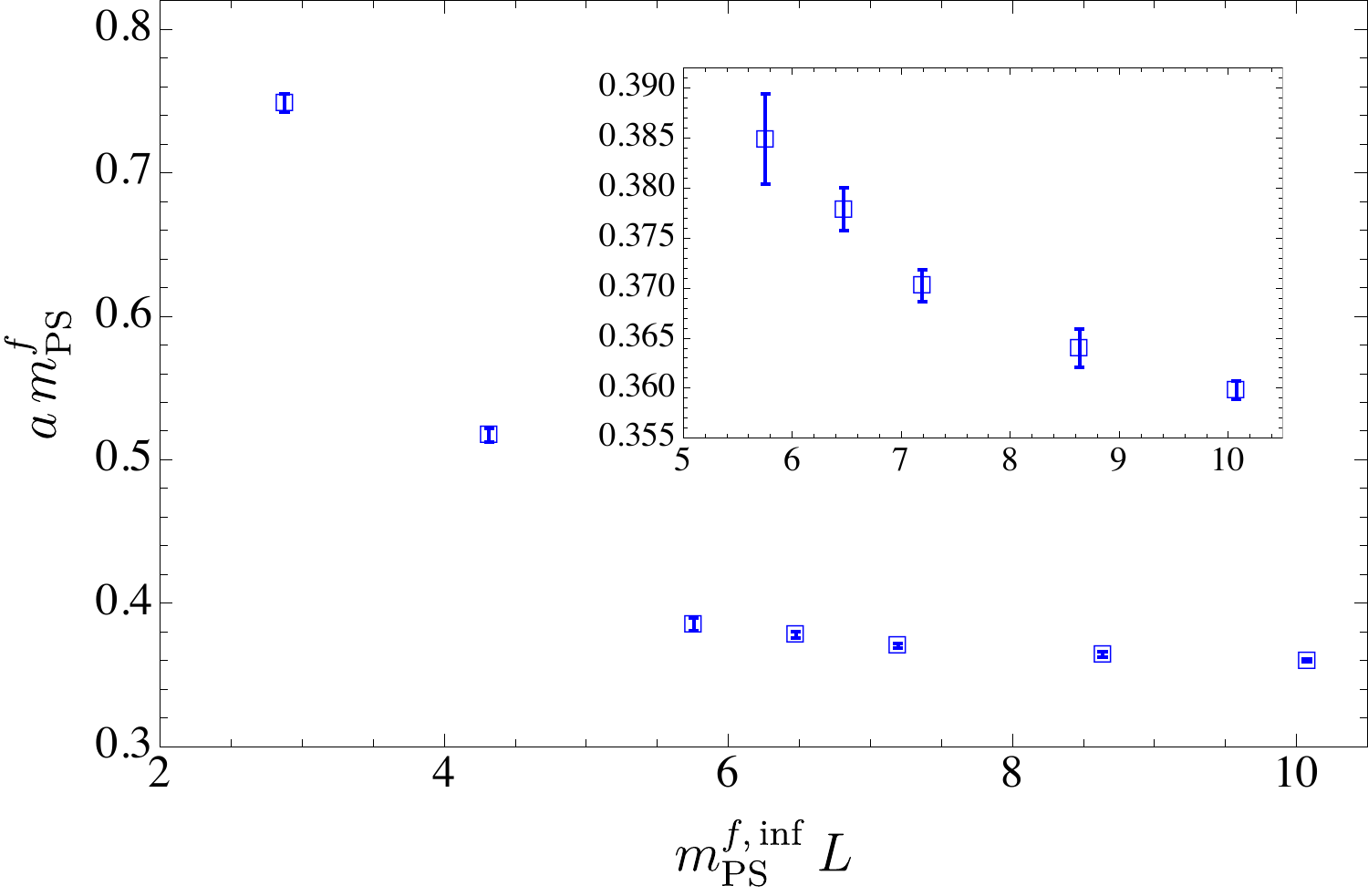}
\includegraphics[width=.59\textwidth]{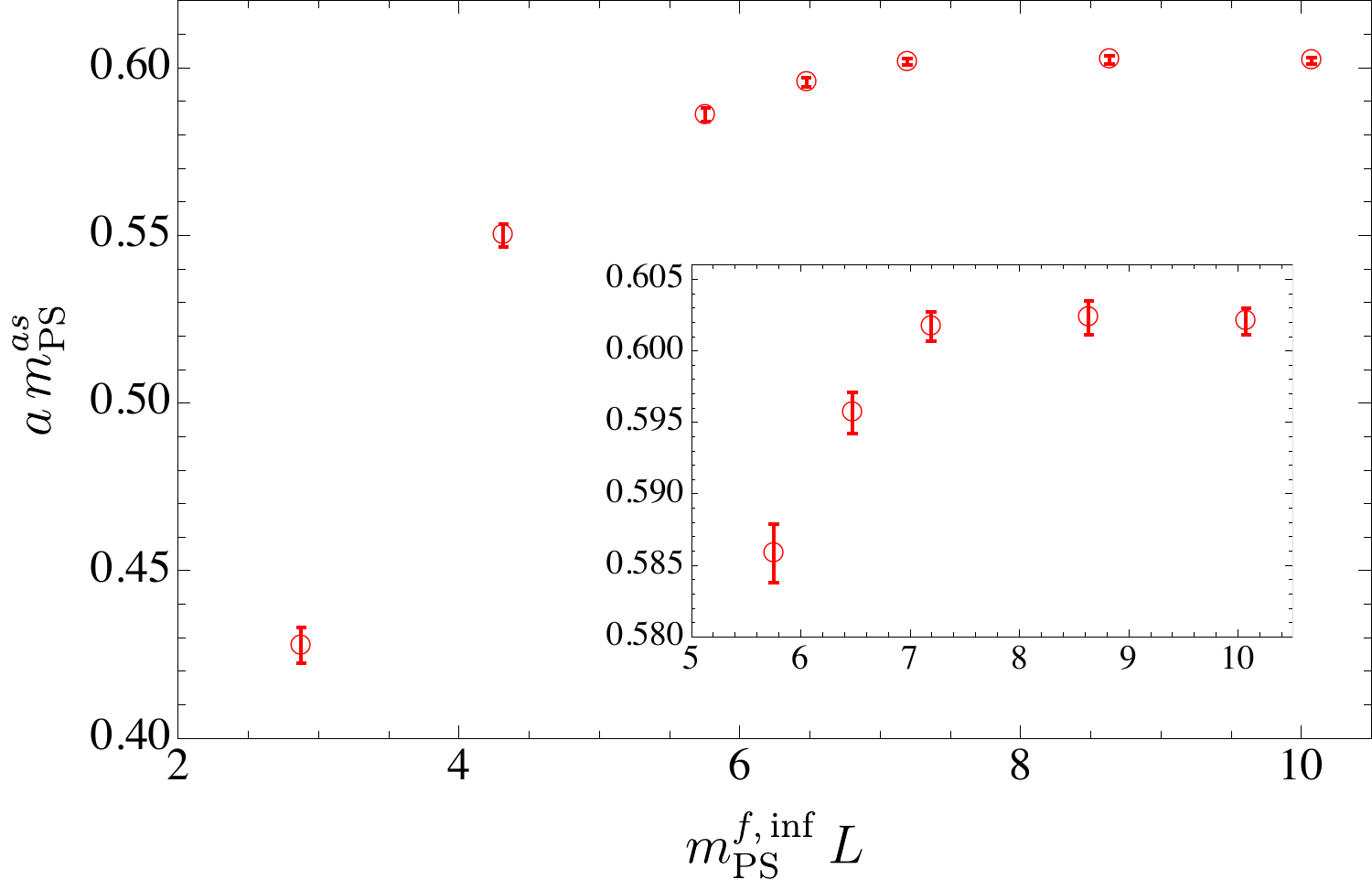}
\caption{%
\label{fig:fv_mps}%
Masses (in lattice units) of pseudoscalar mesons composed of constituent fermions 
transforming in the fundamental (top panel) and antisymmetric (bottom panel) representations,
as a function of the combination $m_{\rm PS}^{f,\,{\rm inf}} L$.
We denote by $m_{\rm PS}^{f,\,{\rm inf}} a$ the mass of the pseudoscalar  extracted from the largest
available lattice, with lattice of volume $54\times28^3$. The lattice parameters $\beta=6.5$, $am_0^{f}=-0.71$, $am_0^{as}=-1.01$
are held fixed, and repeat the measurement of the mass of the pseudoscalar while we vary the size of the lattice.
The smaller inset plots display a detail of the enclosing figures, 
with the range on the vertical axis restricted to highlight the plateaux in the rightmost points.
}
\end{center}
\end{figure}

In this section, we show the results of our numerical investigations of finite volume 
effects in our measurements. Following lattice QCD lore, 
we start by studying the volume dependence of the mass of pseudoscalar mesons, 
the lightest states in the spectrum of composite objects.
 In the upper and lower panels of \Fig{fv_mps} we show our results for
  the masses (in lattice units) of pseudoscalar mesons with $(f)$ and $(as)$ fermion constituents, respectively,
for varying $m_{\rm PS}^{f,\,{\rm inf}} L$. 
We use seven different lattice sizes, six of them have
time-like extent  $N_t=T/a=48$ and space-like extent 
$N_s=L/a=8,\,12,\,16,\,18,\,20,\,24$; 
the largest lattice has size $54 \times 28^3$. 
Details and numerical results are displayed in \App{C},
and are also available in machine-readable form in Ref.~\cite{datapackage}.
The  mass measured from the largest lattice 
has been identifyied with $m_{\rm PS}^{f,\,{\rm inf}}$. 
We fix the lattice coupling to $\beta=6.5$, so that 
the data points are well inside the weak coupling regime. 
The bare masses are $am_0^{f}=-0.71$ and $am_0^{as}=-1.01$.
 The pseudoscalar composed of $(f)$ fermions are lighter than those composed of $(as)$ fermions.
  As shown in the lower panel of \Fig{fv_mps}, we  find that finite volume corrections 
  to the mass of the pseudoscalar mesons composed of $(as)$ fermions
can be  neglected, compared  to statistical fluctuations, for $m_{\rm PS}^{f,\,{\rm inf}} L \gsim 7$. 
In the case of fundamental fermion constituents, the convergence is rather slow, and the size of finite volume effects becomes
less than one percent and compatible with the statistical errors 
only when $m_{\rm PS}^{f,\,{\rm inf}} L \gtrsim 8.5$. 
Achieving higher precision would require to restrict the analysis
to even larger values $m_{\rm PS}^{f,\,{\rm inf}} L$, yet,
given the precision goals of this paper, this is a sufficient threshold to allow us to safely ignore finite-volume effects.

\begin{figure}
\begin{center}
\includegraphics[width=.49\textwidth]{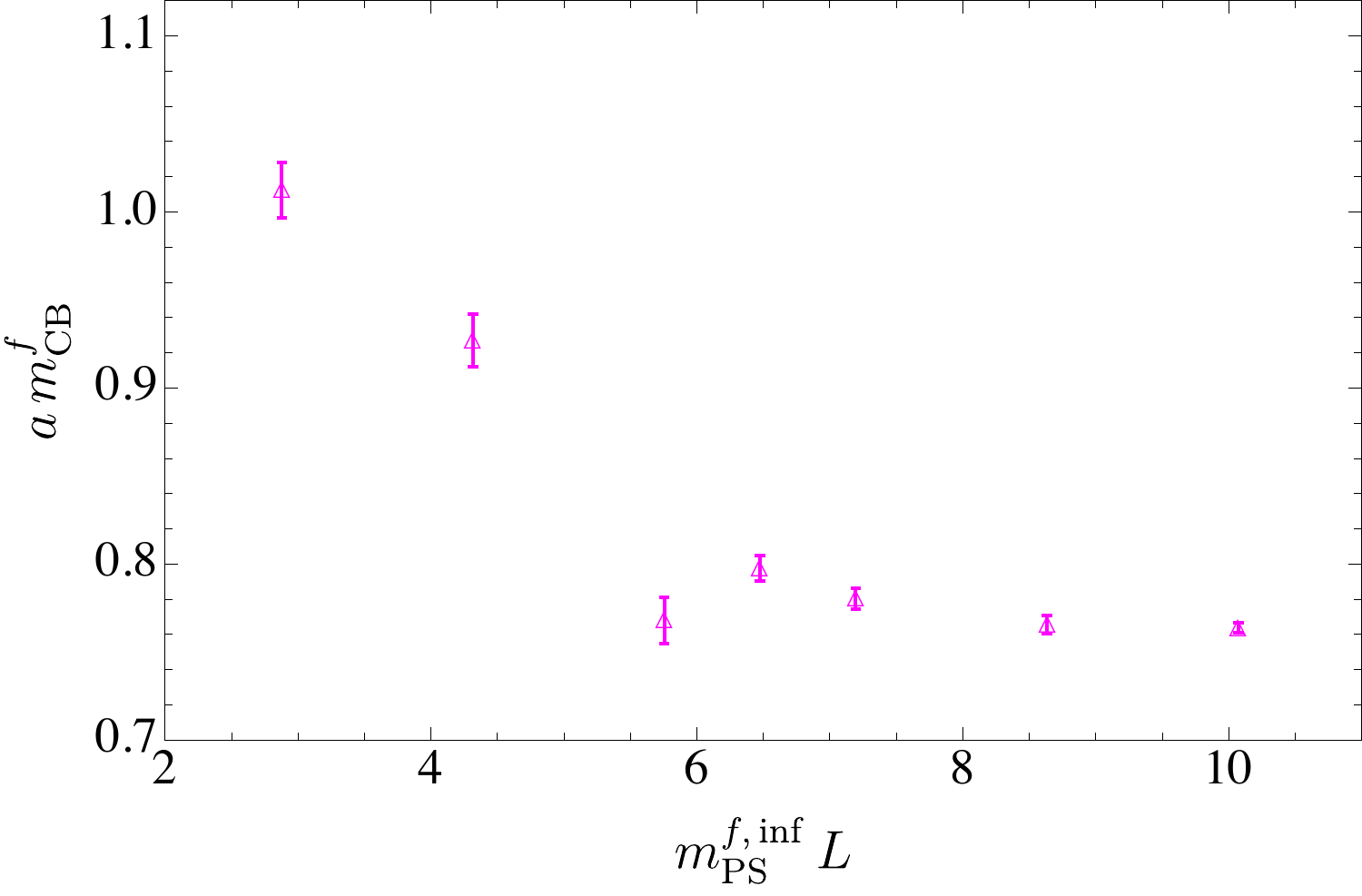}
\includegraphics[width=.49\textwidth]{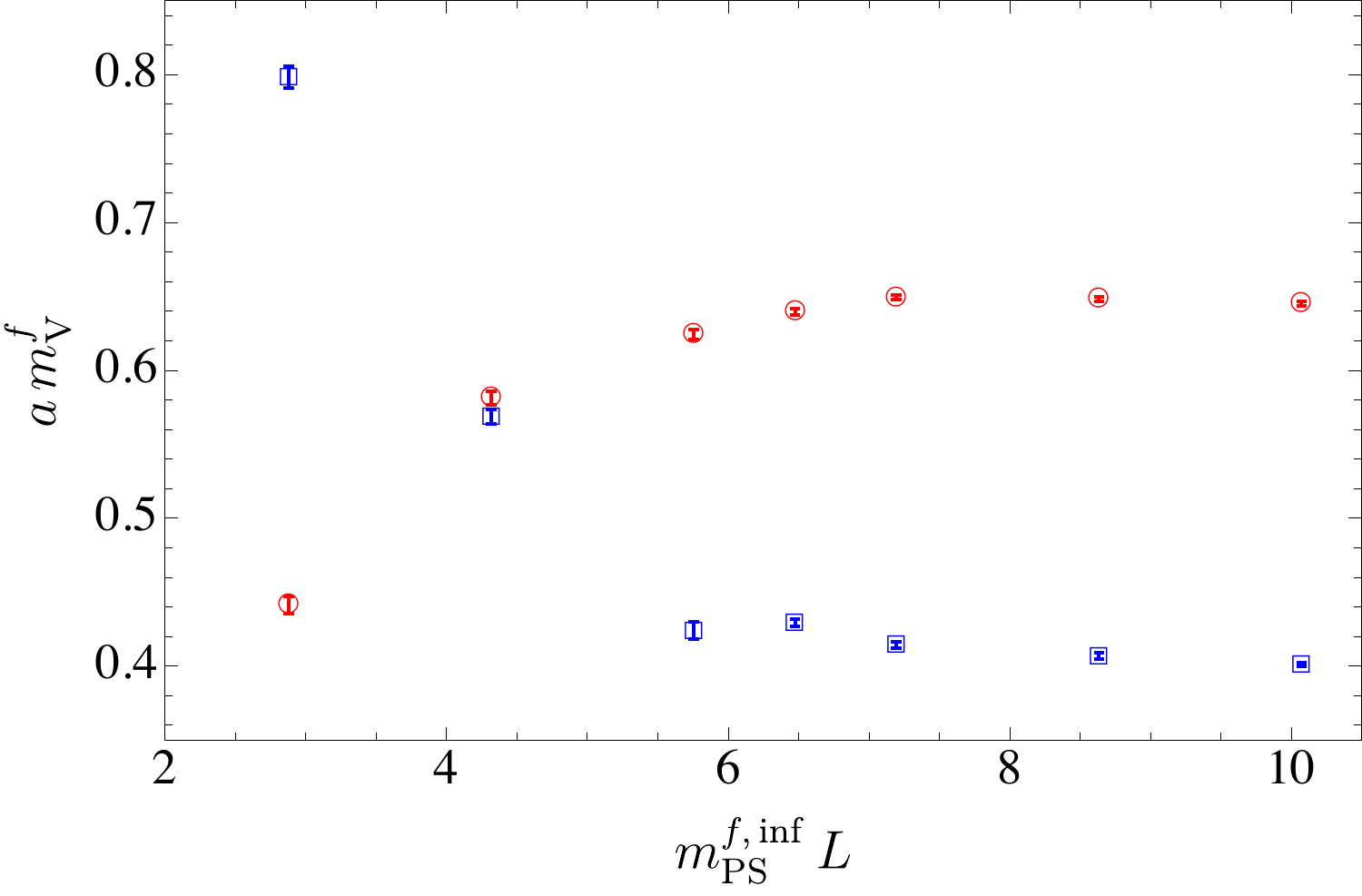}
\includegraphics[width=.49\textwidth]{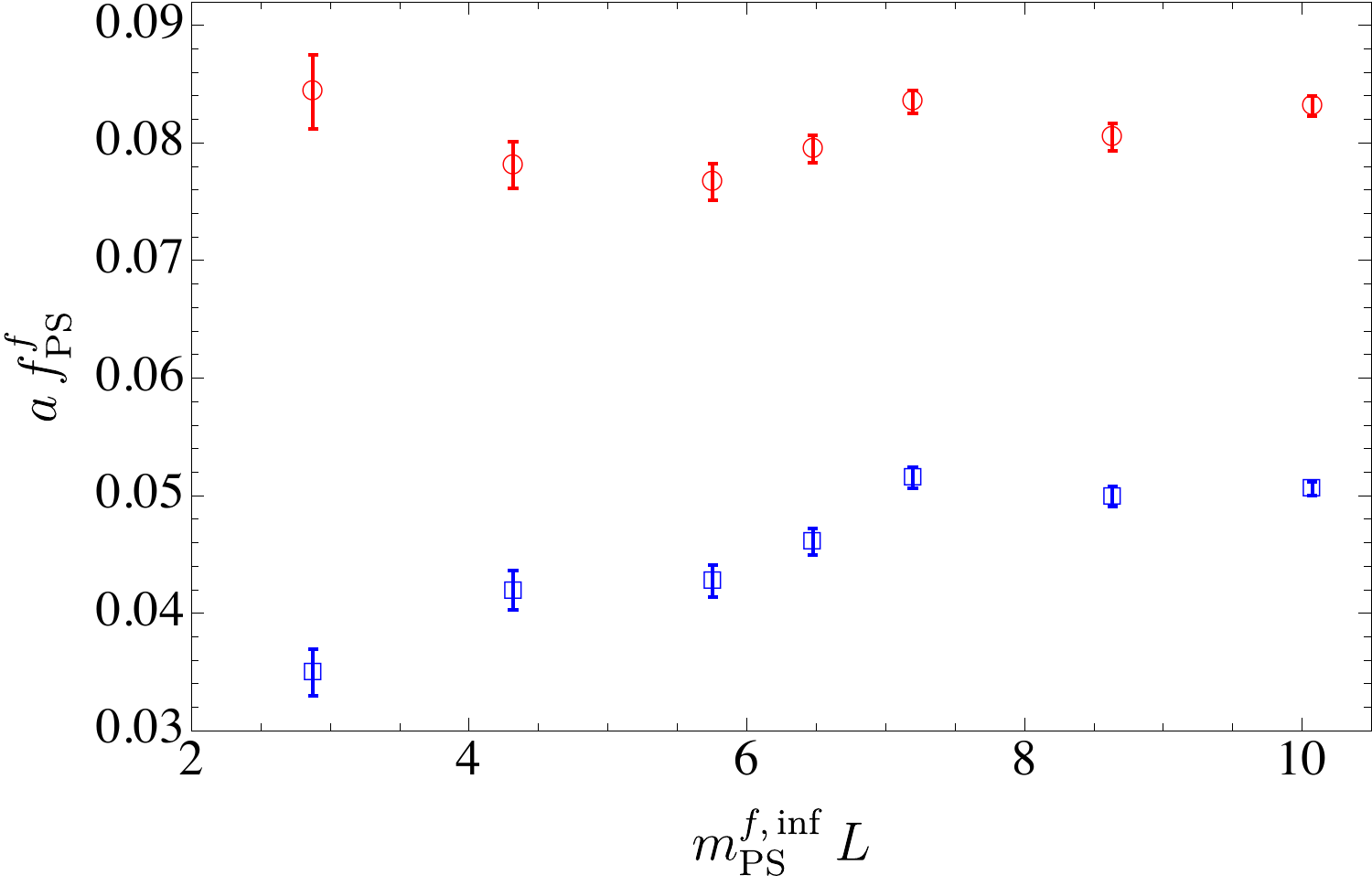}
\includegraphics[width=.49\textwidth]{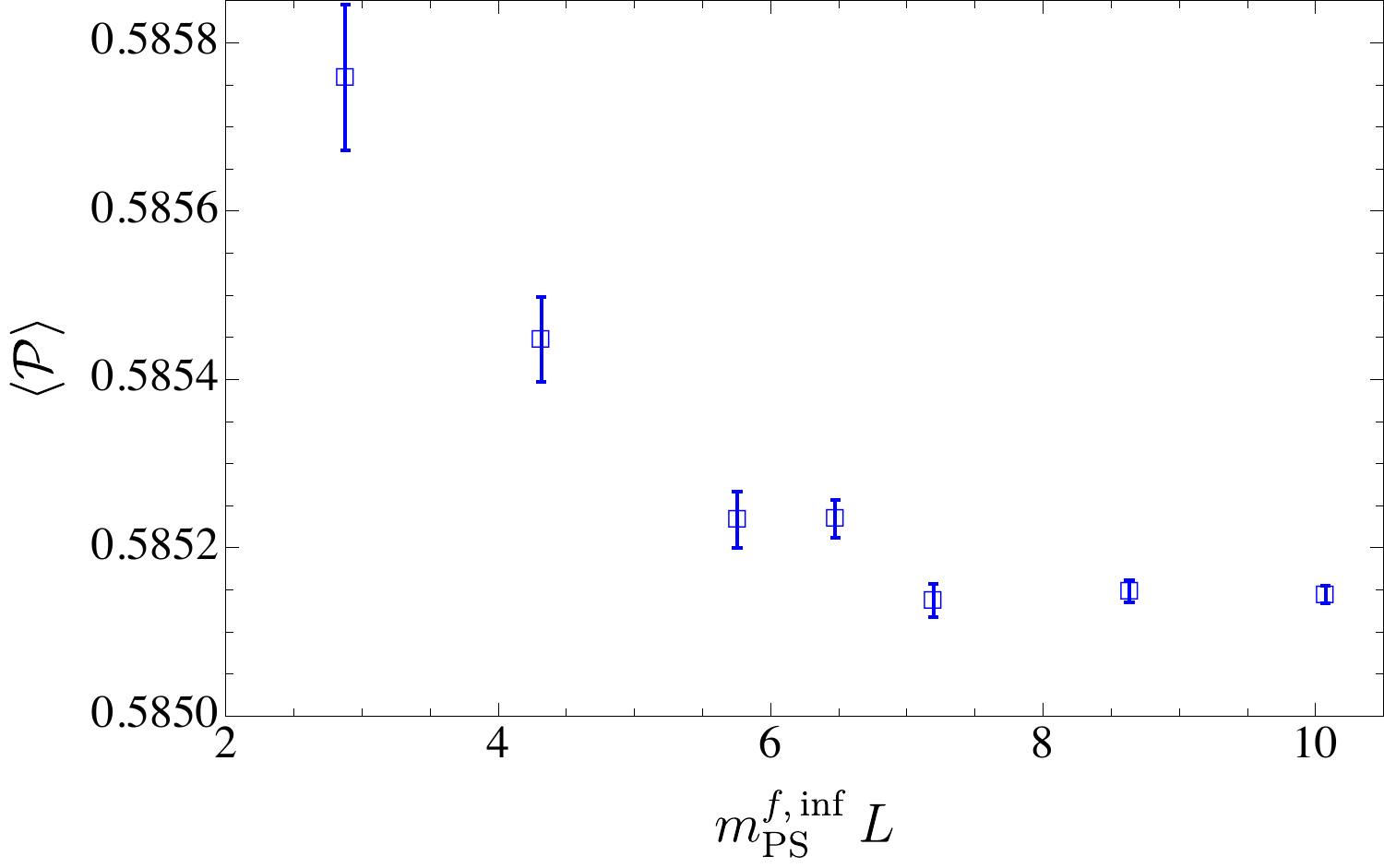}
\caption{%
\label{fig:fv_effects}%
Volume dependence of numerical observables: the mass of the Chimera baryon (top-left), 
the vector meson masses (top-right), the decay constants of pseudoscalar 
mesons (bottom-left), and the average value of plaquettes for the fundamental gauge 
links (top-right). In the cases of meson masses, red and blue colors denote
 the mesons composed of constituent fermions in the antisymmetric and fundamental representations, respectively. 
We denote by $m_{\rm PS}^{f,\,{\rm inf}} a$ the mass of the pseudoscalar  extracted from the largest
available lattice, with the lattice volume $54\times28^3$, as in Fig.~\ref{fig:fv_mps}.
The lattice parameters are $\beta=6.5$, $am_0^{f}=-0.71$, $am_0^{as}=-1.01$. 
}
\end{center}
\end{figure}

We repeat the same exercise for the following observables: 
the masses of the chimera baryons,
 the vector meson masses, the decay constants of the pseudoscalar mesons, 
 and the average plaquette values for the fundamental gauge links. 
 We display the results in \Fig{fv_effects}.
 For all these observables we find that finite volume corrections can be safely neglected, 
 if we constrain the lattice size by imposing the constraint
  $m_{\rm PS}^{f,\,{\rm inf}} L \gtrsim 7$. We could therefore conclude that our conservative 
  estimate of the minimum size of the lattice, such that the finite volume effects are 
  well under control, corresponds to $m_{\rm PS}^{f,\,{\rm inf}} L \simeq 8.5$.

For pseudoscalar and vector meson masses, we observe that the finite volume corrections 
 have opposite signs, depending on the constituent fermions:
 the difference
  $am_{\rm M}^{\rm FV}-am_{\rm M}^{\rm inf}$ between finite- and infinite-volume
  measurements is positive with $(f)$ fermion constituents and negative with
   $(as)$ fermion constituents. 
This  behaviour can be explained
 in the context of chiral perturbation theory ($\chi$PT), 
 as the finite volume corrections arise from pseudoscalar states wrapping around each 
 spatial extent of the lattice. In particular, the next-to-leading order 
 (NLO) expression of the pseudoscalar mass at finite volume is given by
\beq
m_{\rm PS}^2= M^2 \left(
1 + a_M \frac{A(M)+A_{\rm FV}(M)}{F^2} + b_M(\mu) \frac{M^2}{F^2} + \mathcal{O}(M^4)
\right),
\label{eq:finite_nlo_mps}
\eeq
where $M$ and $F$ are the mass and decay constant of the pseudoscalar meson in the 
massless limit,
obtained by replacing the one-loop integrals with finite sums.
 $A(M)$ is the one-loop contribution at infinite volume, known as the chiral 
logarithm, $A(M)=-\frac{M^2}{16 \pi^2} {\rm log}\frac{M^2}{\mu^2}$ with $\mu$ the renormalisation scale. 
The finite-volume contribution $A_{\rm FV}(M)$ arises from a finite sum on a cubic 
box of size $L$ with periodic boundary condition (see,
e.g. the Appendix of Ref.~\cite{Arndt:2004bg}). At the leading-order, 
the difference between the sums and the integrals is 
\beq
A_{\rm FV}(M) \overset{ML\gg 1}{\longrightarrow} -\frac{3}{4\pi^2}
\left(\frac{M\pi}{2L^3}\right)^{1/2}{\rm exp}[-ML].
\label{eq:chiral_logs}
\eeq

The coefficients $A(M)$ and $A_{\rm FV}(M)$ in \Eq{finite_nlo_mps} are independent of the details of the theory,
which are solely encoded in their coefficient $a_M$ \cite{Bijnens:2009qm}:
\beqs
a_M = \begin{cases}
-\frac{1}{2}-\frac{1}{N_f},~~&{\rm for}~SU(2N_f) \rightarrow Sp(2N_f), \\
-\frac{1}{N_f},~&{\rm for}~~SU(N_f)\times SU(N_f)\rightarrow SU(N_f), \\
\frac{1}{2}-\frac{1}{2N_f},~&{\rm for}~~SU(2N_f)\rightarrow SO(2N_f).
\end{cases}
\label{eq:coeff_a}
\eeqs
The first and third classes are particularly relevant to our study: the coefficients $a_M$ 
for two fundamental and three antisymmetric 
Dirac flavours are $-1$ and $+1/3$, respectively. 
Together with the fact that $A_{\rm FV}(M)$ is negative, on the basis of these analytical expressions
we expect   the pseudoscalar mass to receive positive 
 (negative) finite-volume corrections 
 for constituents in the fundamental (antisymmetric) representation, respectively.  
 This  is consistent with our numerical findings as displayed in \Fig{fv_mps},
 though, in the light of the comparatively large mass of the fermions,
one should take a conservative view towards this interpretation.

\subsection{Correlation functions of chimera baryon}
\label{Sec:chimera_prop}

\begin{figure}
\begin{center}
\includegraphics[width=.49\textwidth]{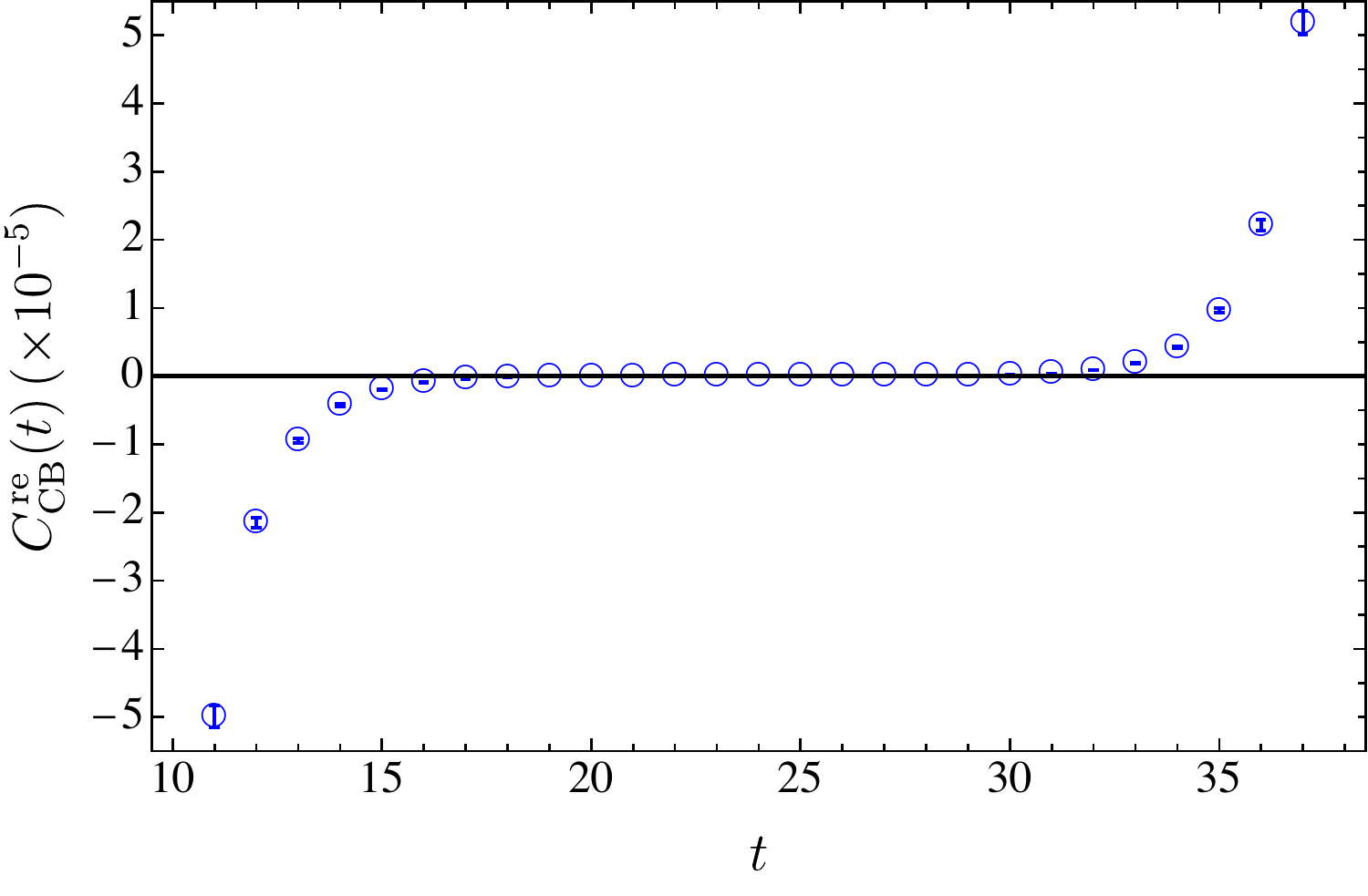}
\includegraphics[width=.49\textwidth]{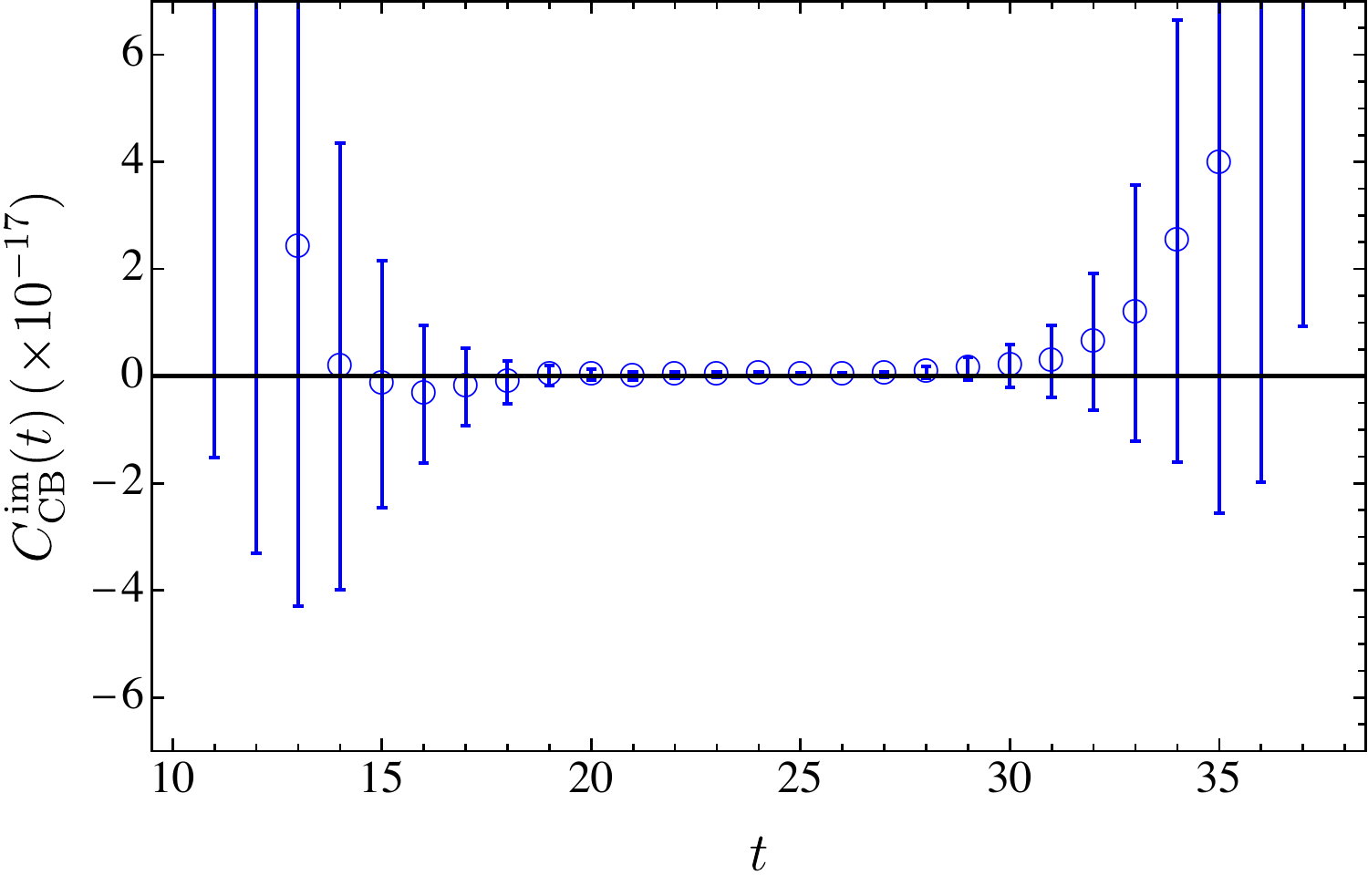}
\includegraphics[width=.49\textwidth]{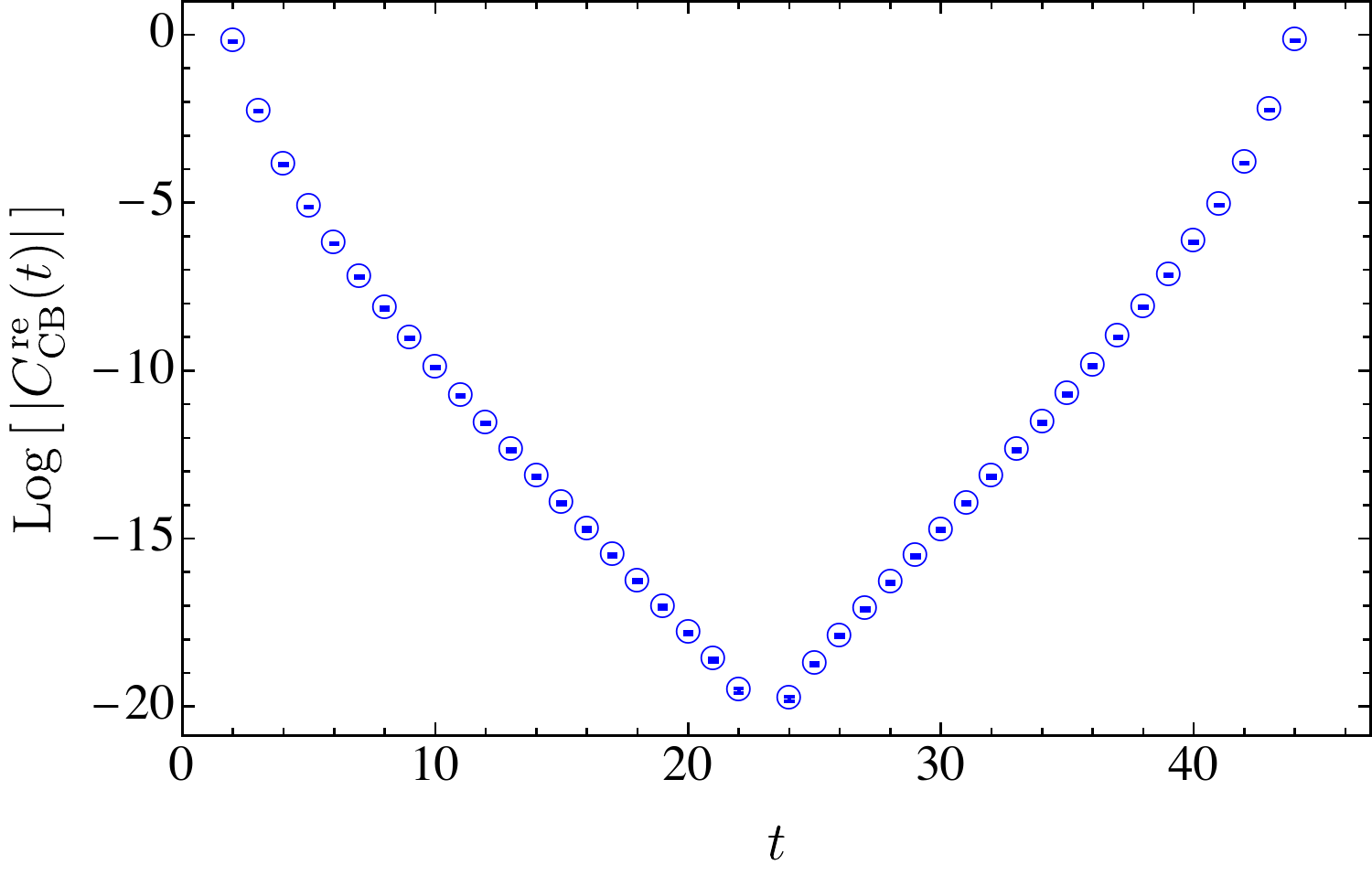}
\includegraphics[width=.49\textwidth]{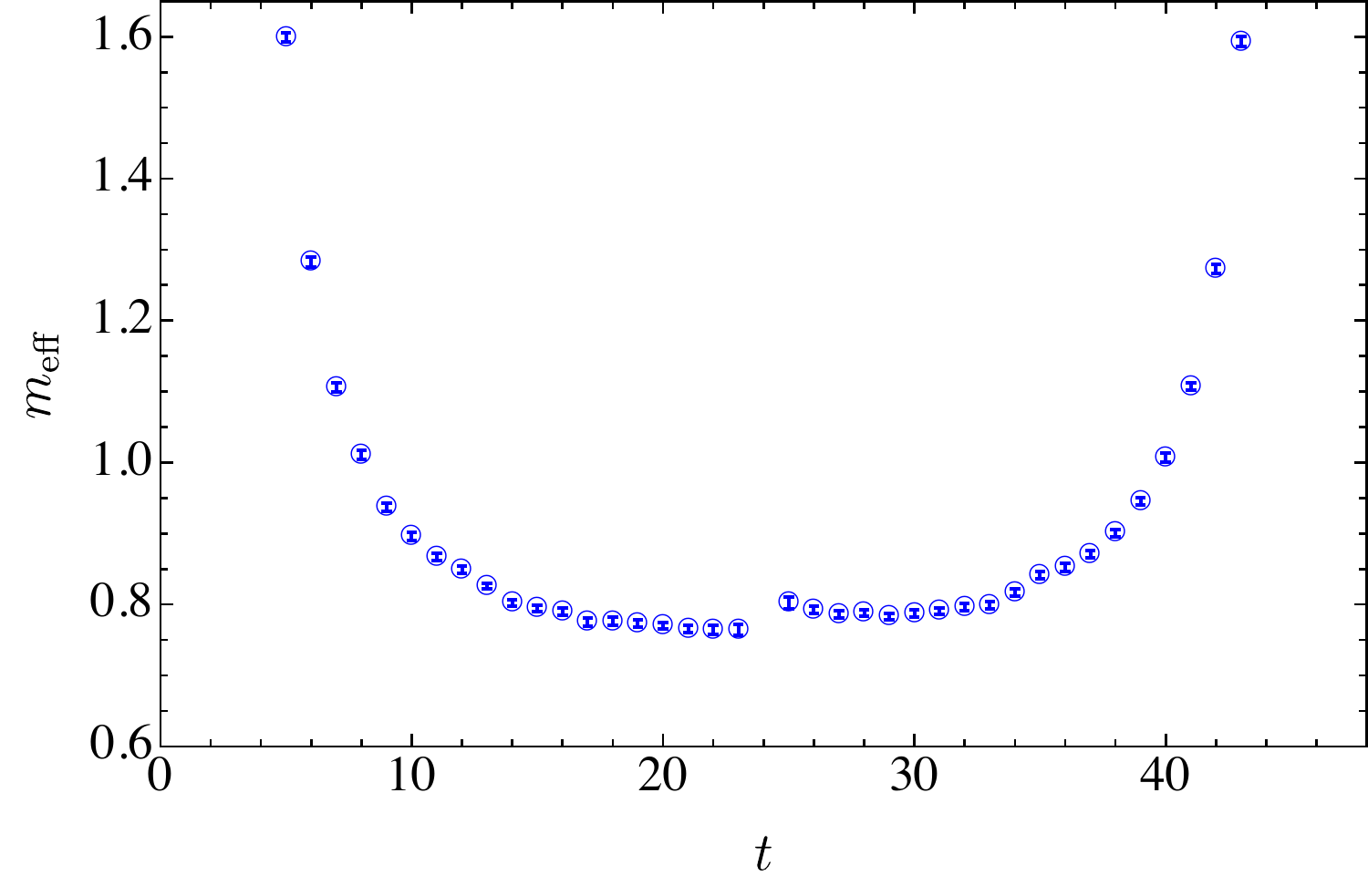}
\caption{%
\label{fig:corr_cb}%
Top panels: real (top-left) and imaginary (top-right) parts 
of the 2-point correlation function of chimera baryons.
Bottom left panel:   logarithm of the absolute value of the real part of the same correlator.
 Bottom-right panel:  the corresponding effective mass plot. 
The errors denote for 1$\sigma$ deviation estimated by using $200$ bootstrap samples. 
The gauge configurations used for the computation are generated 
by using the lattice parameters $\beta=6.5$, $am_0^{as}=-1.01$ and $am_0^f=-0.71$ 
on a lattice with size $48\times 24^3$.  
}
\end{center}
\end{figure}

\begin{figure}
\begin{center}
\includegraphics[width=.49\textwidth]{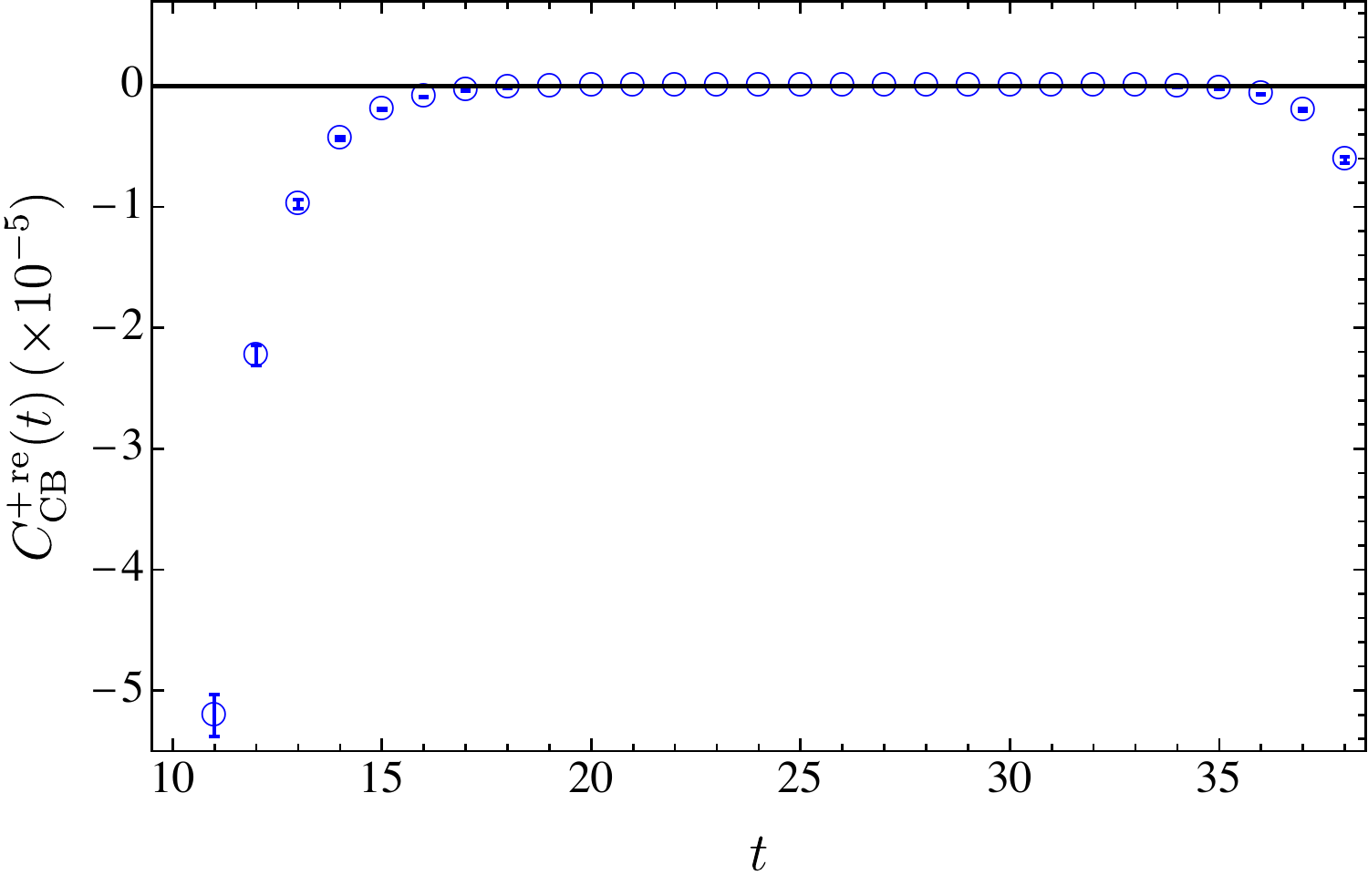}
\includegraphics[width=.49\textwidth]{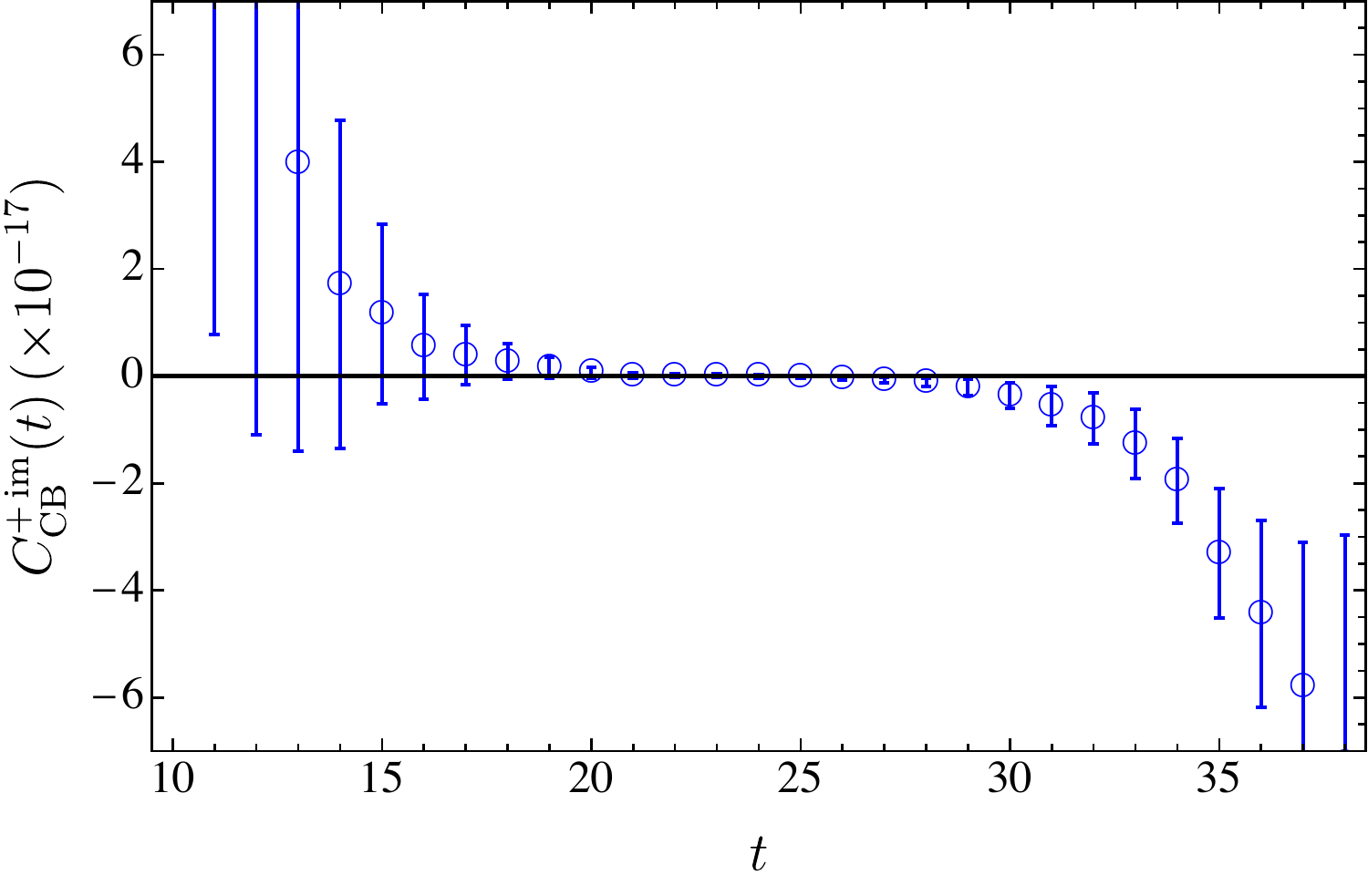}
\includegraphics[width=.49\textwidth]{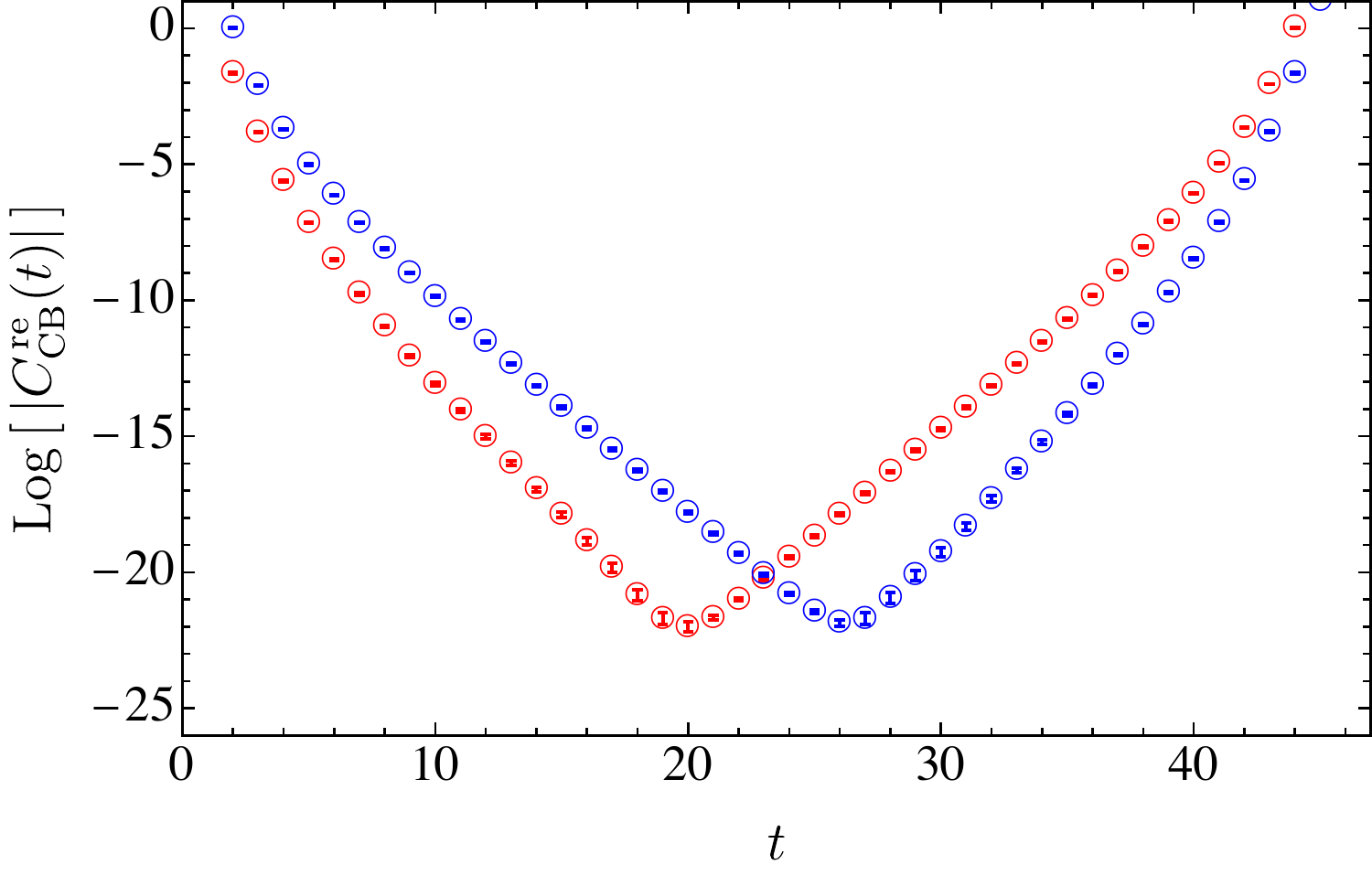}
\includegraphics[width=.49\textwidth]{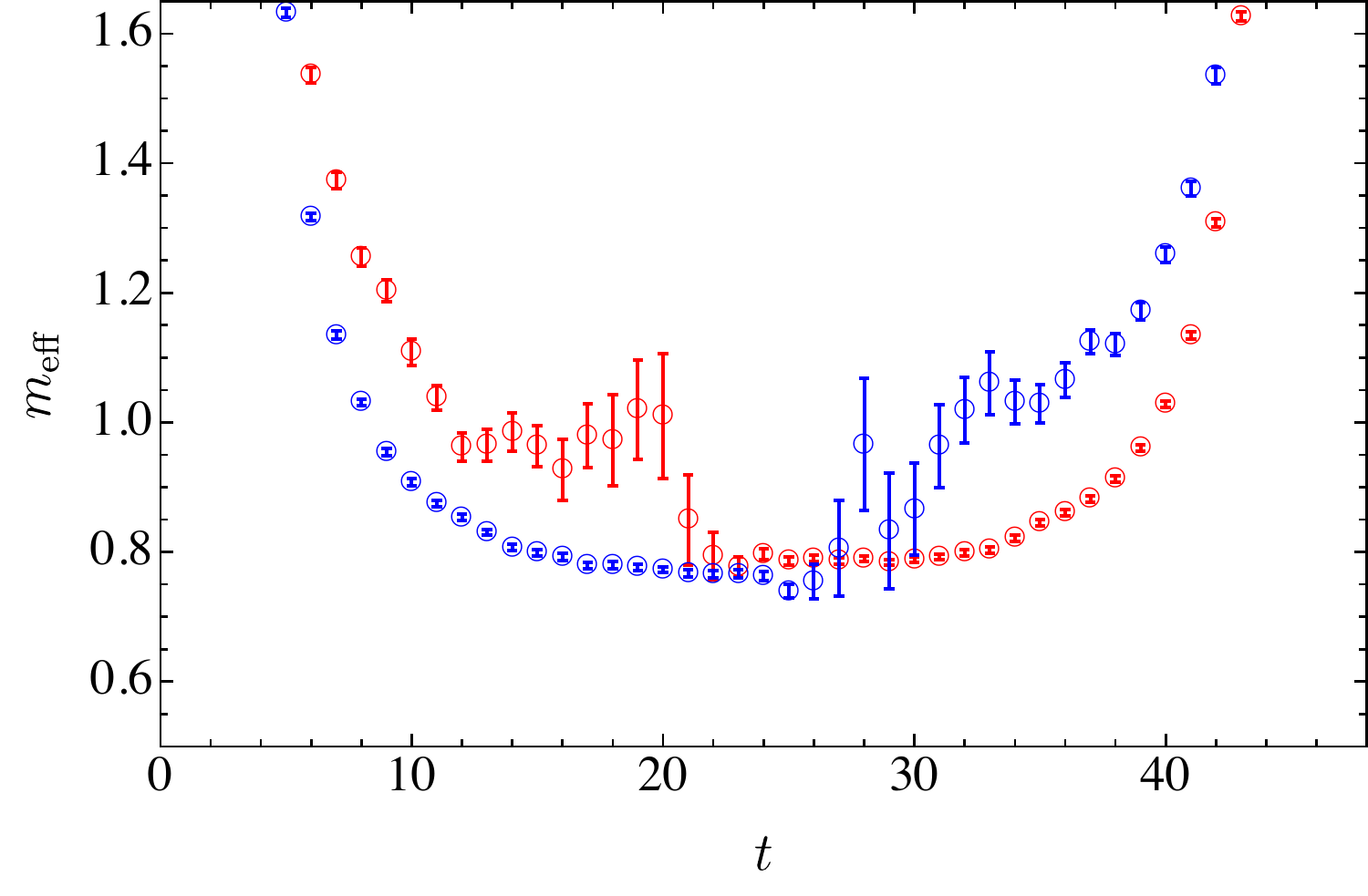}
\caption{%
\label{fig:corr_cb_parity}%
Top panels:  real (top-left) and imaginary (top-right) parts 
of the 2-point correlation function of chimera baryon after positive-parity projection.
Bottom left panel:   logarithm of the absolute value of the real part of the same correlator.
 Bottom-right panel:  the corresponding effective mass plot. 
 In the bottom panels, blue empty circles denote the results with positive-parity projection,
 while red empty circles are obtained with odd-parity projection. 
The errors denote for 1$\sigma$ deviation estimated by using $200$ bootstrap samples. 
The gauge configurations used for the computation are generated 
by using the lattice parameters  $\beta=6.5$, $am_0^{as}=-1.01$ and $am_0^f=-0.71$ 
on a lattice with size $48\times 24^3$.  
}
\end{center}
\end{figure}

We perform the first numerical calculation of the mass spectrum of chimera baryons in the
 $Sp(4)$ gauge theory with two $(f)$  and three $(as)$ Dirac fermions in the sea. 
 Since this type of calculation has never been done before for $Sp(2N)$ gauge theories, 
 we carry out several non-trivial tests using interpolating operators  with 
 and without parity projection, as in Eqs.~(\ref{eq:cb_ops}) and~(\ref{eq:cb_ops_parity}). 
We notice from the outset the comparatively large values of the ratios
  $m_{\rm PS}^{f}/m_{\rm V}^{f}\simeq 0.9$ and  $m_{\rm PS}^{as}/m_{\rm V}^{as}\simeq 0.93$.  
 
 We first present the numerical results without projection, in \Fig{corr_cb}.
 We focus on one of the gauge ensembles already used for the study of finite volume effects in \Sec{fv}. 
 We find that the real part of the correlation function shows a clear signal of exponential decay, 
 while the imaginary part shows large statistical fluctuations,
 being of the order of the machine numerical precision and consistent with zero at every Euclidean time $t$. 
 A  symmetry is visible, in the top-left and bottom-left panels,
 between forward and backward propagation, that differ by  
 having  opposite sign at late Euclidean times. 
 This is
 consistent with our expectations for the 
 asymptotic behaviour of the $2$-point 
 correlation function in \Eq{cb_corr_no_parity}. 
 As is customary, we also define the effective mass as
\beq
m_{\rm eff} = {\rm arcosh} \left(\frac{C_{\rm CB}(t+1)+C_{\rm CB} (t-1)}{2\,C_{\rm CB}(t)}\right).
\eeq
An example of the resulting effective mass plot is shown in the bottom-right panel of \Fig{corr_cb}.
The plateau over several time slices centered in the middle of the temporal extent, 
whose average value is smaller than the effective mass at earlier time, indicates that the exponential 
decay of the correlator is dominated by the ground state, as expected.

We present in \Fig{corr_cb_parity}
 the numerical results for chimera baryon correlators defined with even and odd parity projections. 
 In the top-left and top-right panels, we show the real and imaginary parts of the correlation function 
 obtained from the interpolating operator projected onto its positive parity component. 
 Again, the former shows a clear signal of exponential decay, while the latter is dominated by  statistical noise,
 and is consistent with zero. 
 In contrast with the results without the parity projection, however, we find that 
 the real part is negative and asymmetric in time, which is further evidenced by the logarithmic plot 
 in the bottom-left panel. 
  
  This result is consistent with the analytical expression for the asymptotic behaviour in \Eq{cb_corr_parity}: the forward 
  and backward propagators at late time result in a single exponential decay 
  whose decay rates are the masses of the lightest parity even and odd states, respectively. 
  Also, when we apply the negative parity projection, 
  which yields the results  denoted by red empty circles in the bottom-left and bottom-right panels, 
  we find that the forward and backward propagators exchange their roles, again as expected. 
  Up to the half of the temporal extent, furthermore, we find that the signal is stable even at later time for the positive parity case, 
  while we lose it at relatively earlier time, after a faster decay, in the negative parity  case. 
  
When looking at the effective mass plots, we cannot identify a clear plateau for the negative 
parity case. Yet, the combination of all these results 
 indicates unambiguously that the positive parity state is lighter than the negative one. 
We conclude that the ground state found in the case without  parity projection
 corresponds to the chimera baryon with positive parity, 
as we find that the masses associated with the plateaux in the effective mass 
 plots in Figs.~\ref{fig:corr_cb} and \ref{fig:corr_cb_parity} agree with each other. 
For the purposes of this paper, the discussion of the chimera baryon stops here, yet
we will follow up with more thorough investigations of the spectrum 
 in  forthcoming publications.

\subsection{Spectrum of composite states}
\label{Sec:spectrum}

\begin{figure}
\begin{center}
\includegraphics[width=.59\textwidth]{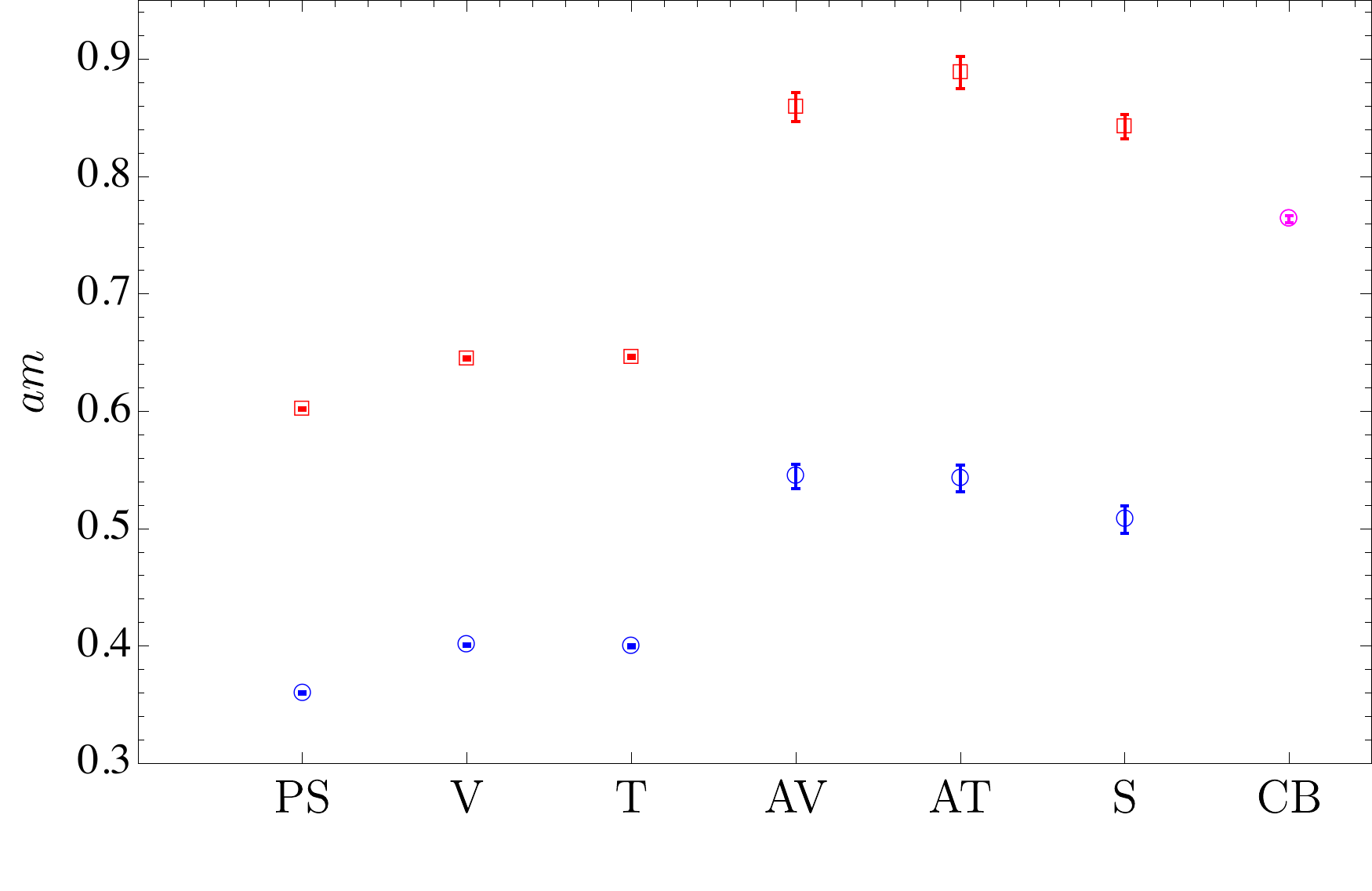}
\caption{%
\label{fig:spectrum}%
Masses $a m$, in lattice units,  of the lightest composite states
in the  $Sp(4)$ gauge theory coupled to $N_f=2$ fundamental and $n_f=3$ antisymmetric fermions. 
The blue and red colors denote 
the mesons for which the  fermion constituents are in the fundamental and antisymmetric representations, 
respectively. The magenta color denotes the chimera baryon (CB),
for which the  constituents are two fermions in the fundamental and one in the antisymmetric reresentation. 
The lattice parameters used are $\beta=6.5$, $a m_0^{as}=-1.01$, $a m_0^{f}=-0.71$, 
while the lattice volume is $N_t\times N_s^3=54\times 28^3$. 
}
\end{center}
\end{figure}

In \Fig{spectrum}, we finally present the mass spectrum of composite states, 
for an illustrative  choice of  parameters, 
in the fully dynamical $Sp(4)$
 lattice gauge theory with $N_f=2$ fundamental 
 and $n_f=3$ antisymmetric Dirac fermions, which improves a similar, preliminary plot, in Refs.~\cite{Lucini:2021xke,Bennett:2021mbw}. 
The lattice parameters are the same adopted earlier on, for the study of finite volume 
effects, restricted to the available largest volume.
Following the discussions in Secs.~\ref{Sec:mesons} and 
\ref{Sec:chimera}, we compute the masses of flavoured spin-0 and spin-1 mesons 
with fermion constituents  in the fundamental and antisymmetric representation, 
as well as the mass of the chimera baryon with positive parity. The numerical values of the results displayed in \Fig{spectrum} can be found in \App{C}.

We observe that, at least for these choices of parameters, the overall 
behaviour of the masses of the lightest states sourced by meson operators with different 
quantum numbers (PS, V, T, AV, AT) is quite similar, when comparing mesons composed 
of $(f)$ and $(as)$ fermions.
Yet, at least in this region of parameter space,  the masses of the latter are much heavier than those in the former.
The lightest chimera baryon is not light, yet its mass is slightly smaller than that of the scalar meson 
composed of constituent fermions in the antisymmetric representation, which is encouraging,
in view of future extensions of this study and possible phenomenological applications.

\section{Discussion and outlook}
\label{Sec:outlook}

This paper reports on a major step in the development
 of the extensive programme of exploration of the dynamics of 
$Sp(2N)$ gauge theories on the lattice~\cite{Bennett:2017kga,Bennett:2019jzz,
Bennett:2019cxd,Bennett:2020hqd,Bennett:2020qtj,AS}.
We considered the lattice field theory with gauge group $Sp(4)$, with matter field content 
consisting of two Wilson-Dirac fermions transforming in the fundamental representation, 
together with three transforming in the 2-index antisymmetric representation.
Due to the odd number of fermions, the contribution of matter fields to the 
non-perturbative dynamics is included by implementing a combination of
HMC and RHMC algorithms, both of which are supported by the HiRep code~\cite{DelDebbio:2008zf},
which we adapted to the treatment of $Sp(2N)$ groups and to the
simultaneous  handling of fermions in multiple representations.
The continuum limit is the minimal  theory---amenable to lattice numerical studies~\cite{Barnard:2013zea}---that
provides a UV completion for
the strongly-coupled sector of extensions of the standard model which combine
composite Higgs and partial top compositeness.

We performed the first scan of the 3-dimensional parameter space of the lattice theory,
finding evidence of the existence of a surface with boundaries separating the strong and 
weak phases. The theory admits first- as well as second-order (bulk) phase-transitions,
and we identified values of the lattice parameter space (the coupling $\beta$ and the masses of the two species of fermions
$a m_0^{f}$ and $a m_0^{as}$) that safely ensure that the lattice theory is connected to  the correct  continuum one.
We tested our algorithms, verifying explicitly that spectrum of the Dirac operator 
reproduces the expectations for the chiral symmetry breaking pattern predicted by (chiral) random matrix theory,
as done in Ref.~\cite{Cossu:2019hse} for a $SU(4)$ theory.
We assessed the size of finite-volume effects in low-lying composite state masses, and identified criteria 
that can be imposed to ensure that such lattice artefacts are negligibly small, in
comparison with statistical uncertainties.
For one choice of lattice parameters, we computed the mass spectra of the lightest mesons 
with different quantum numbers, as well as those  of chimera baryons---see Fig.~\ref{fig:spectrum}.

The combination of all of the above demonstrates that our lattice programme is now ready to start an
intensive process of numerical studies focused on the spectra of mesons and chimera baryons in this theory,
making contact with the model-building literature.
While for  the purposes of this publication we used point-like and stochastic $Z_2$ wall sources for the measurements 
of the $2$-point correlation functions,  to improve the signal to noise ratio in the numerical studies
we will use smearing techniques, both for the sources and for the 
dynamical configurations, and both of which 
have been  tested successfully on this model~\cite{Lucini:2021xke,Bennett:2021mbw}.
By further combining these techniques with the implementation of an appropriate basis for the variational problem,
and of a scale-setting process based on the Wilson flow,
such studies will provide access also to some of the excited states in the theory,
and we will be able, by varying the lattice parameters, to extrapolate 
our spectroscopy results
towards the continuum limit, in the large region of parameter space with intermediate 
fermion masses that is of direct interest for models of composite Higgs and partial top compositeness.

\vspace{1.0cm}
\begin{acknowledgments}

We would like to thank Hwancheol Jeong for useful discussion on the computation of the Dirac spectrum. 

The work of EB has been funded in part by the Supercomputing Wales project, 
which is part-funded by the European Regional Development Fund (ERDF) via Welsh Government,
and by the UKRI Science and Technology Facilities Council (STFC)
 Research Software Engineering Fellowship EP/V052489/1

The work of DKH was supported by the National Research Foundation of Korea (NRF) grant funded by the Korea government (MSIT) (2021R1A4A5031460) and also by Basic Science Research Program through the National Research Foundation of Korea (NRF) funded by the Ministry of Education (NRF-2017R1D1A1B06033701).

The work of JWL is supported by the National Research Foundation of Korea (NRF) grant funded 
by the Korea government(MSIT) (NRF-2018R1C1B3001379). 

The work of CJDL is supported by the Taiwanese MoST grant 109-2112-M-009-006-MY3. 

The work of BL and MP has been supported in part by the STFC 
Consolidated Grants No. ST/P00055X/1 and No. ST/T000813/1.
 BL and MP received funding from
the European Research Council (ERC) under the European
Union’s Horizon 2020 research and innovation program
under Grant Agreement No.~813942. 
The work of BL is further supported in part 
by the Royal Society Wolfson Research Merit Award 
WM170010 and by the Leverhulme Trust Research Fellowship No. RF-2020-4619.

DV acknowledges support from the INFN HPC-HTC project.

Numerical simulations have been performed on the 
Swansea University SUNBIRD cluster (part of the Supercomputing Wales project) and AccelerateAI A100 GPU system,
on the local HPC
clusters in Pusan National University (PNU) and in National Yang Ming Chiao Tung University (NYCU),
and the DiRAC Data Intensive service at Leicester.
The Swansea University SUNBIRD system and AccelerateAI are part funded by the European Regional Development Fund (ERDF) via Welsh Government.
The DiRAC Data Intensive service at Leicester is operated by 
the University of Leicester IT Services, which forms part of 
the STFC DiRAC HPC Facility (www.dirac.ac.uk). The DiRAC 
Data Intensive service equipment at Leicester was funded 
by BEIS capital funding via STFC capital grants ST/K000373/1 
and ST/R002363/1 and STFC DiRAC Operations grant ST/R001014/1. 
DiRAC is part of the National e-Infrastructure.

{\bf Open Access Statement-} For the purpose of open access, 
the author has applied a Creative Commons Attribution (CC BY) 
licence to any Author Accepted Manuscript version arising.

\end{acknowledgments}

\vspace{1.0cm}

\appendix
\section{Notation and conventions}
\label{Sec:A}

In this appendix, we summarise some of
the conventions adopted
 in the construction of the continuum and lattice theories.
We display some technical  relations which are used
  in the main text. In particular, we present 
  the chiral representation of the gamma matrices, 
  both in Minkowski and Euclidean space-time,
and  a choice of generators
   for the fundamental and antisymmetric representations 
   of $Sp(4)$, which are
   required to compute the MD forces in the HMC/RHMC algorithms. 

\subsection{Gamma matrices in Minkowski space}
\label{Sec:A1}

In \Sec{model}, the continuum model relevant for phenomenological applications
is presented in Minkowski space-time.
The metric $\eta^{\mu\nu}$ is given by
\beq
\eta=\left(
\begin{array}{cccc}
1 & 0 & 0 & 0  \\
0 & -1 & 0 & 0  \\
0 & 0 & -1 & 0  \\
0 & 0 & 0 & -1 
\end{array}
\right)\,,
\label{Eq:minkowski_metric}
\eeq
where $\mu,\,\nu=0,\,\cdots,\,3$ are space-time indexes.
The Dirac gamma matrices satisfy the anticommutation relations,\footnote{In this appendix, we 
denote with a subscript $M$ or $E$ the gamma matrices in Minkowski or Euclidean space, respectively.
We suppress this subscript in the main body of the paper, hence, in \Sec{model},
we write $\gamma^{\mu}\equiv\gamma_M^{\mu}$.
Similarly, in \Sec{model} we denote $\gamma^5\equiv \gamma^5_M$ and $ C\equiv C_M$.}
\beq
\{\gamma^{\mu}_M,\gamma^{\nu}_M\}=2\eta^{\mu\nu}\mathbb{1}_4\,,
\label{Eq:anticom}
\eeq
where $\mathbb{1}_4$ is the unit matrix $\delta ^{\alpha\beta}$ in spinor space, with $\alpha,\beta=1,\,\cdots,\,4$.
Hence,  $\gamma^{0}_M=\gamma^{0\,\dagger}_M$, while $\gamma^{i}_M=-\gamma^{i\,\dagger}_M$
for $i=1,\,2,\,3$, and the
Hermiticity condition $\gamma^{\mu\,\dagger}_M=\gamma^{0}_M\gamma^{\mu}_M\gamma^{0}_M$
holds.
We adopt the chiral basis, and write explicitly the matrices as follows:

\beqs
\gamma^0_M&=&\left(\begin{array}{cc}
\mathbb{0}_2 & \mathbb{1}_2 \cr
\mathbb{1}_2 & \mathbb{0}_2
\end{array}\right)\,,~~
\gamma^i_M\,=\,\left(\begin{array}{cc}
\mathbb{0}_2 & -\tau^i \cr
\tau^i & \mathbb{0}_2
\end{array}\right)\,,
\eeqs
with the Pauli matrices $\tau^i$,
\beqs
\tau^1&=&\left(\begin{array}{cc}
0 & 1 \cr
1 & 0 \end{array}\right)\,,\,
\tau^2\,=\,\left(\begin{array}{cc}
0 & -i \cr
i & 0 \end{array}\right)\,,\,
\tau^3\,=\,\left(\begin{array}{cc}
1 & {0} \cr
0 & -1 \end{array}\right)\,.
\eeqs
With this choice, the $\gamma^5_M$ matrix and the charge-conjugation matrix $C_M$ are defined as
\beqs
\gamma^5_M\,=i\gamma^0_M\gamma^1_M\gamma^2_M\gamma^3_M=\,\left(\begin{array}{cc}
\mathbb{1}_2 & \mathbb{0}_2 \cr
\mathbb{0}_2& -\mathbb{1}_2
\end{array}\right)\,,~~
C_M=i\gamma^2_M \gamma^0_M = \,\left(\begin{array}{cc}
 -i\tau^2 & \mathbb{0}_2 \cr
\mathbb{0}_2 & i \tau^2 
\end{array}\right)\,,
\eeqs
where the former defines the chirality as in \Eq{chiral_projector} and satisfies $\{\gamma^\mu_M,\gamma^5_M\}=0$, 
while the latter obeys the defining relations $C\gamma^{\mu}_M C^{-1}=-\gamma^{\mu\, T}_M$
and $CC^{\dagger}=\mathbb{1}_4=-C^2$.

\subsection{Gamma matrices in Euclidean space}
\label{Sec:A2}

$Sp(2N)$ lattice gauge theories are defined
 in four-dimensional Euclidean space-time.
The anticommutators of the (Hermitian) Euclidean gamma matrices satisfy the  relations
\beq
\{\gamma^{\mu}_E,\gamma^{\nu}_E\}=2\delta^{\mu\nu}\mathbb{1}_4\,.
\label{eq:anticommute}
\eeq
The chiral representation of the gamma matrices has the following explicit form\footnote{In 
Sects.~\ref{Sec:latticesetup} and~\ref{Sec:observables},  we omit the subscription $E$,
  and denote $\gamma^\mu \equiv \gamma^\mu_E$, $\gamma^5\equiv
  \gamma^5_E$, and $C\equiv C_E$.  }
\beqs
\gamma^0_E&=&\left(\begin{array}{cc}
\mathbb{0}_2 & -\mathbb{1}_2 \cr
-\mathbb{1}_2 & \mathbb{0}_2
\end{array}\right)\,,~~
\gamma^i_E\,=\,\left(\begin{array}{cc}
\mathbb{0}_2 & -i \tau^i \cr
i \tau^i & \mathbb{0}_2
\end{array}\right)\,.~~
\eeqs
In this basis, the $\gamma^5_E$ and the charge-conjugation $C_E$ matrices are
\beqs
\gamma^5_E\,=\gamma^0_E\gamma^1_E\gamma^2_E\gamma^3_E\,=\left(\begin{array}{cc}
\mathbb{1}_2 & \mathbb{0}_2 \cr
\mathbb{0}_2& -\mathbb{1}_2
\end{array}\right),\,~~
C_E=\gamma^0_E \gamma^2_E = \,\left(\begin{array}{cc}
 i\tau^2 & \mathbb{0}_2 \cr
\mathbb{0}_2 & -i \tau^2 
\end{array}\right).\,
\eeqs

The following relations are used in the algebraic manipulations
of Sects.~\ref{Sec:latticesetup} and~\ref{Sec:observables}:
\beqs
\{\gamma^{\mu}_E,\gamma^5_E\}&=&0,
\label{eq:rel0}
\\
\gamma^{5\dagger}_E &=& \gamma^5_E, 
\label{eq:rel1}
\\
C_E^\dagger = C_E^{-1} = C_E^T &=& -C_E, 
\label{Eq:rel2}
\\
\gamma^0_E C_E^\dagger \gamma^0_E &=& C_E, 
\label{eq:rel3}
\\
\gamma^5_E \gamma^{\mu\,\dagger}_E \gamma^5_E &=& -\gamma^{\mu}_E,
\label{eq:rel4}
\\
C^{-1}_E \gamma^\mu_E C_E = -\gamma_E^{\mu *}&=& -\gamma_E^{\mu T},
\label{eq:rel5}
\\
\left(\gamma^5_E\right)^2=-C_E^2&=&1
\label{eq:rel6}
\eeqs

In particular, by using  \Eq{rel4} and \Eq{DiracF}---or \Eq{inverse_prop}---one can prove the
 $\gamma^5$-hermiticity of the Wilson-Dirac operator $D$, 
or equivalently  the $\gamma^5$-hermiticity of the fermion propagator $S(x,y)$, as follows,

\beqs
\gamma^5_E D_{xy}^{R,\dagger} \gamma^5_E  &=&
\gamma^5_E (S_R(x,y)^{-1})^\dagger \gamma^5_E  \nn \\
&=& \left(4+a m_0^R\right)\delta_{xy}
-\frac{1}{2}\sum_{\mu} \left((1+\gamma_{E\,\mu})U^{(R),\dagger}_\mu(x)\delta_{x+\mu,y}
+(1-\gamma_{E\,\mu})U_\mu^{(R)}(y)\delta_{x,y+\mu}\right) \nn \\
&=& S_R(y,x)^{-1} = D_{y,x}^R.
\eeqs
For the combinations of gamma matrices 
$\left(\Gamma^1,\Gamma^2\right)=\left(C\gamma^5,\mathbb{1}_4\right)$,  
that appear  in Sect.~\ref{Sec:observables}, 
 the following useful relation, which enters the derivation of \Eq{cb_ops_conjugate}, holds:
\beq
\left(\gamma^0_E\Gamma^{1\ast}\gamma^0_E\right)^{\alpha\beta}
\left(\gamma^0_E\Gamma^{2\dagger}\gamma^0_E\right)^{\gamma\delta}
=\left(\Gamma^1\right)^{\alpha\beta} \left(\Gamma^2\right)^{\gamma\delta}\,,
\label{eq:DG_rel}
\eeq
which descends from the fact that that  $\gamma_E^0$, $\gamma_E^2$, $\gamma_E^5$ are real and Hermitian.

\subsection{A basis of generators for $(f)$ and $(as)$ representations of $Sp(4)$}
\label{Sec:A3}

In the HMC/RHMC algorithms it is necessary to have an explicit expression 
for the generators for a given representation $R$ of $Sp(4)$ in order to compute the MD 
forces associated with the HMC/RHMC Hamiltonian.
We make an explicit choice of basis, and report it here, for completeness.
For the fundamental representation $T^A_{(f)}$, with $A=1,2,\cdots,10$, 
our choice is the following:
\beq
T^1_{(f)}=\frac{1}{2}\left(
\begin{array}{cccc}
1 & 0 & 0 & 0  \\
0 & 0 & 0 & 0  \\
0 & 0 & -1 & 0  \\
0 & 0 & 0 & 0  
\end{array}
\right),~
T^2_{(f)}=\frac{1}{2}\left(
\begin{array}{cccc}
0 & 0 & 0 & 0  \\
0 & 1 & 0 & 0  \\
0 & 0 & 0 & 0  \\
0 & 0 & 0 & -1  
\end{array}
\right), ~
T^3_{(f)}=\frac{1}{2\sqrt{2}}\left(
\begin{array}{cccc}
0 & i & 0 & 0  \\
-i & 0 & 0 & 0  \\
0 & 0 & 0 & i  \\
0 & 0 & -i & 0  
\end{array}
\right), \nonumber
\eeq
\beq
T^4_{(f)}=\frac{1}{2\sqrt{2}}\left(
\begin{array}{cccc}
0 & 1 & 0 & 0  \\
1 & 0 & 0 & 0  \\
0 & 0 & 0 & -1  \\
0 & 0 & -1 & 0  
\end{array}
\right),~
T^5_{(f)}=\frac{1}{2}\left(
\begin{array}{cccc}
0 & 0 & 1 & 0  \\
0 & 0 & 0 & 0  \\
1 & 0 & 0 & 0  \\
0 & 0 & 0 & 0  
\end{array}
\right), ~
T^6_{(f)}=\frac{1}{2}\left(
\begin{array}{cccc}
0 & 0 & i & 0  \\
0 & 0 & 0 & 0  \\
-i & 0 & 0 & 0  \\
0 & 0 & 0 & 0  
\end{array}
\right), \nonumber
\eeq
\beq
T^7_{(f)}=\frac{1}{2}\left(
\begin{array}{cccc}
0 & 0 & 0 & 0  \\
0 & 0 & 0 & 1  \\
0 & 0 & 0 & 0  \\
0 & 1 & 0 & 0  
\end{array}
\right),~
T^8_{(f)}=\frac{1}{2}\left(
\begin{array}{cccc}
0 & 0 & 0 & 0  \\
0 & 0 & 0 & i  \\
0 & 0 & 0 & 0  \\
0 & -i & 0 & 0  
\end{array}
\right), ~
T^9_{(f)}=\frac{1}{2\sqrt{2}}\left(
\begin{array}{cccc}
0 & 0 & 0 & i  \\
0 & 0 & i & 0  \\
0 & -i & 0 & 0  \\
-i & 0 & 0 & 0  
\end{array}
\right), 
\eeq
\beq
T^{10}_{(f)}=\frac{1}{2\sqrt{2}}\left(
\begin{array}{cccc}
0 & 0 & 0 & 1  \\
0 & 0 & 1 & 0  \\
0 & 1 & 0 & 0  \\
1 & 0 & 0 & 0  
\end{array}\nonumber
\right). 
\eeq

As discussed in \Sec{action}, the generators for the antisymmetric representation of $Sp(4)$ 
appear in the infinitesimal transformation 
of the antisymmetric link variable in \Eq{U_AS}. We adopt the conventional basis
of matrices $e_{(as)}$  given by
\beq
e^{(12)}_{(as)}=\frac{1}{\sqrt{2}}\left(
\begin{array}{cccc}
0 & -1 & 0 & 0  \\
1 & 0 & 0 & 0  \\
0 & 0 & 0 & 0  \\
0 & 0 & 0 & 0 
\end{array}
\right),~
e^{(23)}_{(as)}=\frac{1}{\sqrt{2}}\left(
\begin{array}{cccc}
0 & 0 & 0 & 0  \\
0 & 0 & -1 & 0  \\
0 & 1 & 0 & 0  \\
0 & 0 & 0 & 0 
\end{array}
\right), ~
e^{(14)}_{(as)}=\frac{1}{\sqrt{2}}\left(
\begin{array}{cccc}
0 & 0 & 0 & -1  \\
0 & 0 & 0 & 0  \\
0 & 0 & 0 & 0  \\
1 & 0 & 0 & 0 
\end{array}
\right), 
\eeq
\beq
e^{(24)}_{(as)}=\frac{1}{2}\left(
\begin{array}{cccc}
0 & 0 & 1 & 0  \\
0 & 0 & 0 & -1  \\
-1 & 0 & 0 & 0  \\
0 & 1 & 0 & 0 
\end{array}
\right), ~
e^{(34)}_{(as)}=\frac{1}{\sqrt{2}}\left(
\begin{array}{cccc}
0 & 0 & 0 & 0  \\
0 & 0 & 0 & 0  \\
0 & 0 & 0 & -1  \\
0 & 0 & 1 & 0 
\end{array}
\right), \nonumber
\eeq
With this convention we find the following expressions of the generators for the antisymmetric representation of $Sp(4)$:
\beq
T^1_{(as)}=\frac{1}{2}\left(
\begin{array}{ccccc}
1 & 0 & 0 & 0 & 0 \\
0 & -1 & 0 & 0 & 0 \\
0 & 0 & 1 & 0 & 0 \\
0 & 0 & 0 & 0 & 0 \\
0 & 0 & 0 & 0 & -1
\end{array}
\right),~
T^2_{(as)}=\frac{1}{2}\left(
\begin{array}{ccccc}
1 & 0 & 0 & 0 & 0 \\
0 & 1 & 0 & 0 & 0 \\
0 & 0 & -1 & 0 & 0 \\
0 & 0 & 0 & 0 & 0 \\
0 & 0 & 0 & 0 & -1
\end{array}
\right), \nonumber
\eeq
\beq
T^3_{(as)}=\frac{-i}{2}\left(
\begin{array}{ccccc}
0 & 0 & 0 & 0 & 0 \\
0 & 0 & 0 & -1 & 0 \\
0 & 0 & 0 & -1 & 0 \\
0 & 1 & 1 & 0 & 0 \\
0 & 0 & 0 & 0 & 0
\end{array}
\right),~
T^4_{(as)}=\frac{1}{2}\left(
\begin{array}{ccccc}
0 & 0 & 0 & 0 & 0 \\
0 & 0 & 0 & -1 & 0 \\
0 & 0 & 0 & 1 & 0 \\
0 & -1 & 1 & 0 & 0 \\
0 & 0 & 0 & 0 & 0
\end{array}
\right), \nonumber
\eeq
\beq
T^5_{(as)}=\frac{1}{2}\left(
\begin{array}{ccccc}
0 & -1 & 0 & 0 & 0 \\
-1 & 0 & 0 & 0 & 0 \\
0 & 0 & 0 & 0 & 1 \\
0 & 0 & 0 & 0 & 0 \\
0 & 0 & 1 & 0 & 0
\end{array}
\right),~
T^6_{(as)}=\frac{-i}{2}\left(
\begin{array}{ccccc}
0 & 1 & 0 & 0 & 0 \\
-1 & 0 & 0 & 0 & 0 \\
0 & 0 & 0 & 0 & -1 \\
0 & 0 & 0 & 0 & 0 \\
0 & 0 & 1 & 0 & 0
\end{array}
\right), 
\eeq
\beq
T^7_{(as)}=\frac{1}{2}\left(
\begin{array}{ccccc}
0 & 0 & 1 & 0 & 0 \\
0 & 0 & 0 & 0 & -1 \\
1 & 0 & 0 & 0 & 0 \\
0 & 0 & 0 & 0 & 0 \\
0 & -1 & 0 & 0 & 0
\end{array}
\right),~
T^8_{(as)}=\frac{-i}{2}\left(
\begin{array}{ccccc}
0 & 0 & -1 & 0 & 0 \\
0 & 0 & 0 & 0 & 1 \\
1 & 0 & 0 & 0 & 0 \\
0 & 0 & 0 & 0 & 0 \\
0 & -1 & 0 & 0 & 0
\end{array}
\right),\nonumber
\eeq
\beq
T^9_{(as)}=\frac{-i}{2}\left(
\begin{array}{ccccc}
0 & 0 & 0 & 1 & 0 \\
0 & 0 & 0 & 0 & 0 \\
0 & 0 & 0 & 0 & 0 \\
-1 & 0 & 0 & 0 & -1 \\
0 & 0 & 0 & 1 & 0
\end{array}
\right),~
T^{10}_{(as)}=\frac{1}{2}\left(
\begin{array}{ccccc}
0 & 0 & 0 & -1 & 0 \\
0 & 0 & 0 & 0 & 0 \\
0 & 0 & 0 & 0 & 0 \\
-1 & 0 & 0 & 0 & 1 \\
0 & 0 & 0 & 1 & 0
\end{array}
\right). 
\nn
\eeq

\section{More about chimera baryons on the lattice}
\label{Sec:B}

In the discussion in the main text, 
we wrote the correlation function $C_{\rm CB}(t-t_0)$ 
involving the chimera baryon operator appearing in $\mathcal{O}_{{\rm CB},\,4}$ 
and  $\mathcal{O}_{{\rm CB},\,5}$ in Eq.~(\ref{Eq:top}). 
It is worth checking that $C_{\rm CB}(t-t_0)$ built from a different choice of 
element of $\mathcal{O}_{\rm CB}\sim 4$ of the global $Sp(4)$ symmetry
gives rise to the same results. 
To this purpose, let us consider the combination of the interpolating operators 
$\frac{1}{2}\left(\mathcal{O}_{{\rm CB},1}+i\mathcal{O}_{{\rm CB},2}\right)$:
\beq
\mathcal{O}^{k\,\gamma}_{\rm CB} (x) = (
\overline{Q^{1\,a}}(x)  \gamma^5 Q^{2\,b}(x))\Omega_{bc} \delta^{\gamma\delta} \Psi^{k\,ca}_\delta(x)\,,
\eeq
and its Dirac conjugate
\beq
\overline{\mathcal{O}^k_{\rm CB}}^{\,\gamma}(x) = 
\delta^{\gamma\delta} \overline{\Psi^{k\,ca}}_\delta(x) \Omega^{cb} (\overline{Q^{2\,b}}(x) \gamma^5 Q^{1\,a}(x)).
\eeq
Then, the corresponding $2$-point correlation function is 
\beqs
\langle \mathcal{O}^{k\,\gamma}_{{\rm CB}}(x) \overline{\mathcal{O}^k_{{\rm CB}}}^{\,\gamma'}(y) \rangle 
&=& 
-\Omega_{bc} \Omega^{c'b'} \delta^{\gamma\delta} \delta^{\gamma'\delta'} 
S^{k\,ca}_{\Psi\,\,\,\,\,c'a'\,\delta\delta'}(x,y) 
S^{2\,b}_{Q\,\,\,b^{\prime}\,\beta\beta'}(x,y) \gamma^{5\,\beta'\alpha'} 
S_{Q\,\,\,\,a\,\alpha'\alpha}^{1\,a'}(y,x) \gamma^{5\,\alpha\beta}\,.
\label{eq:cb_ps} 
\eeqs
To see the equivalence between Eqs.~(\ref{eq:cb_ps}) and (\ref{eq:cb_corr}) with the choice of $(\Gamma^1,\Gamma^2)=(C\gamma^5,\mathbb{1})$, we will use the following properties. 
First of all, for a symplectic unitary matrix $\mathcal{U}\in Sp(4)$:
\beq
\Omega^{-1} \mathcal{U} \Omega = \mathcal{U}^*.
\label{eq:symplectic}
\eeq
We next consider the inverse of the fermion propagator in the Wilson-Dirac formalism
\beqs
S_Q(x,y)^{-1} &=& \langle Q(x) \overline{Q}(y)\rangle^{-1}  \nn\\
&=& \left(4+a m_0^f\right)\delta_{xy} 
-\frac{1}{2}\sum_{\mu} \left((1-\gamma_\mu)U_\mu^{(f)}(x)\delta_{x+\mu,y}
+(1+\gamma_\mu)U_\mu^{(f),\dagger}(y)\delta_{x,y+\mu}\right)\,. 
\label{eq:inverse_prop}
\eeqs
By applying the transpose and the charge conjugation operator to $S_Q^{-1}$, we have
\beqs
C^T (S_Q(x,y)^{-1})^T C &=& \left(4+a m_0^f\right)\delta_{xy} -\frac{1}{2}\sum_{\mu} \left((1+\gamma_\mu)U^{(f),T}_\mu(x)\delta_{x+\mu,y}
+(1-\gamma_\mu)U_\mu^{(f),\ast}(y)\delta_{x,y+\mu}\right)\,.
\label{eq:inverse_prop_inter}
\eeqs
Using Eq.~(\ref{eq:symplectic}), we arrive at
\beqs
\Omega^{-1} C^T (S_Q(x,y)^{-1})^T C \Omega &=& \left(4+a m_0^f\right)\delta_{xy}-\frac{1}{2}\sum_{\mu} \left((1+\gamma_\mu)U^{(f),\dagger}_\mu(x)\delta_{x+\mu,y}
+(1-\gamma_\mu)U_\mu^{(f)}(y)\delta_{x,y+\mu}\right) \nn \\
&=& S_Q(y,x)^{-1},
\eeqs
which in turn implies that
\beq
\Omega^{-1} C^T S_Q^T(x,y) C \Omega = 
S_Q(y,x).
\eeq
Using this result, with $\Gamma^1=C\gamma^5$ and $\Gamma^2=\mathbb{1}$, we can rewrite \Eq{cb_corr} as
\beqs
\langle { \mathcal{O}^{k}_{{\rm CB}}}^{\gamma}(x) \overline{\mathcal{O}^{k}_{{\rm CB}}}^{\,\gamma'}(y) \rangle  
&=& 
\Omega_{da} \Omega^{d'a'} \delta^{\gamma\delta} \delta^{\gamma'\delta'}\nonumber
S^{k\,cd}_{\Psi\,\,\,\,\,c'd'\,\delta\delta'}(x,y)\, \times \nn\\
&&
~~~~~~~~~~~~~~~~~~~~~~~\times\Tr_{\hspace{-2pt}s} \left[
S_{Q\,\,\,a'}^{2\,a}(x,y) \gamma^5 \left( \Omega^{-1} C^T \left(S_Q^1(x,y)\right)^T C \Omega \right)^{c'}_{\,\,\,c} \gamma^5
\right] \nonumber \\
&=&
\Omega_{da} \Omega^{d'a'} \delta^{\gamma\delta} \delta^{\gamma'\delta'}
S^{k\,cd}_{\Psi\,\,\,\,\,c'd'\,\delta\delta'}(x,y) \,~\Tr_{\hspace{-2pt}s} \left[
S_{Q\,\,\,a'}^{2\,a}(x,y) \gamma^5 
S^{1\,c'}_{Q\,\,\,c}(y,x)
\gamma^5 \right]. 
\label{eq:cb_dq2}
\eeqs
Comparing Eqs.~(\ref{eq:cb_ps}) and (\ref{eq:cb_dq2}), we conclude that the chimera propagators built out of $\mathcal{O}_{{\rm CB},1(2)}$ and $\mathcal{O}_{{\rm CB},4(5)}$ are identical to one another.

\section{Tables of  numerical results}
\label{Sec:C}

\begin{table}[h]
\caption{%
\label{tab:ensembles}%
Ensembles generated for the numerical study of finite volume effects
reported  in \Sec{fv}. $N_t$ and $N_s$ are the temporal and spatial extents of the lattice, while  $N_{\rm conf}$ and $\delta_{\rm traj}$ denote the number of configurations and the length of the Monte Carlo trajectory between adjacent configurations. In the last column, we show  the average plaquette value $\langle {\cal P} \rangle$. 
}
\begin{center}
\begin{tabular}{|c|c|c|c|c|}
\hline
~~Ensemble~~ & ~~$N_t\times N_s^3$~~ & ~~$N_{conf}$~~ & ~~$\delta_{\rm traj}$~~ & $\langle \cal P \rangle$ \\
\hline
E1 & $36\times8^3$ & 160 & 24 & ~~0.585758(87)~~\\
E2 & $48\times12^3$ & 130 & 24 & 0.585447(51)\\
E3 & $48\times16^3$ & 140 & 20 & 0.585233(34)\\
E4 & $48\times18^3$ & 180 & 12 & 0.585234(22)\\
E5 & $48\times20^3$ & 130 & 12 & 0.585137(20)\\
E6 & $48\times24^3$ & 165 & 8 & 0.585148(13)\\
E7 & $54\times28^3$ & 180 & 12 & 0.585144(11)
\\\hline
\end{tabular}
\end{center}
\end{table}

In this appendix, we tabulate some numerical information relevant to the
discussions in Sections~\ref{Sec:fv}-\ref{Sec:spectrum}. 
The parameters of the lattice theory are  $\beta=6.5$, $am_0^{as}=-1.01$ and $am_0^{f}=-0.71$. 
The baryonic and mesonic observables are measured using point and stochastic wall sources, respectively. The numerical results are presented in lattice units.

In \Tab{ensembles}, we list the details characterising the  ensembles used for our
 investigations of finite volume effects. 
 The ensembles denoted by E6 and E7 are also used for numerical studies of the chimera baryon 
 and the combined spectrum, respectively. 
 We save configurations separated by $\delta_{\rm traj}$ trajectories, 
 after discarding  a sufficient large number of initial trajectories to allow for the thermalisation, 
so that those are independent to each other. 
We determine $\delta_{\rm traj}$ by monitoring the average plaquette values $\langle {\cal  P} \rangle$,
and chose it to be comparable  to one autocorrelation length. 

\begin{table}[h]
\caption{%
\label{tab:finite_V_F}%
Numerical results for the masses and decay constants of pseudoscalar mesons (PS), 
and the masses of vector mesons (V), used to investigate finite volume effects in \Sec{fv}. 
The constituent fermions are in the fundamental representation. 
The pseudoscalar mass at infinite volume, $a m_{\rm PS}^{f,\,{\rm inf}}$,
 is the one extracted from the ensemble with the largest volume, $54\times 28^3$.
}
\begin{center}
\begin{tabular}{|c|c|c|c|c|}
\hline
~~Ensemble~~ & $am_{\rm PS}^f$ & $am_{\rm V}^f$ & $af_{\rm PS}^f$ & $m_{\rm PS}^{f,\,{\rm inf}}\,L$\\
\hline
E1 & ~~$0.7488(64)$~~ & ~~$0.7982(72)$~~ & ~~$0.0349(20)$~~ & ~~$2.8783(76)$~~\\
E2 & $0.5171(48)$ & $0.5685(51)$ & $0.0419(17)$ & $4.317(11)$\\
E3 & $0.3849(45)$ & $0.4238(58)$ & $0.0427(14)$ & $5.757(15)$\\
E4 & $0.3778(22)$ & $0.4290(24)$ & $0.0461(11)$ & $6.476(17)$\\
E5 & $0.3702(16)$ & $0.4142(22)$ & $0.05151(88)$ & $7.196(19)$\\
E6 & $0.3640(19)$ & $0.4067(20)$ & $0.04992(87)$ & $8.635(23)$\\
E7 & $0.35979(95)$ & $0.4009(11)$ & $0.05058(61)$ & $10.074(27)$\\
\hline
\end{tabular}
\end{center}
\end{table}

\begin{table}[h]
\caption{%
\label{tab:finite_V_AS}%
Numerical results for the masses and decay constants of pseudoscalar mesons (PS), 
and the masses of vector mesons (V), used to investigate finite volume effects in \Sec{fv}. 
The constituent fermions are in the antisymmetric representation. 
We also list, in the last column,  the mass of chimera baryons with positive parity. 
}
\begin{center}
\begin{tabular}{|c|c|c|c|c|}
\hline
~~Ensemble~~ & $am_{\rm PS}^{as}$ & $am_{\rm V}^{as}$ & $af_{\rm PS}^{as}$ & $am^+_{\rm CB}$\\
\hline
E1 & ~~$0.4277(53)$~~ & ~~$0.4411(60)$~~ & ~~$0.0843(32)$~~ & ~~$1.012(16)$~~\\
E2 & $0.5499(35)$ & $0.5814(47)$ & $0.0781(20)$ & $0.927(15)$\\
E3 & $0.5858(21)$ & $0.6241(33)$ & $0.0767(15)$ & $0.768(13)$\\
E4 & $0.5956(14)$ & $0.6395(21)$ & $0.0794(12)$ & $0.7974(72)$\\
E5 & $0.6017(10)$ & $0.6491(15)$ & $0.08349(96)$ & $0.7803(60)$\\
E6 & $0.6023(12)$ & $0.6481(14)$ & $0.0805(12)$ & $0.7654(50)$\\
E7 & $0.60205(92)$ & $0.6450(15)$ & $0.08313(88)$ & $0.7636(28)$\\
\hline
\end{tabular}
\end{center}
\end{table}

\begin{table}[h]
\caption{%
\label{tab:multirep_spectrum}%
Numerical results for the masses of mesons in additional spin-0 and spin-1 channels, 
sourced by the interpolating operators  in \Eq{meson_channels}. 
The representation of the constituent fermions
 are denoted by superscripts $f$ and $as$. 
 The measurements are performed on the ensemble with the largest-volume, E7.}
\begin{center}
\begin{tabular}{|c|c|c|c|c|c|c|c|c|}
\hline
~~Ensemble~~ & $am_{\rm T}^{f}$ & $am_{\rm AV}^{f}$ & $am_{\rm AT}^{f}$ & $am^{f}_{\rm S}$ & $am_{\rm T}^{as}$ & $am_{\rm AV}^{as}$ & $am_{\rm AT}^{as}$ & $am^{f}_{\rm S}$\\
\hline
E7 & ~~$0.3995(13)$~~ & ~~$0.544(10)$~~ & ~~$0.543(11)$~~ & ~~$0.508(12)$~~ & ~~$0.6461(14)$~~ & ~~$0.859(12)$~~ & ~~$0.889(14)$~~ & ~~$0.843(11)$~~\\
\hline
\end{tabular}
\end{center}
\end{table}

In Tables~\ref{tab:finite_V_F} and \ref{tab:finite_V_AS}, we present the results of the 
measurements of the masses of the pseudoscalar (PS) and vector (V) mesons composed of 
fermionic constituents in the fundamental and antisymmetric representations, and
the decay constant of the pseudoscalar meson.
We also show the mass of the
 chimera baryon (CB) with positive parity,  and $m_{\rm PS}^{f,\,{\rm inf}}\,L$---$a m_{\rm PS}^{f,\,{\rm inf}}$
 is  extracted from the measurement on the ensemble with the largest available lattice. 
 
 In \Tab{multirep_spectrum}, we present the numerical results for the masses of the other mesons
  in the spin-0 and spin-1 channels, besides to the ones we have 
  already presented in Tables~\ref{tab:finite_V_F} and \ref{tab:finite_V_AS}.
  These are sourced  by the tensor (T), axial-vector (AV), axial-tensor (AT) and scalar (S) 
  interpolating operators defined with the gamma structures in \Eq{meson_channels}. 
  These measurements have been carried out by using ensemble E7, the one that has the largest volume.


\end{document}